\documentstyle[preprint,eqsecnum,aps]{revtex}
\tightenlines

\begin{document}

\newcommand{\beq}{\begin{equation}}
\newcommand{\eeq}{\end{equation}}
\newcommand{\beqa}{\begin{eqnarray}}
\newcommand{\eeqa}{\end{eqnarray}}
\newcommand{\boldsigma}{\mbox{\boldmath$\sigma$}}
\newcommand{\boldtau}{\mbox{\boldmath$\tau$}}
\newcommand{\ott}{\boldtau_i \cdot \boldtau_j}
\newcommand{\oss}{\boldsigma_i\cdot\boldsigma_j}
\newcommand{\os}{S_{ij}}
\newcommand{\ols}{({\bf L\cdot S})_{ij}}
\newcommand{\hovm}{\frac{\hbar^2}{m}}
\newcommand{\hovtm}{\frac{\hbar^2}{2m}}
\newcommand \beqas{\begin{eqnarray*}}
\newcommand \eeqas{\end{eqnarray*}}
\newcommand {\veb}[1]{\bf{#1}}
\newcommand \la{\raisebox{-.5ex}{$\stackrel{<}{\sim}$}}
\newcommand \ga{\raisebox{-.5ex}{$\stackrel{>}{\sim}$}}
\newcommand \degs{^{\circ}}
\newcommand \nuc[2]{$^{#2}$#1}
\newcommand {\msol}{$M_\odot \;$}

\draft

\title{The equation of state of nucleon matter and neutron star structure}

\author{A. Akmal\cite{aa}, V. R. Pandharipande\cite{vrp}
and D. G. Ravenhall\cite{dgr}}
\address{Department of Physics, University of Illinois at Urbana-Champaign,
         1110 W. Green St., \\
Urbana, Illinois 61801}

\date{\today}
\maketitle
\begin{abstract}

Properties of dense nucleon matter and the structure of neutron stars 
are studied using variational chain summation methods and
the new Argonne $v_{18}$ two-nucleon interaction, which 
provides an excellent fit to all of the nucleon-nucleon scattering data 
in the Nijmegen data base. 
The neutron star gravitational mass limit obtained with this 
interaction is 1.67\msol.
Boost corrections to the two-nucleon interaction, which give the leading
relativistic effect of order $(v/c)^2$, as well as three-nucleon 
interactions, are also included in the nuclear Hamiltonian.  Their 
successive addition increases the mass limit to 1.80 and 2.20 \msol. 
Hamiltonians including a three-nucleon interaction predict a transition
in neutron star matter to a phase with neutral pion condensation at a
baryon number density of $\sim 0.2$ fm$^{-3}$.  
Neutron stars predicted by these 
Hamiltonians have a layer with a thickness on the order of tens of
meters, over which the density changes rapidly from that of the normal
to the condensed phase.  The material in this thin layer is a mixture
of the two phases.  We also investigate the possibility of dense nucleon 
matter having an admixture of quark matter, described 
using the bag model equation
of state. Neutron stars of 1.4\msol do not appear to have
quark matter admixtures in their cores. However, the heaviest stars
are predicted to have cores consisting of a quark and 
nucleon matter mixture.  These
admixtures reduce the maximum mass of neutron stars from 2.20
to 2.02 (1.91)~\msol for bag constant 
$B = 200$ (122) MeV/fm$^3$.  Stars with 
pure quark matter in their cores are found to be unstable.  We also 
consider the possibility that matter is maximally incompressible above 
an assumed density, and show that realistic models of nuclear forces 
limit the maximum mass of neutron stars to be below 2.5\msol.  
The effects of the phase transitions on the composition of neutron star
matter and its adiabatic index $\Gamma$ are discussed. 

\end{abstract}

\newpage

\section{INTRODUCTION}

The significant influence of nuclear forces on neutron star structure
is by now firmly established by a large body of theoretical and observational
evidence \cite{st}.
In the absence of these forces, the maximum possible mass of
neutron stars composed of non-interacting neutrons is $\sim 0.7$ solar
masses (\msol) \cite{OV39}. Since most observed neutron star masses 
are above 1.3\msol \cite{nstarm}, they must be supported against
gravitational collapse by pressure originating from nuclear forces.
In the present work, we study neutron star structure using one
of the most realistic models of nuclear forces currently available. A
brief outline of previous calculations leading to this work is presented
below.

Shortly after the discovery of pulsars, calculations of the equation of state 
(EOS) of neutron star matter with realistic models of the two nucleon 
interaction (NNI), obtained by fitting the nucleon-nucleon (NN) scattering data 
then available, were carried out using the lowest order constrained 
variational method \cite{Pan71,BJ74}.  The results demonstrated that nuclear 
forces increase the mass limit of stable neutron stars beyond 1.4\msol.

By the late 1970s it had become clear that the NNI alone could not
account for the properties of nuclear matter or few-body nuclei.
Variational \cite{PW79} and Brueckner calculations \cite{DW85},
including higher order cluster contributions, established that
nuclear matter with realistic NNI saturates at too high a density.
In addition, these interactions were known to underbind $^3$H.
Ignoring the latter problem, plausible density
dependent terms were added to the Urbana $v_{14}$ (U14) model of NNI
\cite{LP81a,LP81b} to reproduce the observed equilibrium properties
of nuclear matter.  The resulting density dependent (U14-DDI) model of
nuclear forces predicted stable neutron stars having masses up to 1.8\msol
\cite{FP81,LRP93}.

Since nucleons are made up of quarks and have internal degrees of freedom, 
we can expect interactions among three (and perhaps four or more) nucleons,
in addition to the NNI.  
The Urbana three nucleon interaction (TNI) models contain only 
two terms, with strengths fixed by the saturation density of nuclear 
matter and the binding energy of $^3$H.  Wiringa, Fiks and Fabrocini (WFF) 
\cite{WFF88} used the U14 and the subsequent Argonne $v_{14}$ 
(A14) \cite{WSA84} models of NNI, 
together with the Urbana VII (UVII) model of TNI, to study neutron star 
structure and obtained mass limits of 2.19\msol and 2.13\msol
with the U14+UVII and A14+UVII, respectively. 
They also found that pure neutron matter (PNM) undergoes a 
transition to a phase having spin-isospin order, attributed to neutral pion 
condensation, at a density 
of $\sim 0.2$ fm$^{-3}$ with the A14+UVII,
but not with the U14+UVII. Neither
of these models results in a phase transition in symmetric nuclear 
matter (SNM), which is composed of equal numbers of neutrons and protons.

In the early 1990s the Nijmegen group \cite{SKR93} examined carefully all 
NN scattering data at energies below 350 MeV published between 1955 
and 1992.  They extracted 1787 proton-proton (pp) and 2514 proton-neutron (np) 
``reliable'' data, and demonstrated that these data determine
all NN phase shifts and mixing parameters quite accurately.  
The NNI models which fit this Nijmegen
data base with a $\chi^2/N_{data} \sim 1$ are called ``modern''.  These 
include the Nijmegen models \cite{SKT94}: Nijmegen I, II and Reid-93, 
the Argonne $v_{18}$ \cite{WSS95}, denoted here by A18, and the CD-Bonn 
\cite{MSS96}.  In order to fit the pp and np data simultaneously and 
accurately, these models include a detailed description of the electromagnetic 
interactions and terms that break the isospin symmetry of nuclear forces. 
All include the long range one-pion exchange potential, but follow
different treatments of the intermediate and short range parts of the NNI.
The differences among the predictions of these models for the properties 
of many-body systems are much smaller than those among the predictions of 
older models, presumably because all modern potentials 
accurately fit the same scattering data.  
For example, the $^3$H binding energies predicted by the modern 
Nijmegen models and A18 are between -7.62 to -7.72 MeV \cite{FPS93}, 
while that of CD-Bonn is -8.00 MeV \cite{MSS96}.  The difference between 
these results and the experimental value of -8.48 MeV
is used to fix one of the parameters of Urbana TNI models.

Detailed studies of the energies of dense nucleon matter were carried out 
recently by Engvik {\em et al.} \cite{Eng97} using all the modern models of NNI 
and the lowest order Brueckner (LOB) method.
According to these studies,
the results with the modern potentials are all quite similar up to densities 
relevant to neutron stars.  For example, the energies predicted with 
the LOB method for the energy of neutron matter at 5$\rho_0$, where 
$\rho_0 = 0.16$ fm$^{-3}$ is the equilibrium (saturation)
density of nuclear matter, 
range from 80 to 93 MeV per nucleon.  The spread of 13 MeV in these 
energies is small compared to the possible errors in the LOB method and the 
expected contributions of TNI at this density.  This model independence 
results from the fact that the mean 
interparticle distance at $\rho \sim 5\rho_0$ is greater than 1~fm, 
and the predicted matter energy is therefore not sensitive to the 
details of the interaction at $r_{ij} < 1$~fm.

In this paper we study the structure of neutron stars with the A18 model 
using variational chain summation (VCS) methods, which  hopefully include
all leading many-body correlation effects.  
The Urbana model IX (UIX) \cite{Pud95} 
is used to estimate the effect of TNI.  Previous studies of nucleon 
matter with A18 and UIX interactions have indicated the possibility 
of a transition to a neutral pion condensed phase for both PNM and SNM 
\cite{AP97}.  The effects of such a transition on the structure of neutron 
stars are studied here in detail.  The effect of relativistic boost 
corrections \cite{FPF95} to the A18 interaction is also examined. At 
high densities, we consider the possibility of matter
becoming maximally incompressible, as well as that of a transition to 
mixed phases of quark and nucleon matter \cite{Gle92,HPS93}.

The relativistic mean field (RMF) approximation \cite{Wal74} 
has been used in many studies of high density matter and neutron stars.  
There exists a vast amount of literature on this topic, some of which has been 
reviewed by Glendenning \cite{Gle97}.  While the RMF approximation is 
very elegant and pedagogically useful, 
it is not valid in the context of what is 
known about nuclear forces, which is the theme of this work. 
For example, using the meson parameters of the CD-Bonn S-wave potentials 
in the RMF approximation leads to unbound SNM. (In order to accurately fit 
the NN scattering data, the phenomenological scalar meson parameters in 
the CD-Bonn model are allowed to depend on partial wave quantum numbers.)
At a density of $\rho_0$, for SNM, the RMF approximation yields 
an energy per nucleon of $\sim +20$ MeV, 
while the LOB method gives $\sim -18$ MeV.  In the 
mean field Hartree approximation, which is implicit in the RMF calculations, 
the A18 NNI gives energies per nucleon of $\sim$ +30 (+37)~MeV 
at $\rho_0$ and +155 (+204)~MeV 
at 5$\rho_0$ for SNM (PNM), while the variational calculations presented here 
give $\sim$ -18 (+12) and +25 (+88) MeV, respectively.

The main problem is that the mean field approximation for meson fields is 
only valid for $\mu \bar{r}\ll 1$, 
where $\mu$ is the inverse Compton wave length 
of the meson and $\bar{r}$ is the mean interparticle spacing. 
Over the 1-5 $\rho_0$ density range, $\bar{r}$ estimated using a
body centered cubic lattice ranges from 2 to 1.2 fm.  Thus $\mu \bar{r}$ 
is in the range 1.4 to 0.8 for the pion and 7.8 to 4.7 for vector mesons. 
The mean field approximation is not applicable, since these values are 
obviously far from being much smaller than one.  The RMF approximation 
can be based on effective values of the coupling constants that take
into account the correlation effects.  
However, these coupling 
constants then have a density dependence, and a microscopic theory is 
needed to calculate them.

This paper is organized as follows.  Section II contains a summary of the 
non-relativistic calculations with A18 and A18+UIX models of nuclear 
forces, while section III describes the calculations including the 
relativistic boost interaction, denoted by $\delta v$, without and with 
the TNI model UIX$^*$.  The beta equilibrium of neutron star matter is 
discussed in section IV and results for neutron star structure are presented
in section V, where we also discuss the effects of the possible transition 
to mixed nucleon and quark matter phases.  The adiabatic index and 
sound velocities in neutron star matter are given in section VI, and 
conclusions are presented in section VII.

\section{NONRELATIVISTIC CALCULATIONS}

Nonrelativistic calculations of SNM and PNM with the 
A18 and UIX interactions were carried out using
Variational Chain Summation (VCS) techniques
described in detail in \cite{AP97}. 
Energies are calculated by evaluating the expectation value of
the Urbana-Argonne Hamiltonian with a variational wavefunction
composed of a product of pair correlation operators acting on a
Fermi gas wavefunction. The pair correlation operators are written
as a sum of eight radial correlation functions, each multiplied by
one of the two-body operators: 
$\left(1,\oss,\os,\ols\right)\otimes\left(1,\ott\right)$.
The wavefunction depends
on three variational parameters:
the range of the tensor correlations, $d_t$, the range of all other
correlations, $d_c$ and a quenching parameter $\alpha$,
meant to simulate medium effects.
In this section we discuss results obtained for four cases, namely
SNM and PNM with and without the three-nucleon interaction, in order to indicate
their sensitivity to various terms in the nuclear force, 
and to extend them to higher densities, beyond the range
covered in \cite{AP97}.

The optimum values of the parameters $d_t$ and $d_c$ in matter
without and with the three-nucleon interaction are shown in
Fig.~\ref{fig:fig1}. Some of the noise
in the variation of these parameters with matter density is due to the
insensitivity of the energy to their values at the variational minimum.
The large increase in $d_t$ of PNM without $V_{ijk}$ at
$\rho\sim 0.5$~fm$^{-3}$
is due to a transition to  a phase with $\pi^0$-condensation 
as discussed in \cite{AP97}.
With $V_{ijk}$ there are sudden changes in the $d_c$ and $d_t$
of PNM at $\rho\sim 0.2$~fm$^{-3}$, and in the $d_t$ of SNM at
$\rho\sim 0.32$~fm$^{-3}$. These are associated with the same
phase transition. We note that the present variational 
wavefunction is not fully adequate to describe the long range order
in the $\pi^0$-condensed phases. However, since the change in the energy
due to the $\pi^0$-condensation appears to be small, we expect our
estimates of the energies of matter with $\pi^0$-condensation
to be useful.

Our nonrelativistic Hamiltonian, $H_{NR}$, is comprised of the 
nonrelativisic kinetic energies  and the two-body A18 and three-body UIX 
interactions. The NNI
includes a static, long-range one-pion exchange part, with
short-range cutoff, and phenomenological intermediate
and short-range parts, which depend on the six static
two-body operators $(1,\oss,\os)\otimes(1,\ott)$ and eight
momentum-dependent (MD) two-body operators:
$(\ols,L^2_{ij},\oss L^2_{ij},\ols^2) \otimes(1,\ott)$.
The A18 includes an additional isovector
operator, and three isotensor operators, which distinguish between
pp, np and nn interactions. 
These isovector and isotensor terms are small, and give zero
contribution to the energy of SNM to first order. They are therefore
neglected in SNM calculations. In the case of PNM all
the isospin operators can be eliminated and the full A18
with isovector and isotensor terms becomes the sum of a static part
with operators $(1,\oss,\os)$, and a MD part with operators
$\ols,L^2_{ij},\oss L^2_{ij}$ and $\ols^2$.
The UIX  model of $V_{ijk}$ contains two static terms; 
the two-pion exchange Fujita-Miyazawa interaction,
$V^{2\pi}_{ijk}$, and a phenomenological, intermediate range
repulsion $V^R_{ijk}$. 
The strength of the $V^{2\pi}_{ijk}$ interaction was determined
by reproducing the binding energy of the triton
via Greens Function Monte Carlo (GFMC) calculations 
\cite{Pud95}, while that of $V^R_{ijk}$ was adjusted to reproduce
the saturation density of SNM.

Expectation values of the various interactions are calculated in
the VCS framework by summing terms in their
cluster expansions. The one-body, two-body and many-body 
contributions to the kinetic and NNI energies 
are listed in Tables~\ref{tab:t1}-\ref{tab:t4}.
The one-body cluster contribution includes only the Fermi gas
kinetic energy, $T_F$. The remainder of the kinetic energy
is separated into the contribution from the two-body
cluster $\langle T\rangle_{2B}$ and that coming from the many-body clusters,
$\langle T\rangle_{MB}$.
The kinetic energy can be calculated using different expressions
related by integration by parts. If all MB contributions are calculated,
these expressions yield the same result. However, they yield different
results when only selected parts of the MB clusters are summed by VCS
techniques. We have calculated the many-body kinetic energy using 
expressions due to Pandharipande and Bethe (PB), and to
Jackson and Feenberg (JF) \cite{Zab77}.
The averages of the PB and JF results appear under the column 
$\langle T\rangle_{MB}$, and the differences $T_{PB}-T_{JF}$
are listed under $\Delta T$.
In studies of atomic helium liquids the exact energies, calculated via
Monte Carlo (MC) methods, lie between the PB and JF values evaluated using
VCS methods \cite{UFP82}. Although exact MC calculations are as yet 
not practical for nucleon matter, we believe the average of the
two expressions to be more accurate than either one, and that the difference
provides a measure of the uncertainty in the many-body calculation.
The $\langle T\rangle_{MB}$ is quite small in nucleon matter, due
to cancellations between various many-body terms, and therefore the
difference $\Delta T$ is a better indication of this uncertainty.
 
The two-body cluster contribution to the static and MD parts of the 
NNI energy are listed
under $\langle v^s \rangle_{2B}$, and $\langle v^{MD} \rangle_{2B}$. 
The $\langle v^s \rangle_{2B}$ is negative and large enough to bind
SNM, though not PNM. 
The $\langle v^{MD} \rangle_{2B}$ increases 
rapidly with density, and is proportional to $\rho T_F$ at $\rho > 
2 \rho_0$.
The many-body contributions to the static and MD parts of the
NNI energy are listed separately as
$\langle v^s\rangle_{MB}$ and $\langle v^{MD}\rangle_{MB}$.
Previously, great efforts were made to improve upon the 
accuracy of the calculation of $\langle v^s\rangle_{MB}$ \cite{PW79,Wir83};
however, at large densities the $\langle v^{MD} \rangle_{MB}$ 
grows rapidly, and becomes larger in magnitude than 
$\langle v^s\rangle_{MB}$.
The MD contribution is 
more difficult to calculate because of the gradients in $v^{MD}_{ij}$, 
which may operate on the correlations of nucleons $i$ and $j$ with other 
nucleons.  All the leading terms are calculated as discussed in 
\cite{AP97}, and the corresponding errors should therefore be much smaller 
than the reported values.
At higher densities the $\langle v^{MD} \rangle_{MB}$ becomes proportional 
to $\rho \langle T \rangle_{2B}$, as expected.
An additional perturbative correction to the two-body energy  is listed as 
$\delta E_{2b}$. This small correction is due to an improvement in the
variational wavefunction, which occurs when correlation functions
are calculated separately in each $l,S,J$ channel \cite{AP97}.

The expectation values of $v^{\pi}_{ij}$, $v^R_{ij}$, $V^{2\pi}_{ijk}$ and 
$V^R_{ijk}$ for SNM and PNM are listed in Table~\ref{tab:t5}.  Here, $v^R_{ij}$ 
is the phenomenological part, $v_{ij}-v^{\pi}_{ij}$, of the NNI.
Since the UIX $V_{ijk}$ is purely static, the error in the calculation of its 
expectation value is likely to be small.  In SNM, with only the A18 
interaction, the $v^{\pi}$ gives more than half of the total NNI 
energy at all densities considered.  The corresponding calculation
with A18+UIX interactions, shows a significant increase in 
the magnitudes of the negative contribution of  
$V^{2\pi}_{ijk}$ between $\rho = 0.32$ and 0.4 fm$^{-3}$, associated with 
pion condensation.  In PNM these pion exchange interactions make
relatively small contributions at densities below the phase transition, 
occurring at $\rho \sim 0.2$ (0.5) fm$^{-3}$ with (without) UIX.
However, at the densities above the 
transition they make large negative contributions comparable to those in 
SNM.  At the highest densities the contributions of $V^{2\pi}_{ijk}$ and 
$V^R_{ijk}$ become very large, and the validity of this 
purely nonrelativistic approach becomes 
questionable.  As discussed in the following section, approximately
40 \% of the 
contribution of $V^R_{ijk}$ is due to relativistic boost corrections to 
the NNI , and a more plausible theory is therefore obtained by 
removing this boost contribution from the UIX interaction.  

The total energies, calculated in the manner previously described for SNM 
and PNM, appear
in Tables~\ref{tab:t6}-\ref{tab:t7}, and in Figs.~\ref{fig:fig2} 
and \ref{fig:fig3}.
The pronounced kink in the $E(\rho)$ of SNM with $V_{ijk}$,
at $\rho=0.32$~fm$^{-3}$, is due to the 
phase transition; the corresponding feature in PNM
is a somewhat more subtle change in the slope of the curve at
$\rho=0.2$~fm$^{-3}$.

It is evident from the $E(\rho)$ figures for SNM that without the 
$V_{ijk}$, the present calculation cannot explain the empirical 
saturation density $\rho_0$, of nuclear matter.  As previously noted, the 
strength of $V^R_{ijk}$ is adjusted to obtain the correct 
equilibrium density in calculations with $V_{ijk}$. However, 
the present calculations with $V_{ijk}$ underbind 
SNM at saturation density, giving $E(\rho_0) \sim -12$~MeV per 
nucleon instead of the empirical value of $\sim -16$~MeV.
This discrepancy is presumably due to use of imperfect 
variational wavefunctions,
which do not include, for example, three- and higher-body correlations.
It is known from comparison of the results of variational Monte Carlo (VMC)
and exact GFMC calculations \cite{Pud97,Wir98a}, 
that variational wave functions of the present form
underbind the light p-shell nuclei. The variational energy 
of $^8$Be for example, is above the exact GFMC result by $\sim$~12~\%, 
even after incorporating into the wave function
some of the three-body correlations, which we have neglected here.

\section{RELATIVISTC BOOST CORRECTION TO THE NN INTERACTION}

In all analyses, the NN scattering data is reduced to 
the center of mass frame and 
fitted using phase shifts calculated from the NNI, $v_{ij}$, 
in that frame.  The $v_{ij}$ obtained by this procedure 
describes the NN interaction in that frame, in which the total momentum 
${\bf P}_{ij} = {\bf p}_i + {\bf p}_j$, is zero.  In general, the 
interaction between particles depends upon their total momentum,  
and can be written as
\begin{equation}
v({\bf P}_{ij}) = v_{ij} + {\delta}v({\bf P}_{ij}),
\label{eq:vofp}
\end{equation}
where $v_{ij}$ is the interaction for ${\bf P}_{ij} = 0$, and 
${\delta}v({\bf P}_{ij})$ is the boost interaction \cite{FPF95} which
is zero when ${\bf P}_{ij} = 0$.

It is useful to consider a familiar example.  The Coulomb-Breit 
electromagnetic interaction 
\cite {BS57} between two particles of mass {\em m} and charge {\em Q},
ignoring spin dependent terms for brevity, is given by
\begin{equation}
v({\bf p}_i,{\bf p}_j) = 
\frac{Q^2}{r_{ij}} \left( 1 - \frac{{\bf p}_i \cdot {\bf p}_j}{2m^2} 
- \frac{{\bf p}_i \cdot {\bf r}_{ij} {\bf p}_j \cdot {\bf r}_{ij}}
{2 m^2 r^2_{ij}}\right),
\label{eq:vemt}
\end{equation}
up to terms quadratic in the velocities of the interacting particles. 
In our notation it is expressed as
\beq
v({\bf p}_i,{\bf p}_j)  =  v_{ij} + {\delta}v({\bf P}_{ij}),
\label{eq:vemg}
\eeq
with
\beqa
v_{ij} & = & \frac{Q^2}{r_{ij}} \left( 1 + \frac{p^2_{ij}}{2m^2} 
+ \frac{({\bf p}_{ij} \cdot {\bf r}_{ij})^2}{2 m^2 r^2_{ij}}\right),
\label{eq:vemtd} \\
{\delta}v({\bf P}_{ij}) & = & -\ \frac{Q^2}{r_{ij}} \left( 
\frac{P^2_{ij}}{8m^2} + \frac{({\bf P}_{ij} \cdot {\bf r}_{ij})^2}
{8 m^2 r^2_{ij}}\right),
\label{eq:vemb}
\eeqa
where ${\bf p}_{ij} = ({\bf p}_i - {\bf p}_j)/2$ is the relative momentum. 

In all realistic models of $v_{ij}$, such as the A18, the dependence on
${\bf p}_{ij}$ is included in the momentum-dependent part of the
interaction, $v^{MD}_{ij}$.
However, we have neglected the ${\delta}v({\bf P}_{ij})$ 
in the calculations presented in the previous section.  Even though 
contributions of the boost interaction to the binding energy of SNM and
$^3$H were estimated by Coester and coworkers years ago \cite{CPS74,GLC86}, 
these contributions have been neglected in most subsequent 
studies of dense matter.

Following the work of Krajcik and Foldy \cite{KF74}, Friar \cite{Fri75} 
obtained the following equation relating the boost interaction of order
$P^2$ to the interaction in the center of mass frame:
\begin{equation}
{\delta}v({\bf P}) = -\frac{P^2}{8m^2} v 
+\frac{1}{8m^2}[{\bf P \cdot r \; P \cdot {\nabla}},v] 
+\frac{1}{8m^2}[({\bf \sigma}_i - {\bf \sigma}_j) \times 
{\bf P \cdot \nabla}, v] .
\label{eq:friar}
\end{equation}
The general validity of this equation in relativistic mechanics and field 
theory was recently discussed \cite{FPF95}. 
Incorporating the boost into the interaction yields a nonrelativistic 
Hamiltonian of the form:
\begin{equation}
H^\ast_{NR} = \sum \frac{p^2_i}{2m} + \sum (v_{ij} + {\delta}v({\bf P}_{ij})) 
+ \sum V^*_{ijk} + \cdots ,
\label{eq:nrhwb}
\end{equation}
where the ellipsis denotes the three-body boost, and four and higher body 
interactions. This $H^\ast_{NR}$ contains all terms quadratic in the particle 
velocities, and is therefore suitable for complete studies in the 
nonrelativistic limit.

Studies of light nuclei using the VMC method \cite{CPS93,For98} find that
the contribution of the two-body boost interaction to the energy is
repulsive, with a magnitude which is 37\% of the $V^R_{ijk}$ contribution.
The boost interaction thus accounts for a significant part of
the $V^R_{ijk}$ in Hamiltonians which fit nuclear energies
neglecting ${\delta}v$.

In the present calculations we keep only the terms of the boost 
interaction associated with the static part of $v_{ij}$, and neglect the
last term in Eq.~(\ref{eq:friar}).  That term is responsible for Thomas 
precession and quantum contributions that are negligibly 
small here \cite{FPC95}.
Our $\delta v$ is given by
\begin{equation}
{\delta}v({\bf P}) = -\frac{P^2}{8m^2}v^s +\frac{1}{8m^2}
{\bf P \cdot r}\ {\bf P \cdot \nabla} v^s.
\label{eq:afriar}
\end{equation}
The two terms are due to the relativistic energy expression and Lorentz 
contraction, and are denoted $\delta v^{RE}$ and $\delta v^{LC}$,
respectively.  The three-nucleon 
interaction used in the $H^\ast_{NR}$ Eq.~(\ref{eq:nrhwb})
is denoted by $V^\ast_{ijk}$.  Its parameters are obtained by fitting the 
binding energies of $^3$H and $^4$He, and the equilibrium density of 
SNM, including $\delta v$. The strength of $V^{R\ast}_{ijk}$ is 0.63 times that
of $V^R_{ijk}$ in UIX, while that of $V^{2 \pi}_{ijk}$ is unchanged.  The 
resulting model of $V_{ijk}$ is called UIX$^\ast$.

The approximate Hamiltonian $H_{NR}$, containing A18 and UIX interactions 
without $\delta v$, and the more correct Hamiltonian $H^\ast_{NR}$, containing 
A18, UIX$^\ast$ and $\delta v$ interactions, yield very similar results 
for light nuclei up to $^8$Be \cite{For98} and for SNM up to
equilibrium density.
However, the two models differ at higher densities, 
since the contributions of $\delta v$ and $V^R_{ijk}$ have different 
density dependences.

One may also consider relativistic nuclear Hamiltonians of the type
\begin{equation}
H_R = \sum \sqrt{p^2_i + m^2} + \sum ({\tilde v}_{ij} + 
{\delta}v({\bf P}_{ij})) + \sum {\tilde V}_{ijk} + \cdots  ,
\label{eq:rhwb}
\end{equation}
which require re-fitting the two-nucleon scattering data 
to determine the two-body interaction,
${\tilde v}_{ij}$, using relativistic kinetic energies 
\cite{CPS93}. In light nuclei, the $\delta v$ contribution accounts for 
most of the difference between the energies obtained with $H_R$ and $H_{NR}$,
since the difference between the contributions of the nonrelativistic and 
relativistic kinetic energies is largely cancelled 
by the difference in interaction energy contributions from $v_{ij}$ 
and ${\tilde v}_{ij}$.  
The results obtained with $H^\ast_{NR}$ are very
close to those from $H_R$, indicating that the former 
represents a significant improvement over $H_{NR}$ 

\subsection{Calculation of Boost Interaction Energy}

The relativistic boost contributions are calculated by evaluating 
terms in the cluster expansion of $\langle \delta v^{RE}_{ij}\rangle$
and $\langle \delta v^{LC}_{ij}\rangle$. 
In addition to the dominant two-body cluster, we have calculated
dressed three-body separable diagrams and central chain diagrams.

In the case of the two-body cluster, the gradients in the center of
mass momentum operator, $P_{ij}=-i(\nabla_i+\nabla_j)$, can act only
on the Fermi gas part of the wavefunction, since the correlations
$f_{ij}$ depend only on the relative coordinate.
Thus, the two-body cluster contribution to $\langle\delta v^{RE}_{ij}\rangle$ 
is 
\beqa
C_{2B}^{RE}(dir) &=& -\frac{k_F^2}{8m^2}\frac{3}{5}\rho\sum_{pmp^\prime}
\int d^3r_{ij} \left(f^pv^mf^{p^\prime}\right)_{ij} 
C\left(O^p_{ij}O^m_{ij}O^{p^\prime}_{ij}\right) \\
\label{eq:c2bre}
C_{2B}^{RE}(ex)  &=& \frac{1}{8m^2}\frac{\rho}{s}\sum_{pmp^\prime n}
\int d^3r_{ij} \left(l^{\prime 2}-l\nabla^2l\right)_{ij}
\left(f^pv^mf^{p^\prime}\right)_{ij}
C\left(O^{n}_{ij}O^p_{ij}O^m_{ij}O^{p^\prime}_{ij}\right).
\eeqa
The quantities $C(\cdots)$ in the integrand of these expressions 
represent the spin-isospin-independent
part (or ``C-part'') of the operator product enclosed by the parentheses.
Only the C-part of operator products appear in the cluster integrals,
since the energy expectation value requires a sum over all possible
$\sigma_z$ and $\tau_z$ \cite{PW79}, which average to zero in 
isotropic matter.  In the case of SNM the
indices {\em p,m,p$^\prime$} run over the first six operators, as
we boost only the static interactions and
consider only static correlations in this calculation. 
The index {\em n}=1,4 comes from the exchange operator. 
In the case of PNM the $\ott$ operators are eliminated from the 
Hamiltonian, thus the indices $p,m,p^{\prime}$ and $n$ can 
represent only unit or spin-dependent operators.
The exchange operator also contributes a factor of 1/{\em s}, where
{\em s} is the degeneracy of the system (4 for SNM, 2 for PNM.)

The $C_{2B}^{RE}(dir)$ differs from the corresponding
expression for the two-body
direct contribution to $\langle v^s_{ij}\rangle$ by a factor 
$-(1/8m^2)(6k_F^2/5)$, which is the expectation value of
$-P^2_{ij}/(8m^2)$ in the Fermi gas. The exchange part of the cluster,
$C_{2B}^{RE}(ex)$, has the same form as the corresponding expression
for $\langle v^s_{ij}\rangle_{2B}(ex)$, with $l^2$ replaced by
$-(2)/(8m^2)(l^{\prime2}-l\nabla^2l)$,
where $l\equiv l(k_Fr)$ is the Slater function.
This expression results from from gradients in $P^2_{ij}$
acting on the plane waves
\beqa
\frac{1}{A\Omega}\sum_{ij} 
e^{-i\left( {\bf k}_i\cdot {\bf r}_j+{\bf k}_j\cdot {\bf r}_i\right)}
\left(-\nabla^2_i-\nabla^2_j-2\nabla_i\cdot\nabla_j\right)
e^{+i\left( {\bf k}_i\cdot {\bf r}_i+{\bf k}_j\cdot {\bf r}_j\right)}
= \nonumber \\
& &\!\!\!\!\!\!\!\!\!\!\!\!\!\!\!\!\!\!\!\!\!\!\!\!\!\!\!\!\!\!\!
2\rho\left[\left(l^\prime(k_Fr_{ij})\right)^2-l(k_Fr_{ij})
\nabla^2 l(k_F r_{ij})\right].
\eeqa
Here, {\em A} is the number of nucleons, $\Omega$ is the normalization
volume, and since we are in the thermodynamic limit, $A/\Omega=\rho$.

The two-body cluster contribution to $\langle\delta v^{LC}_{ij}\rangle$ is
\beqa
C_{2B}^{LC}(dir) &=& \frac{k_F^2}{8m^2}\frac{\rho}{5}\sum_{pmp^\prime}
\int d^3r_{ij} \left(f^p r\frac{dv^m}{dr}f^{p^\prime}\right)_{ij} 
C\left(O^p_{ij}O^m_{ij}O^{p^\prime}_{ij}\right) \\
C_{2B}^{LC}(ex)  &=&-\frac{1}{8m^2}\frac{\rho}{s}\sum_{pmp^\prime n}
\int d^3r_{ij} \left(l^{\prime 2}-ll^{\prime\prime}\right)_{ij}
\left(f^p r\frac{dv^m}{dr}f^{p^\prime}\right)_{ij}
C\left(O^{n}_{ij}O^p_{ij}O^m_{ij}O^{p^\prime}_{ij}\right),
\eeqa
which is simply the cluster contribution of a nonrelativistic
potential $r\; dv/dr$, with the direct term multiplied by
$(2/5)(k_F^2/8m^2)$, and with the $l^2$
in the exchange term replaced by 
$(2/8m^2)(l^{\prime2}-ll^{\prime\prime})$.
As in the $\delta v^{RE}$ case, the extra factors in
$\langle\delta v^{LC}\rangle$ result from the gradients
in $\delta v^{LC}$ acting on the Fermi gas part of
the wave function.

The three-body separable diagrams represent the most
significant many-body cluster contributions to the boost
energy. The direct term for a boost $\delta v^x$, where
$x=RE$ or $LC$, has the form:
\beqa
&&\frac{1}{A\Omega^3}\sum_{ijk}\sum_{pmp^\prime qq^\prime}\int d^3r_k\ d^3r_{ij}
\ d^3R_{ij} e^{-i\left({\bf k}_k\cdot {\bf r}_k+{\bf k}_{ij}\cdot {\bf r}_{ij}
+{\bf K}_{ij}\cdot {\bf R}_{ij}\right)} \nonumber \\  
&&C\left(\frac{1}{4}
\left\{f^p_{ij}O^p_{ij},f^q_{ik}O^q_{ik}\right\}
\left(\delta v^x_{ij}\right)^mO^m_{ij}
\left\{f^{p^\prime}_{ij}O^{p^\prime}_{ij},f^{q^\prime}_{ik}O^{q^\prime}_{ik}
\right\} 
-\left(f^pO^p\left(\delta v^x\right)^mf^{p^\prime}O^{p^\prime}\right)_{ij}
\left(f^qO^qf^{q^\prime}O^{q^\prime}\right)_{ik}
\right) \nonumber \\ 
&&e^{i\left({\bf k}_k\cdot {\bf r}_k+{\bf k}_{ij}\cdot {\bf r}_{ij}
+{\bf K}_{ij}\cdot {\bf R}_{ij}\right)}.
\eeqa

The plane waves are written here in terms of the relative momentum,
${\bf k}_{ij}=({\bf k}_i-{\bf k}_j)/2$, and the center of mass momentum,
${\bf K}_{ij}={\bf k}_i+{\bf k}_j$, of the interacting pair.
The interacting exchange $(ij-ex)$ and the passive exchange $(ik-ex)$
expressions are obtained from the above by inserting the
appropriate exchange operators ($(1/s)\sum_{n=1,4} O^{n}_{ij}$
or $(1/s)\sum_{n=1,4} O^{n}_{ik}$) to the far left of
each operator product, and replacing the first plane wave product by
$e^{-i\left({\bf k}_k\cdot {\bf r}_k-{\bf k}_{ij}\cdot {\bf r}_{ij}
+{\bf K}_{ij}\cdot {\bf R}_{ij}\right)}$
or
$e^{-i\left({\bf k}_i\cdot {\bf r}_k-{\bf k}_{jk}\cdot {\bf r}_{ij}
+{\bf K}_{jk}\cdot {\bf R}_{ij}\right)}$.

Following the notation used in the calculation of the MD interaction 
energy \cite{LP80,AP97}, separable diagrams are
classified as K-diagrams and F-diagrams. The
former have gradients in $\delta v^x$ acting on the Fermi gas part
of the wavefunction, and the latter have them
acting on the correlation operators $F_{ik}$. 
As with the two-body cluster contribution $C_{2B}^x$, the K-diagrams
depend linearly on the Fermi kinetic energy. While $C_{2B}^x$ scales as
$\rho \; T_F$, like the $\langle v^{MD} \rangle_{2B}$, 
the K-diagram contributions scale roughly as $\rho^2 T_F$.
The K-diagrams generally make only small
contributions to $\langle\delta v^x\rangle$, the major separable 
contributions coming from F-diagrams. 
The relatively large contribution of the F-diagrams, versus the
K-diagrams, can be understood in the following way. The correlated
particle $k$ in the separable diagram modifies the
center of mass momentum of the interacting pair $ij$ via $F_{ik}$, 
thus enhancing the boost correction. As the form of the
F-diagram integrals suggest, we find that their contributions
exhibit the same scaling behavior as $\langle v^{MD} \rangle_{MB}$, 
namely as $\rho \langle T\rangle_{2B}$.

K-diagram contributions to $\langle\delta v^{RE}_{ij}\rangle$ have been
evaluated for the direct three-body separable diagram and the
interacting exchange diagram. These contributions factorize
into an integral over $r_{ij}$, which is simply the corresponding 
two-body boost diagram, and an integral over $r_{ik}$.
The latter integral is a so-called single-loop vertex correction,
which is included in the more general vertex correction, $M_{dd}-1$,
to $\langle v^s_{ij}\rangle$, defined in \cite{PW79}.
The direct K-diagram contribution is
\beqa
W^{RE}_s(K, dir)=-\frac{k_F^2}{8m^2}\frac{3}{5}\rho\sum_{pmp^\prime q}
\int d^3r_{ij} \left(f^pv^mf^{p^\prime}\right)_{ij} K^{pmp^\prime}A^{p^\prime}
\nonumber \\
& &\!\!\!\!\!\!\!\!\!\!\!\!\!\!\!\!\!\!\!\!\!\!\!\!\!\!\!\!\!\!\!
A^q\frac{\rho}{2}\int d^3r_{ik}\left(f^q_{ik}\right)^2
\left(D_{pq}+D_{mq}+D_{p^\prime q}\right),
\eeqa
where the $K^{pmp^{\prime}}$, $A^p$ and $D_{pq}$ matrices,
defined in \cite{PW79}, give the C-parts taking into
account the non-commutativity of operators $O^p_{ij}$ and $O^q_{ik}$.
The corresponding interacting exchange contribution is given by
\beqa
\lefteqn{W^{RE}_s(K, ij-ex)=} \nonumber \\
& &\frac{1}{8m^2}\frac{\rho}{s}\sum_{pmp^\prime qm^\prime n}
\int d^3r_{ij}\left(l^{\prime 2}-l\nabla^2 l\right)_{ij}
\left(f^pv^mf^{p^\prime}\right)_{ij}\frac{1}{2}\left(K^{npm^\prime}
K^{mp^\prime m^\prime}+K^{pmm^\prime}K^{p^\prime n m^\prime}\right) 
A^{m^\prime}\nonumber \\
& &\rho \int d^3r_{ik}\left(f^q_{ik}\right)^2 A^q\left(
D_{qm^\prime} +\frac{1}{2}D_{qm}+\frac{1}{2}D_{qn}\right).
\eeqa

The F-diagram contributions to $\langle\delta v^{RE}_{ij}\rangle$
are evaluated for the direct, interacting exchange and passive
exchange terms. These contributions also factor into separate
integrals, where the $r_{ij}$ integral has the form of the
two-body contribution to $\langle v^s\rangle$, and the $r_{ik}$ integral
is a new type of vertex correction involving gradients of the $f_{ik}$.
The direct and $ij$-exchange integrals are
\beqa
\lefteqn{W^{RE}_s(F, dir)=
\sum_{pmp^\prime q}\frac{\rho}{2}\int d^3r_{ij}
\left(f^pv^mf^{p^\prime}\right)_{ij}
K^{pmp^\prime}A^{p^\prime}} \nonumber \\
& & \frac{\rho}{4m^2}\int d^3r_{ik} f^q_{ik}
\left(\nabla^2f^q-\frac{6}{r^2}f^q\left(\delta_{q5}+\delta_{q6}\right)
\right)_{ik}\left[\frac{1}{4}\left(D_{pq}+D_{mq}+D_{p^\prime q}\right)
+1\right]A^q
\eeqa
\beqa
\lefteqn{W^{RE}_s(F, ij-ex)=\sum_{pmp^\prime qm^\prime n}-\frac{\rho}{2s} 
\int d^3r_{ij} \left(l^2f^pv^mf^{p^\prime}\right)_{ij} \frac{1}{2}
\left(K^{npm^\prime}K^{mp^\prime m^\prime}+K^{pmm^\prime}
K^{p^\prime nm^\prime}\right)A^{m^\prime}}\nonumber \\ 
& &\frac{\rho}{4m^2}\int d^3r_{ik} f^q_{ik}
\left(\nabla^2f^q-\frac{6}{r^2}f^q\left(\delta_{q5}+\delta_{q6}\right)
\right)_{ik}\left[\frac{1}{2}D_{qm^\prime}+\frac{1}{4}D_{qm}
+\frac{1}{4}D_{qn}+1\right]A^q
\eeqa
The $ik$-exchange contribution is evaluated only to leading order,
namely for the cases where at least one $f_{ik}$ is a central
link. In this approximation, this contribution takes the form:
\beqa
\lefteqn{W^{RE}_s(F, ik-ex)=\sum_{pmp^\prime n}-\frac{\rho}{2}
\int d^3r_{ij}\left(f^pv^mf^{p^\prime}\right)_{ij}K^{pmp^\prime}A^{p^\prime} } 
\nonumber \\
& & \frac{1}{4m^2}\frac{\rho}{s}\int d^3r_{ik} 
l^2_{ik} \Big[\left(f^c\nabla^2f^c\right)_{ik}
+\left(f^c\nabla^2f^n\right)_{ik}A^n\left(1+\frac{1}{2}
D_{np^\prime}\right)+\left(f^n\nabla^2f^c\right)_{ik}
A^n \nonumber \\
& & \left(1+\frac{1}{2}D_{np}\right)\Big].
\eeqa
In this equation, the index {\em n} runs from 2 to 4 only.

The K-diagram separable three-body  contributions to 
$\langle\delta v^{LC}_{ij}\rangle$
have the same general structure as the corresponding contributions
to $\langle\delta v^{RE}_{ij}\rangle$. The direct term and the
interacting exchange term have been evaluated and are presented
below.
\beqa
\lefteqn{W^{LC}_s(K, dir)=} \nonumber \\
& &\sum_{pmp^\prime q}\frac{k_F^2}{8m^2}\frac{\rho}{5}\int d^3r_{ij}
\left(rf^p\frac{dv^m}{dr}f^{p^\prime}\right)_{ij} K^{pmp^\prime}A^{p^\prime}
\ \frac{\rho}{2}\int d^3r_{ik}\left(f^q_{ik}\right)^2
A^q\left(D_{pq}+D_{mq}+D_{p^\prime q}\right)
\eeqa
\beqa
\lefteqn{W^{LC}_s(K, ij-ex)=\sum_{pmp^\prime qm^\prime n}} \nonumber \\
& &-\frac{1}{8m^2}\frac{\rho}{s}
\int d^3r_{ij}\left(l^{\prime 2}-ll^{\prime\prime}\right)_{ij}
\left(rf^p\frac{dv^m}{dr}f^{p^\prime}\right)_{ij}
\frac{1}{2}\left(K^{npm^\prime}K^{mp^\prime m^\prime}
+K^{pmm^\prime}K^{p^\prime nm^\prime}\right)A^{m^\prime} \nonumber \\
& &\rho\int d^3r_{ik}\left(f^q_{ik}\right)^2
A^q\left(D_{qm^\prime}+\frac{1}{2}D_{qm}+\frac{1}{2}D_{qn}\right)
\eeqa

The separable three-body F-diagrams have a more complicated structure
for the $\delta v^{LC}_{ij}$. The direct diagram has the general form
\beqa
\sum_{pmp^\prime qq^\prime}
\lefteqn{\frac{\rho^2}{8m^2}\int d^3r_{ij}d^3r_{ik}} \nonumber \\
& &\left( -\frac{1}{4}\right)C\left[
\left\{f^p_{ij}O^p_{ij},f^q_{ik}O^q_{ik}\right\}
\left( r_{ij}\cdot\nabla_R\right) 
\left(\nabla \left(v^m_{ij}O^m_{ij}\right)\cdot\nabla_R\right)
\left\{f^{p^\prime}_{ij}O^{p^\prime}_{ij},
f^{q^\prime}_{ik}O^{q^\prime}_{ik}\right\}\right].
\eeqa
The integrand can be written as a sum of four terms having the gradients
in $\delta v^{LC}_{ij}$ acting on different parts of the correlations:
\beqa
-\frac{1}{4}\left(rf^p\frac{dv}{dr}^mf^{p^\prime}\right)_{ij}f^q_{ik}\left(
\left[\left(f^{q^\prime}\right)^{\prime\prime}-\frac{1}{r}
\left(f^{q^\prime}\right)^\prime\right]
\cos^2\theta_i+\frac{1}{r}\left(f^{q^\prime}\right)^\prime\right)_{ik}
\!\!\!\! C\left[\left\{O^p_{ij},O^q_{ik}\right\}O^m_{ij}\left\{
O^{p^\prime}_{ij},O^{q^\prime}_{ik}\right\}\right] \nonumber\\
-\frac{1}{2}\left(f^p\frac{dv}{dr}^mf^{p^\prime}\right)_{ij}
\left(f^q\left(f^{q^\prime}\right)^\prime\right)_{ik}
\cos\theta_i \ C\left[
\left\{O^p_{ij},O^q_{ik}\right\}O^m_{ij}\left\{O^{p^\prime}_{ij},r_{ij}\cdot
\nabla O^{q^\prime}_{ik}\right\}\right]\nonumber \\
-\frac{1}{4}\left(f^p\frac{1}{r}\frac{dv}{dr}^mf^{p^\prime}\right)_{ij}
\left(f^qf^{q^\prime}\right)_{ik}
C\left[\left\{O^p_{ij},O^q_{ik}\right\}O^m_{ij}
\left\{O^{p^\prime}_{ij},(r_{ij}\cdot\nabla)r_{ij}\cdot
\nabla O^{q^\prime}_{ik}\right\}\right] \nonumber\\
-\frac{1}{4}\left(f^pv^mf^{p^\prime}\right)_{ij}f^q_{ik}
C\left[\left\{O^p_{ij},O^q_{ik}\right\}
(r_{ij}\cdot\nabla_R)\nabla_R\cdot\left(f^{q^\prime}_{ik}\nabla O^m_{ij}
\left\{O^{p^\prime}_{ij},O^{q^\prime}_{ik}\right\}\right)\right]
\eeqa
The C-parts in the above expressions depend on the cosine of
the azimuthal angle, $\theta_i$, and therefore cannot be expressed
exclusively in terms of the $D_{pq}$ matrices, which include an
implicit average over that angle. The C-parts must then be
individually evaluated for each operator product. 

The contribution of the direct diagram is dominated by the first term 
in the above sum. In evaluating the interacting exchange and
passive exchange diagrams, only the term corresponding to that
dominant part has been included. For example, the interacting exchange
contribution is approximated by
\beqa
W^{LC}_s(F, ij-ex)=
\sum_{pmp^\prime qq^\prime n}\frac{1}{8m^2}\frac{\rho^2}{s}\int d^3r_{ij}
d^3r_{ik}\left(l^2f^p\frac{dv}{dr}^mf^{p^\prime}r\right)_{ij} \nonumber \\
f^q_{ik}\left(\left[\left(f^{q^\prime}\right)^{\prime\prime}
-\frac{1}{r}\left(f^{q^\prime}\right)^\prime\right]
\cos^2\theta_i + \frac{1}{r}\left(f^{q^\prime}\right)^\prime\right)_{ik}
\frac{1}{4}C\left[O^n_{ij}\left\{O^p_{ij},O^q_{ik}\right\}
O^m_{ij}\left\{O^{p^\prime}_{ij},O^{q^\prime}_{ik}\right\}\right].
\eeqa
The corresponding passive exchange contribution was calculated with 
the additional simplification of considering only leading term contributions,
having at least one central $f^c_{ik}$ link.

We have also evaluated central-chain diagram contributions, 
denoted by $W_{cch}$, to the
boost interaction expectation values. These diagrams are obtained by
dressing two-body cluster diagrams with hypernetted central chains.
Their contribution was found to be significantly
smaller than the contribution from separable three-body terms.

The combined results for the two-body clusters and for the many-body
clusters appear in Table~\ref{tab:t8}.
The first four columns contain
the boost contributions to the energy of SNM and PNM,
calculated using the optimal
wavefunctions for the A18 interaction alone, while the 
next four columns contain the corresponding boost contributions,
calculated with the optimal wavefunctions for the A18+UIX.
As expected, the many-body contributions to the boost interaction 
energy are comparable to the two-body contributions, because $T_F$ 
and $\langle T \rangle_{2B}$ have similar magnitudes.

Detailed breakdowns of the boost contributions in
the A18+UIX$^\ast$ model are presented in Tables~\ref{tab:t9} and 
\ref{tab:t10} for $\delta v^{RE}$ and $\delta v^{LC}$.
In all cases the bulk of the contributions come from the direct terms. 
Since we can integrate $\langle \delta v^{LC} \rangle$ by parts to 
obtain $\langle \delta v^{RE} \rangle$ plus additional terms,
their expectation values should be similar in magnitude.
The ratio of the contributions of $\delta v^{LC}$ and $\delta v^{RE}$ 
in SNM is found to be $\sim$ 0.7 at all densities, while in PNM
it is $\sim$ 0.75 at $\rho < 0.2$ fm$^{-3}$, and $\simeq 1$ at 
higher densities.  Results of VMC 
calculations \cite{CPS93} have shown that this ratio is $\sim$ 0.5
in $^3$H and $^4$He. 

The $\delta v^{RE}$ has two factors, $P^2/8m^2$ and $v^s$, suggesting 
the approximation
\beqa
\langle \delta v^{RE} \rangle &\approx & - \langle \frac{P^2}{8m^2} \rangle \ 
\langle v^s \rangle,  \nonumber     \\
&=& - \frac{1}{2m}\ \langle T \rangle \ \langle v^s \rangle .
\label{eq:approx}
\eeqa
Our results show that this approximation has errors of only $\sim$ 10 \% 
in SNM, but can be wrong by $\sim$ 50 \% in PNM.  Nevertheless 
it may be used to estimate the order of magnitude of the contribution 
of relativistic boost interactions.

\subsection{Nucleon Matter Energies}

We evaluate the boost interaction contributions as a first order 
perturbation.  Thus the energies of nucleon matter with the A18+$\delta v$
interactions are obtained by simply adding the $\delta v$ 
contributions listed in Table~\ref{tab:t8} to the A18 energies.  The 
results are listed in Tables~\ref{tab:t6} and \ref{tab:t7}, and shown in 
Figs.~\ref{fig:fig2} and \ref{fig:fig3}.

The three-body interaction, UIX$^*$, to be used with A18+$\delta v$, contains 
the term $V^{R*}_{ijk}$ that is 0.63 times the $V^R_{ijk}$ in the UIX. 
Since the boost effects are treated only in first order, 
the energies for the A18+$\delta v$+UIX$^*$ interaction 
are obtained by adding
$\langle [\delta v_{ij} - 0.37~V^R_{ijk}] \rangle$
to the energies for A18+UIX interactions.  The results are 
listed in Tables~\ref{tab:t6} and \ref{tab:t7}, and 
shown in Figs.~\ref{fig:fig2} and \ref{fig:fig3}.
At low densities the A18+UIX and A18+$\delta v$+UIX$^*$ interactions 
yield similar results. However the energies predicted by the latter 
model are lower than  those of the former 
at higher densities, where 0.37$\langle V^R \rangle$ is 
much larger than $\langle \delta v \rangle$.  The difference 
between the energies predicted by A18 + $\delta v$ and A18+$\delta v$+UIX$^*$
interactions at higher densities is due to three-body forces; 
it is smaller than 
that between the energies obtained from A18 and A18+UIX by almost a 
factor of two.

The results obtained with the Urbana density-dependent interaction
(U14-DDI) \cite{FP81} are also 
shown in Figs.~\ref{fig:fig2} and \ref{fig:fig3} for comparison.  
Since Skyrme-type 
interactions based on the U14-DDI $E(\rho)$ explain nuclear binding energies 
quite accurately, it is likely to provide
a reliable representation of phenomena at $\rho \leq \rho_0$.

The variation of the kinetic and interaction energies with nucleon density 
in our most realistic model, with A18+$\delta v$+UIX$^*$, 
interactions is shown in Figs.~\ref{fig:fig4} and \ref{fig:fig5}. 
Due to a large cancellation between the contributions of $V^{2\pi}$ 
and $V^{R\ast}$, the total $V_{ijk}$ contribution is now smaller than 
that of $v_{ij}$, and the $\delta v$ contribution is also small. 
An order of magnitude estimate of the boost correction to the three body 
interaction is given by 
$\langle T \rangle \langle V_{ijk} \rangle /3m$, 
using generalizations of Eq.~(\ref{eq:approx}).
This estimate is less than 10 MeV at the highest density considered. 

The most significant remaining problem appears to be our neglect of the 
boost corrections to 
the $v^{MD}$, the unboosted contribution of which 
is quite large at high densities. 
Such corrections involve terms of order (velocity)$^4$, which are 
beyond the scope of the present work.  The kinetic energy also has 
corrections of that order, which we cannot include because the 
A18 interaction is fitted to data using nonrelativistic kinetic energy. 
However, the correction to the Fermi gas kinetic energy 
is $3k_F^4/56m^3$, which is less than 10 MeV in PNM at 
the density of 6$\rho_0$. 

The Dirac-Brueckner approximation provides another way to 
estimate relativistic effects using realistic models of 
nuclear forces.  The energies of our 
A18+$\delta v$+UIX$^*$ model are lower than those of  
lowest order Dirac-Bruekner calculations with the 
Bonn A potential \cite{LMB92}.
For example, at 4$\rho_0$ we obtain 58 and 128 MeV per nucleon, 
while the Dirac-Bruekner  
calculation gives 76 and 164 MeV per nucleon, for SNM and PNM respectively.
However, the difference between the results of the two methods
is significantly smaller than that between 
either of the two and the uncorrected A18 energies of 3 and 59 MeV per nucleon.
(Tables \ref{tab:t6} and \ref{tab:t7}).

In the last column of Table \ref{tab:t6} we list SNM energies obtained by 
adding a correction $\gamma_2 \rho^2 e^{-\gamma_3 \rho}$ to the 
A18+$\delta v$+UIX$^*$ results.  The empirical binding energy 
and density of SNM are reproduced with $\gamma_2 = 2822$ MeV fm$^6$, 
and $\gamma_3 = 18.34$ fm$^3$.  This correction has a maximum value 
of 4.5 MeV at 0.11 fm$^{-3}$, and the ``corrected" $E(\rho)$ of SNM is 
a better representation of known nuclear properties
at lower densities.  It becomes identical to the $E(\rho)$ obtained using the
A18+$\delta v$+UIX$^*$ model at higher densities.

\section{Cold Catalyzed Nucleon Matter}

In this section we use the results of the earlier sections to calculate the 
equation of state and composition of cold, catalyzed matter,
{\em i.e.} matter at zero temperature in its lowest energy state.
Here the matter is assumed to be made up of neutrons, protons and leptons;
the possible admixture of quark matter is considered later.
Since the boost interaction is clearly an integral part of the 
two-nucleon interaction, we regard the models A18+$\delta v$ and 
A18+$\delta v$+UIX$^*$ as realistic, and discuss their results in 
detail.  The difference between the two models demonstrates the effect of the 
three-nucleon interaction.  Some of the results obtained with 
the less realistic models without the boost interaction are also
presented for comparison.

Matter at zero pressure, at the surface of a neutron star,
is made up of atoms of $^{56}$Fe, just as in terrestrial iron. 
This is the most stable form of electrically neutral matter 
composed of neutrons, protons and electrons.
Below the neutron star surface, as a result 
of the increased pressure of the matter caused by the gravitational
attraction, 
the atoms become completely ionized and the electrons form a relativistic
Fermi gas, whose Fermi energy becomes competitive in magnitude with
nuclear energies \cite{PR95}. Consequently, electron capture by 
the protons can occur.
As a result, with increasing depth below the stellar surface
the nuclei become more neutron rich, and cross the 
neutron-drip line.  At this point the most energetic neutron orbitals 
have become
unbound, and the matter consists of neutron-rich nuclei immersed in
a neutron gas, whose density also increases as the pressure is 
increased further.  The baryon number density at which this transition occurs
is about $2 \times 10^{-4}$ fm$^{-3}$.  
As the pressure and density continue to increase, 
the charge number $Z$ of the nuclei 
remains in the
range $30 - 40$ but the mass number $A$ grows steadily, and
the distance between nuclei decreases. 
In the density range above about $0.06$ fm$^{-3}$, where the volumes occupied
by nuclei and by the surrounding neutron gas become comparable,
the matter may undergo an inversion to bubbles of neutron gas surrounded
by nuclear matter by going through a progression of phases involving
non-spherical shapes, although this aspect of the crustal structure
is somewhat model dependent \cite{LRP93,PRL95}.  
At a density above about
$0.1\;$fm$^{-3}$ there no longer occur nuclei or other clumps of 
proton-containing matter, and cold, catalyzed matter becomes 
a uniform fluid of neutrons with a small fraction of protons.
 
A reliable discussion of the properties of matter over the crustal density
range  
requires a nuclear model that can describe inhomogeneous matter in the 
geometries occurring there.  The difficulties encountered cause
the model dependence mentioned.  However, for
neutron stars with masses  $\ga$1.4\msol, 
the mass fraction contained in the crust of the star
is less than about 2\%.  We have therefore used results of earlier
work \cite{LRP93,PRL95} for matter at densities $\le 0.1$ fm$^{-3}$.
Since over that density range our present matter energies agree well
with ones used earlier, this substitution causes negligible 
inconsistency in our conclusions about the total mass of the neutron star.
 
At densities of  0.1 fm$^{-3}$ and greater, we require properties of 
charge neutral uniform matter
made up of neutrons, protons, electrons and 
muons in beta equilibrium.
At a given baryon number density $\rho = \rho_n + \rho_p$, the conditions to be
satisfied by the components are charge neutrality,
\beq
 \rho_p = \rho_e + \rho_{\mu} ,
\eeq
and beta equilibrium,
\beq
 \mu_n = \mu_p + \mu_e, \qquad \mu_{\mu} = \mu_e .  \label{betamu}
\eeq
Here $\rho_i$ is the number density of the species indicated by the
subscript {\em i}, and $\mu_i$ is its chemical potential including rest 
mass. After achieving a solution
to these conditions, one obtains the total pressure,
\beq
 P = P_N + P_e + P_{\mu},
\eeq
and total energy density 
\beq
\varepsilon 
  = \varepsilon_N + \varepsilon_e + \varepsilon_{\mu},
\eeq
of the matter, and
thereby the EOS, in the form
$P = P(\tilde{\rho})$, where $\tilde{\rho} = \varepsilon / c^2$ is
the mass-energy density of the matter.  The composition and
other thermodynamic variables are also determined in this process.

The baryonic part of the matter consists of strongly interacting neutrons
and protons with a given proton fraction $x_p = \rho_p/\rho$.
The minimum processing necessary in order 
to obtain an EOS for any of the sets of SNM and PNM energies 
of earlier sections is to fit the energies to a smooth function of
density, so that the derivatives needed to obtain 
chemical potentials ($\mu_i = \partial \varepsilon / \partial \rho_i )$ 
and pressure ($P = \sum_i \rho_i \mu_i - \varepsilon)$ 
can be calculated.  One then needs
to interpolate between the $x_p = 0.5$ SNM and
$x_p = 0$ PNM results in order to find the value of $x_p$
required by beta equilibrium.  

The fitting procedure used earlier \cite{PR89}, and
which we employ here, 
introduces more than the minimum processing just described.  It 
uses a generalized Skyrme interaction containing a momentum and 
density-dependent delta function interaction.
By a well-known procedure \cite{VB72}, for an uniform
system of infinite extent the matrix elements of this interaction
can be written as an energy-density functional in the form
\beqa
H_{eff} =  \left(\frac{\hbar^2}{2m}+ f(\rho, x_p) \right) \tau_p
  +  \left(\frac{\hbar^2}{2m}+ f(\rho, 1-x_p) \right) \tau_n 
  +  g(\rho, x_p) +  H_{gradient} .\label{fpsk}
\eeqa
The number and kinetic densities $\rho_{n(p)}$ and $\tau_{n(p)}$ 
are defined in terms of the
neutron or proton orbitals $\phi_{\alpha,n(p)}({\bf r})$ by
$\rho_i({\bf r}) = \sum_{\alpha}n_i(\alpha,T) |\phi_{\alpha,i}({\bf r})|^2$ and
$\tau_i({\bf r}) = \sum_{\alpha}n_i(\alpha,T) 
|\nabla \phi_{\alpha,i}({\bf r})|^2$, where the Fermi-statistics
density of states $n_i(\alpha,T) = [1+\exp((e_{\alpha,i}-\mu_i) /T)]^{-1}$ 
at $T = 0$ become a step function $\Theta(e_{\alpha,i}-\mu_i)$.
The function $f$ in Eq.~(\ref{fpsk}) comes from the assumed momentum 
dependence of the interaction, and is parametrized in the form
\beq
f(\rho, x_p) = f_0(\rho) + x_p f_1(\rho). \label{fs}
\eeq
It produces density- and isospin-dependent effective nucleon masses
$m_i^*(\rho, x_p)$, 
\beq
\frac{\hbar^2}{2m_p^*(\rho, x_p)} = \frac{\hbar^2}{2m}+ f(\rho, x_p),\qquad 
\frac{\hbar^2}{2m_n^*(\rho, x_p)} = \frac{\hbar^2}{2m}+ f(\rho, 1-x_p). 
  \label{m*}
\eeq
The term $H_{gradient}$ in Eq.~(\ref{fpsk}) depends on gradients of the
densities, and thus affects finite systems; it is determined uniquely
by the functions $f$ \cite{PRL95}.  
The relation between $\tau_i$ 
and $\rho_i$ depends on temperature via the Fermi functions.  
Therefore knowledge of the matter energies
at finite temperatures, as was given 
in Ref.\cite{FP81}, permits a determination of $f(\rho, x_p=.5)$
and $f(\rho, x_p=0)$, {\em i.e.} $f_0(\rho)$ and $f_1(\rho)$ in Eq.~(\ref{fs}). 
The potential-energy terms $g(\rho, x_p=.5)$ and $g(\rho, x_p=0)$
may then be obtained by comparison with the zero-temperature energies.
The functional forms used for $f(\rho, x_p=.5, 0)$ and $g(\rho, x_p=.5, 0)$
are chosen to represent with appropriate accuracy the calculated
energies; it is not sufficient, nor necessary, that they have
the extremely simple power dependence on $\rho$ of customary
Skyrme interactions. The effective interaction FPS obtained 
in this manner from the SNM and PNM energies calculated with the
U14-DDI interaction has been described and used elsewhere 
\cite{PR89,PRL95,LRP93}.

In the present work, we have studied the matter energies only
at zero temperature, and cannot therefore make a new determination
of the momentum-dependent $(\tau$-containing) terms in $H_{eff}$.
Rather than omitting them, we have kept intact 
the $\tau$-dependent terms
from the earlier determination \cite{PR89}, and have modified
only the potential energy term $g(\rho, x_p)$.  
To facilitate handling of the two different phases, 
we make separate fits to the normal low density phase (LDP)
and the high density phase (HDP) with pion condensation.
The analytic forms used for the fitting are given in Appendix A.  
They are chosen solely to provide an economical fit to the calculated
energies.

The interpolation between $x_p=0.5$ and $x_p=0$ is carried out 
assuming a $(1 - 2 x_p)^2$ dependence of the energy at a given density.
It is well known that matter energies, as a function of $x_p$, 
can be expanded in powers of $(1 - 2 x_p)$ about $x_p=0.5$.  
Previous studies \cite{LP81c}, 
using cluster expansion techniques \cite{PW79},
have found that the quartic terms are small, and that
the quadratic terms determined from results at $x_p=0$ and $0.5$ 
are sufficient to obtain few percent accuracy in the interpolation.
The potential energy terms $g$ in the effective interaction
Eq.~(\ref{fpsk}) obtained by fitting to PNM and SNM
are therefore interpolated by means of the expression
\beq
 g(\rho, x_p) = g(\rho, x_p=.5) \left(1 - (1 - 2 x_p)^2\right)
             + g(\rho, x_p=0) (1 - 2 x_p)^2   .  \label{gx}
\eeq

\subsection{The Phase Transition}

In fitting the results of the calculations described in Sections~I-III,
we find that the SNM and PNM energies coming from the models 
that include a three-nucleon interaction
have a clear discontinuity in slope, associated with the
phase transition, so that different analytic forms
are needed above and below the critical densities,
$\rho_t = 0.20$~fm$^{-3}$ for PNM, $0.32$~fm$^{-3}$ for SNM.
In Fig.~\ref{fig:fig6} the energies and the fits are shown for the
A18+$\delta v$+UIX$^*$ model.   (Here, and for the rest of the
paper, we use for the SNM energies of this model the ``corrected''
values given in the last column of Table~\ref{tab:t6}.)
The A18+$\delta v$ and A18 models
do not show such a discontinuity in any marked way.

An important assumption of our treatment of the two phases of nuclear
matter and the phase transition, as exhibited in Fig.~\ref{fig:fig6}, is that
the analytic forms fitted to the PNM and SNM energies of each phase may be
extrapolated beyond the density region in which they are determined. 
Since the equilibrium is a two-dimensional phenomenon, the 
energy as a function of $\rho$ and $x_p$
can be represented by a surface above the $\rho, x_p$ plane. 
The interpolation described in the last subsection 
makes the energy surface of the LDP
a valley parabolic in the $x_p$ dimension 
extending from $x_p = 0$ to $x_p = 1$, with its minimum at $x_p=0.5$.
A similar interpolation between the energies of the 
HDP produces another parabola.  Note that because of the 
charge symmetry breaking terms in the A18 interaction only the $x_p \le 0.5$ 
side of the parabolas is useful, and that the parabolic shape approximation 
is tested only in the LDP \cite{LP81c}.

A transition to a neutral pion condensed phase of PNM was obtained 
earlier with the older Argonne $v_{14}$ (A14) NNI and 
Urbana VII (UVII) TNI interactions by WFF
\cite{WFF88}.  However, with those interactions SNM has a normal ground 
state.  WFF estimated the properties of the A14+UVII 
model of cold catalyzed matter by interpolating between the normal 
SNM and the pion condensed PNM.   For the
A18+$\delta v$+UIX$^*$
interaction, we find that such a procedure overestimates
the symmetry energy of the HDP
by $\sim 10$ to $20$\% in the density range $0.2$ to $0.5$~fm$^{-3}$.
Fortunately we see the transition for both PNM and SNM, and
can avoid that problem.

In Fig.~\ref{fig:fig7} we show for the
A18+$\delta v$+UIX$^*$
model the density at which the LDP and HDP
$E(\rho,x_p)$ surfaces intersect, {\em i.e.} where the
interpolated phases have the same energy.
The curve obtained is not
necessarily a parabola in $x_p$, since the density dependence of 
the curvature of the parabolas
is more complicated than quadratic.
Also shown is the proton fraction of beta-stable matter in each phase.

The transition of matter from the LDP to the HDP probably occurs 
via a mixed phase region in which
each phase need not be charge neutral \cite{Gle92},
as discussed in the following section. 
Here, for simplicity, we assume that each phase is a charge-neutral fluid
and make a Maxwell construction to obtain the change
in density due to the phase transition. The baryon chemical potentials,
which for beta equilibrium are equal to the neutron chemical
potentials,
and total pressures of the two phases are equated to obtain the densities
and proton fractions of the LDP and HDP in equilibrium.
They are $(0.204$~fm$^{-3}, 0.073)$ and $(0.237$~fm$^{-3}, 0.057)$
respectively.  In the region between these two densities
the most stable form of the matter is a mixture of the two
phases. 

\subsection{Proton fraction and $\mu_e$}

The proton fraction and the related electron chemical potential 
of the matter are important in assessing the cooling rates of neutron stars 
\cite{Pet92} and the possibility of kaon condensation in neutron 
star interiors \cite{BLR94,PPT95}.  
For the models discussed here we plot in Fig.~\ref{fig:fig8} 
the proton fraction $x_p$ against baryon density.  
This figure also shows, for comparison, results obtained for other
models and by other methods: the 
U14+UVII model, a predecessor of the present A18+UIX model, as given by 
WFF with the VCS method; 
the A18 model, results of Engvik {\em et al.} \cite{Eng97} with lowest 
order Brueckner (LOB) calculations; 
the U14-DDI model \cite{FP81} using the FPS effective interaction
\cite{LRP93}. 
Indicated on the plot is the critical value
of $x_p$, as a function of $\rho$, for the onset of the direct Urca cooling 
process in the presence of both electrons and muons.
That process allows energy to be conducted from the
interior of the star by neutrinos generated in binary thermal
collisions, a very efficient process \cite{Pet92}. 
In the models containing
three-nucleon interactions, the discontinuity in $x_p$ 
at $\rho \sim 0.2$~fm$^{-3}$ 
signals the onset of the HDP.  
The corresponding electron chemical potentials are plotted in 
Fig.~\ref{fig:fig9}.

As suggested in Ref.\cite{WFF88}, the determining
quantity for $x_p$ is the symmetry energy,
given approximately by $E_{sym} \approx E_{PNM} - E_{SNM}$.
The LOB and VCS methods yield similar values for $x_p$ with the A18 
interaction up to $\rho = 0.6$~fm$^{-3}$. 
Beyond this density the $x_p$ obtained 
from VCS calculations starts to decrease, while that from LOB 
calculations continues to increase with density.  The VCS and LOB 
energies for the A18 model are compared in Fig.~\ref{fig:fig10}.  They 
are not too different for PNM. For SNM, however, at $\rho > 0.6$~fm$^{-3}$ the 
LOB energy is much lower than that of the VCS.  The VCS $E_{sym}$ 
saturates for the A18 model at $\rho \sim 0.6$ fm$^{-3}$, while that 
of LOB calculations continues to increase with density, causing the 
$x_p$ to do the same.  It will be interesting to see if three hole-line 
and higher terms \cite {DW85,SBG97} neglected in the LOB calculations 
reduce the difference between LOB and VCS energies.

Engvik {\em et al.} \cite{Eng97} have calculated $x_p$ in cold 
matter using the LOB method with
all five modern NN potentials that provide high 
precision fits to the Nijmegen NN scattering data base.
At a density of one nucleon per fm$^3$ the CD-Bonn model \cite{MSS96}
gives the largest $x_p$ of 0.15, while the Nijmegen I model \cite{SKT94} gives 
the smallest value of 0.10.  The spread in these values is 
comparable with the difference between VCS and LOB results for A18
of $x_p =$ 0.09 and 0.14 at this density.

The $\delta v$ term and the three-nucleon interaction increase the 
symmetry energy, and push the $x_p$ barely above 
the Urca limit at high densities.  For the A18+$\delta v$+UIX$^*$ 
model the threshold is at a density of $\rho =0.78$~fm$^{-3}$, 
and, as discussed in the next section,
stars must have a mass $ >$2.0\msol to achieve
such a density.  However, this density is at the limit of our calculations
and of the input physics.  For example, admixtures of quark matter 
with hadronic matter, considered in the next section,  
may affect the Urca process in matter at such densities.  

The U14-DDI (FPS) model predicts values for $x_p$ that are much smaller than
those predicted by all other models considered here, and in fact
go to zero for $\rho \sim 1$ fm$^{-3}$.
It is based on the U14 NN interaction, also used 
in the U14+UVII model.  However, instead of adding the UVII three nucleon 
interaction to obtain the empirical saturation 
density of nuclear matter, it uses a density dependent modification 
(U14-DDI) of the U14 NN interaction \cite{LP81b} chosen to reproduce 
the energy, density and compressibility of equilibrium nuclear matter.
Unlike the UVII interaction, this modification reduces the symmetry energy,
and thus the $x_p$, at high density.
The main advantage of using three-nucleon
interactions, instead of density dependent modifications of the two-nucleon 
interaction, is that the former can be tested via accurate 
calculations of the light nuclei.
Unfortunately, the available results \cite{Pud97} indicate that the 
UIX model may be overestimating the repulsion between three 
neutrons, thus overestimating the $x_p$; an improved version of the UIX 
model is currently being developed.

\section{Neutron stars}

Using the methods just described we obtain for each model the EOS
for cold, catalyzed beta-stable matter.  At a baryon number density of
0.1~fm$^{-3}$ they are joined onto an earlier EOS in which
properties of the crust material has been treated more
accurately \cite{LRP93}. 
The Oppenheimer-Volkoff general relativistic equations for 
a spherically symmetric (non-rotating) neutron star \cite{st} are
\beq
\frac{d P}{d r} = - \frac{(\tilde{\rho}+P/c^2) G \left(m(r)+
4 \pi r^3 P/c^2\right) \Lambda(r)}{r^2} , \qquad
 m(r) = \int_0^r 4 \pi r^2 \tilde{\rho}\; dr , \label{dpdr}
\eeq
where $\Lambda (r) = (1 - 2 G m(r)/ r c^2)^{-1}$.
The corresponding equations for obtaining the moment of inertia,
for a slowly rotating star, are given in Appendix~B.
Starting from some central mass-energy density 
$\tilde{\rho}_c$, or equivalently from
a central number density $\rho_c$, these equations are integrated 
outwards to a radius $r = R$, at which  $P$ is zero, 
thus yielding the stellar radius, {\em R}, the gravitational
mass of the star, $m(R) = M$, and the moment of inertia $I$.

The dependence of the neutron star mass on central baryon density $\rho_c$ for
the four models is shown in Fig.~\ref{fig:fig11}.
In order to estimate the effect of beta-stability on these
results, we show also the trajectories obtained by using the
pure neutron matter EOS for densities greater than 0.1 fm$^{-3}$,
joined to the crust results of Ref.\cite{LRP93}.
Earlier results with the FPS EOS \cite{LRP93} are included for comparison.
For the same set of results, the neutron star mass is
plotted against the star radius in Fig.~\ref{fig:fig12}. 

The maximum masses for the five models illustrated in 
Figs.~\ref{fig:fig11} and \ref{fig:fig12}
are listed in Table~\ref{tab:t11}.   
While the models based on only two-nucleon
interactions have maximum masses at or below $1.8 M_{\odot}$,
those for the two models containing three-nucleon interactions
have maximum masses well above $2 M_{\odot}$.   The model that we 
believe includes most of the necessary physics is 
A18+$\delta v$+UIX$^*$, which yields a maximum mass of $2.2 M_{\odot}$.
This model achieves its maximum mass for a central
baryon density $\rho_c = 1.14$~fm$^{-3}$, which is not far beyond
the calculated VCS
energies, so that only slight numerical extrapolation is involved.

The moments of inertia given in Table~\ref{tab:t11} are the maximum
values for each model.   They occur for slightly lower central
densities than do the maximum masses.  The effect of the three-nucleon
interactions on the maximum moment of inertia is in general
considerably greater than on the maximum masses, because TNI tend
to increase stellar radii also.

The relativistic correction to the Fermi energies, $-3k_F^4/56m^3$, 
mentioned at the end of section~III-B, when inserted (with
its attendant contributions to chemical potentials and pressure) 
into the EOS of the model A18+$\delta v$+UIX$^*$, 
produces at $\rho = 0.1$~fm$^{-3}$ a reduction in
pressure of 6.9\%, but at $\rho = 1.0$~fm$^{-3}$  the reduction is only
1.1\%.  The effect on the maximum mass is a 0.2\% reduction, smaller 
than the digits quoted in  Table~\ref{tab:t11}.   

The mass limits obtained with PNM EOS are within 
$\sim 0.5$\% of the values determined from the EOS of matter in beta 
equilibrium, and the radii are within $ \sim 5$\%.  
Thus the uncertainties in the proton fraction of matter 
discussed in the last section do not have a large effect on these 
aspects of neutron star structure.

\subsection{Superluminality and Maximally Incompressible Matter}

Indicated on the curves in Fig.~\ref{fig:fig11}
are the densities $\rho_{sl}$ at which
the sound speed $c_s = \sqrt{\partial P/\partial \tilde{\rho}}$ 
becomes greater than the speed of light, $c$.
Superluminal behavior would
not occur with a fully relativistic theory, and it is necessary to
gauge the magnitude of the effect it introduces at the higher densities.
Kalogera and Baym \cite{KB96} provide one method
for doing this.  
Following Rhoades and Ruffini \cite{RR74}, they assume that the stiffest
physically allowable EOS produces matter with a sound speed equal to
the speed of light, {\em i.e.} 
\beq
 \frac{\partial P}{\partial\tilde{\rho}} = c_s^2 \rightarrow c^2 .
\eeq
Since the matter is at zero temperature, for a single phase
the partial derivative becomes a total derivative.
Thus for higher densities than some $\tilde{\rho}_I$ the EOS is replaced 
by 
\beq
 P(\tilde{\rho}) = P(\tilde{\rho}_I) +  (\tilde{\rho} - \tilde{\rho}_I) c^2,  
    \label{kb}
\eeq
and the matter interior to $\tilde{\rho}_I$ is then maximally incompressible.
Other thermodynamically related quantities may be obtained 
easily \cite{Gle97}; the corresponding baryon number density is
\beq
\rho = \rho_I \; \sqrt \frac{\tilde{\rho} + \frac12(P_I - \tilde{\rho}_I c^2)} 
                        {\tilde{\rho}_I + \frac12(P_I - \tilde{\rho}_I c^2)} \ .
\eeq 
The relatively small reduction in the energy of the
A18+$\delta v$+UIX$^*$
model resulting from this replacement for $\rho_I = \rho_{sl} 
= 0.86$~fm$^{-3}$  is shown in Fig.~\ref{fig:fig13}.
In Fig.~\ref{fig:fig14} we show 
the effect on the mass {\it vs} radius plot.
The maximum mass is reduced
from $2.20 M_{\odot}$ to $2.19 M_{\odot}$, a very small change.
In all models of the present work superluminal behavior occurs in 
stars only very close to the maximum mass limit, as can be seen from 
Fig.~\ref{fig:fig11}.  Therefore replacing superluminal matter 
with maximally incompressible matter has little effect on the 
stellar properties.

It is also possible that due to neglect of four-body and higher 
forces, and relativistic corrections of order (velocity)$^4$ and 
higher, the present work underestimates 
the sound velocity at lower densities.  The effects of this 
possibility can be studied by assuming that the EOS of the 
A18+$\delta v$+UIX$^*$ model is valid up to a chosen density 
$\rho_I$, beyond which it is maximally incompressible.  The 
results obtained for $\rho_I =$ 2, 3 and 4$\rho_0$ are shown 
in Figs.~\ref{fig:fig13} and \ref{fig:fig14}, and Table~\ref{tab:t11}.
The difference 
between the $E(\rho )$ of the matter obtained by assuming that 
$\rho_I =$ 2$\rho_0$ and the $E(\rho)$ of the A18+$\delta v$+UIX$^*$ 
model is several times the contribution of the UIX$^*$ 
interaction (see Fig.~\ref{fig:fig14}). 
It therefore appears unrealistic to assume that 
$\rho_I$ can be as small as 2$\rho_0$.  Results obtained with 
$\rho_I = 3 \rho_0$ provide a better indication of what we can 
expect from the hardest EOS consistent with realistic models of 
nuclear forces.  

\subsection{Transition to Quark Matter}

It is also possible that the present EOS is too hard due to the 
assumption that neutron star matter contains only nucleons and 
leptons.  Should it also contain other hyperons such as $\Lambda, 
\Sigma^{-,0,+}$ and $\Delta^{-,0,+,++}$, the EOS may be softer 
than that obtained with nucleons only \cite{Pan71}. The forces 
between hyperons and nucleons and between hyperons are not as 
well known as nuclear forces, and it is therefore difficult to 
estimate whether such exotic species are present in neutron star matter.

The chemical equilibrium in matter containing nucleons, lambdas, 
sigmas, deltas and leptons is governed by the equations
\beqa
\mu_{\Sigma^-} = \mu_{\Delta^-} = \mu_n + \mu_e ,\\
\mu_{\Lambda} = \mu_{\Sigma^0} = \mu_{\Delta^0} = \mu_n ,\\
\mu_{\Sigma^+} = \mu_{\Delta^+} = \mu_p = \mu_n - \mu_e ,\\
\mu_{\Delta^{++}} = \mu_n - 2 \mu_e .
\eeqa
If we neglect the interaction between nucleons and these hyperons, then
the chemical potential of a hyperon at threshold density is 
given by its mass.  Under such 
an assumption, the negatively charged hyperons appear in the ground 
state of dense matter when $\mu_n + \mu_e$ reaches their mass, while 
the neutral hyperons appear when $\mu_n$ equals their mass.  Since 
$\mu_e$ in dense matter is larger than the mass differences between 
lambdas, sigmas and deltas, the $\Sigma^-$ and $\Delta^-$ will appear 
at lower densities than the neutral $\Lambda$, provided the interaction 
effects are small. The chemical potentials of electons,
and of neutrons in beta-stable matter, are shown in Fig.~\ref{fig:fig15} 
for the A18+$\delta v$+UIX$^*$ and A18+$\delta v$ models.  The threshold 
densities for non-interacting $\Sigma^-$, $\Delta^-$ and $\Lambda$ 
are indicated by horizontal line segments. Given their 
relatively low values, it is clear that in the 
absence of interactions these particles would be present in most 
neutron stars.  Results obtained using crude models of the interactions 
between hyperons and nucleons, and between hyperons \cite{Pan71} indicate 
that the $\Sigma^-$ and $\Delta^-$ have the largest 
effect of all hyperons on the EOS; 
however the magnitude of the effect is very sensitive to the 
interaction model.  

The available hyperon-nucleon scattering data has been reviewed recently
by de Swart, Maessen and Rijken \cite{DMR93,Rij97} along with the status of 
one-boson exchange models of the interactions between hyperons.  Additional 
information on $\Lambda$-nucleon interactions can be obtained from the 
measured $\Lambda$-nucleus binding energies.  These indicate the presence of 
$\Lambda NN$ three-body forces that are as strong as the three-nucleon 
interaction \cite{UPU95,Usm95}.  In view of these uncertainties, particularly 
concerning the important $\Sigma^-$ and $\Delta^-$ interactions, 
we do not attempt 
to estimate the effect of these hyperons on the EOS of neutron star matter.

A transition from hadronic to quark matter is expected at high
densities.  Knowledge of the EOS of both hadronic and quark matter 
is necessary to estimate the possible effects of this transition on 
neutron stars.  Here, we use the present models of 
the EOS of hadronic matter, containing only nucleons and leptons,
and the quark bag model 
with $u, d$ and $s$ quarks for the quark matter.   
The $u$ and $d$ quarks are taken to be massless,
and $s$ quarks to have a mass of 150~MeV. A Fermi gas of quarks of
flavor {\em i} has density $\rho_i = k_{Fi}^3/\pi^2$, due to the three 
color states.  There is no one-gluon exchange interaction energy 
between quarks of different flavor, while that between quarks of 
flavour {\em i} is given by
$(2 \alpha/3 \pi) E_i$ per quark {\em i} \cite{BC76}. Here $E_i$ is the average 
kinetic energy per quark, and $\alpha$ is the strong interaction 
coupling constant, assumed to have a value of 0.5.
The value of the bag constant {\em B} is poorly known, and we
present results using two representative values of
$B=122$~MeV \cite{CLS86} and $B=200$~MeV fm$^{-3}$ \cite{Sat82}.

The beta equilibrium conditions for charge neutral quark matter are
\beq
\mu_u + \mu_e =  \mu_d = \mu_s,  \qquad \mu_{\mu} = \mu_e. 
\eeq
The energy densities of charge neutral quark matter and  
nuclear matter are plotted in Fig.~\ref{fig:fig16}.
In the interesting region of $\rho \sim 1$ fm$^{-3}$ the total energy 
density of quark matter is about 1200 MeV fm$^{-3}$, of which only 
122 or 200 MeV fm$^{-3}$ comes from the bag.

If, during the phase transition from nuclear to quark matter, 
the nuclear and the quark phase are each required to be charge-neutral
beta-stable fluids whose pressures and baryon 
chemical potentials are equilibrated, then for the 
A18+$\delta v$+UIX$^*$
model the transition is found to
extend over the density range $\rho = 0.86 \leftrightarrow 
1.57$~fm$^{-3}$ for the $B = 200$~MeV~fm$^{-3}$ case. For
$B = 122$~MeV~fm$^{-3}$, the range is $0.79 \leftrightarrow 1.20$~fm$^{-3}$.
The matter within this density range, as so treated, is a constant-pressure
mixture of that at the two ends of the range.
Such a constant-pressure mixed phase 
does not occur in the neutron star: as can be seen from
Eq.~(\ref{dpdr}), since the pressure does not change, the density changes 
discontinuously from the lower nuclear matter density to the 
higher quark matter density.

Since for the nuclear component of the matter at 
$\rho \sim 1$~fm$^{-3}$ the electron screening length is $\sim 7$~fm,
while the quark matter component has negligible
electron distribution, the previous assumption
that each fluid retains its electron distribution during the
phase-mixing is not a good approximation at length scales $\la 7$~fm.
As described in Ref.\cite{Gle92,Gle97}, an alternative assumption
that the nuclear and the quark
matter share a common electron distribution in the mixture is more
appropriate.   For a given nucleon matter density $\rho_N$ 
the transition now involves equilibrating the chemical potentials,
\beq
\mu_n = \mu_u + 2 \mu_d, \ \ \ \ \ \mu_p = 2 \mu_u + \mu_d, \label{munq}
\eeq
and the hadronic pressures of nuclear and quark matter.
In the process, the proton fraction of the nucleon matter and
the densities of the quarks are also determined.
The lepton chemical potentials are then given by
\beq
\mu_e = \mu_{\mu} = \mu_n-\mu_p = \mu_d - \mu_u,
\eeq
from which the electron and muon densities are readily found.
The fractional volume, $f_Q$, occupied by the quark component in the
mixture is now chosen so that the
nuclear, quark and lepton components are in sum charge neutral:
\beq
 \rho_e + \rho_{\mu} = \frac{1}{3}f_Q 
 (2 \rho_u - \rho_d - \rho_s) + (1 - f_Q) \rho_p,
  \label{fq}
\eeq
after which the macroscopic baryon density of the
mixture can be calculated:
\beq
\rho = \rho_{mix} = \frac{1}{3}f_Q (\rho_u + \rho_d + \rho_s) 
+ (1 - f_Q) \rho_N.
\eeq
The procedure is carried out provided that $0 \le f_Q \le 1$.
Some numerical details are given in Appendix C.

The mixture of A18+$\delta v$+UIX$^*$ model nucleon and 
the $B = 200$~MeV~fm$^{-3}$ quark matter 
occurs over the density range $\rho_{mix} = 0.74 \leftrightarrow 
1.80$~fm$^{-3}$.  Over this range,
a charge neutral mixture of quark and nucleon matter is more stable
than the constant-pressure mixture of separately neutral nuclear
matter and quark matter.
The quark matter fraction $f_Q$, the charge per baryon of
the nucleon matter, given by the proton fraction $x_p$, 
the charge, $x_c$, and strangeness, $x_s$, per baryon of 
the quark matter and the charge $x_{\ell}$ per baryon carried by the 
leptons in the mixed matter are plotted 
for this case in Fig.\ref{fig:fig17}.

Over the density range occupied by the mixture the nucleon matter
has densities $\rho_N = 0.74 \leftrightarrow 0.96$~fm$^{-3}$, 
while the quark matter densities are $ 1.12 \leftrightarrow
1.82$ baryons fm$^{-3}$. The mixture thus consists of dense negatively
charged quark matter immersed in
somewhat less dense positively charged nucleon matter. 
The neutralizing charge of the leptons decreases in magnitude from its
value in the purely nucleon matter, and tends rapidly to
zero by the time the quark fraction has reached 50\%.  

The lower density end of the mixed phase region is more relevant for 
neutron stars.  Here one expects small drops of 
dense quark matter with $x_c \sim x_s \sim -1$ appropriate for 
matter made up of $\Sigma^-$ hyperons.  Ground states of dense matter 
can have such a form also if matter made of $\Sigma^-$ has a softer EOS 
than nucleon matter.  Knowledge of the interactions between $\Sigma^-$ 
hyperons is necessary to further explore this possibility.  
In regard to the direct Urca cooling process, the proton fraction of the 
nucleon matter in the mixture 
increases with $f_Q$, but the lepton fraction decreases.  The
required momentum balance among neutrons, protons and electrons
is only achieved at a baryon density of $0.86$~fm$^{-3}$, somewhat larger
than the value $0.78$~fm$^{-3}$ needed for the nucleon matter alone. 
At densities larger than $0.86$~fm$^{-3}$
the momentum balance remains possible, although
the decreasing density of electrons reduces the rate of
the Urca process.
Thus the indirect
effect of the quarks on the nucleons is to delay the start of the Urca process
somewhat.  
At a density of $0.86$~fm$^{-3}$ the quarks, which are about 7\% of the
matter, can if behaving as free particles contribute 
to the Urca process directly. Such   
considerations, however, are complicated  by effects of 
quark matter occurring as droplets.

At the high density end of the mixed phase, 
the quark matter has approximately equal number of 
{\em u}, {\em d} and {\em s} 
quarks, and nearly symmetric nucleon matter occupies a small fraction of 
the total volume.  Here one expects small drops of SNM, 
{\em i.e.} nuclei compressed to a density of $\sim 6 \rho_0$ by the 
pressure of surrounding quark matter.
However, this interesting form of matter at very large densities 
does not seem to occur in stable 
neutron stars according to the present calculations.

The two-component equilibrium suggested in Ref.\cite{Gle92} is
formally the same as the equilibrium between nuclear matter and dripped
neutrons in crustal neutron-star matter.  
Extending that similarity, Ref.\cite{HPS93}
includes the effects on the phase transition
of the energy of the surface dividing 
nucleon and quark matter, as well as the Coulomb energy
associated with the difference between the charge densities of 
nucleon and quark matter.
There then occur in the mixed region varying sizes and 
shapes of matter in one phase surrounded by liquid in the other phase.
The presence of a new parameter of unknown magnitude, the surface
energy, complicates the problem, however, and lacking new information on it,
we have not included surface and Coulomb effects in results given here.  

The energy densities of the mixed region, calculated neglecting its 
surface and Coulomb contributions,
are shown in Fig.~\ref{fig:fig16}. As discussed 
at length in Ref.\cite{Gle92}, the total pressure obtained is now not
constant over the density range of the mixture, although
as can be deduced from Fig.~\ref{fig:fig16}, it is 
reduced from that of the pure nuclear matter.  
The resulting neutron star masses
are shown in Fig.~\ref{fig:fig18} as a function of central baryon density.
Maximum masses are given in Table~\ref{tab:t11}.
For the A18+$\delta v$ model, admixing quark matter has small effects
on neutron star structure; however, the effects are noticeable for models 
containing three-nucleon forces.  Note that by neglecting the surface 
and Coulomb energies we underestimate the energy density of mixed 
matter and thus overestimate the effect of the admixture.
The maximum densities in stable neutron stars remain below those for 
pure quark matter in all the four cases considered.

\subsection{Transition to spin-ordered phase}

When each phase is required to be charge neutral, 
the mixed region between 
the normal nuclear matter LDP and the spin-ordered,
or neutral pion condensed HDP
has constant pressure and constant average nucleon chemical potential.   
In the neutron star such a mixed region would not occur,
and the density at this pressure would change discontinuously 
by $0.033$~fm$^{-3}$ for the A18+$\delta v$+UIX$^*$ model, 
a 15\% jump.  However, we should use the approach described in the
last subsection for this transition also, and consider  
mixtures of charged LDP and HDP in a common lepton sea.  
On applying the equilibrium 
conditions of nuclear pressure and neutron and proton chemical
potentials between the LDP and HDP, one finds that 
the density limits between which a mixture occurs is only slightly
extended compared to those with the charge-neutral-fluid 
equilibrium quoted earlier, but the mixture does now have 
a pressure that changes with density. However, due to
the similarity in charge character of the two phases the change in 
pressure in going from pure LDP to pure HDP is rather small, and 
therefore the thickness of the LDP-HDP mixed region in the star is 
also fairly small.  

Selected neutron star density profiles for the
A18+$\delta v$+UIX$^*$
model are shown in Fig.~\ref{fig:fig19}.  Although it is not 
detectable from the figure,
the mixed region associated with pion condensation now 
extends over a finite region of the star. 
Its radial thickness, 
$\Delta r_T$, is $\sim 14$~m 
for the $2.10$M$_{\odot}$ case, and $\sim 40$~m
for the $1.41$M$_{\odot}$ case. 
The 2.10 and 2.00M$_{\odot}$ stars shown in Fig.~\ref{fig:fig19} 
have the same central density of 0.86 fm$^{-3}$; their mass difference 
is due to the admixture of quark matter considered only in the 2.00 
M$_{\odot}$ star.  The 1.41 M$_{\odot}$ stars cannot have quark 
matter admixtures in the present models.

It is interesting to know
the structure of matter having a mixture of LDP and HDP.
One possibility is that suggested for
the nuclear matter-quark matter transition by Heiselberg, Pethick
and Staubo \cite{HPS93},
that there occur in the mixed region varying sizes and 
shapes of matter in one phase surrounded by liquid in the other phase.
If we apply their estimates to this phase transition,
assuming a surface tension
for the interface in the range $\sigma = 1$ to $10$~MeV~fm$^{-2}$, 
we find that the characteristic length scale $a$ of the entities involved
is $\sim 60$ to $120$~fm.  The electron screening length,
which for this approximation to be valid needs to be also of this magnitude,
is only $\sim 10$~fm, however.  This indicates that neither assumption 
about the mixture,
that of two charge-neutral fluids (which results in $\Delta r_T = 0$)
nor of two nuclear components with 
a common electron fluid (which gives the $\Delta r_T$ values given above)
is correct, and we need to look at a mixture
that is somewhere in between these extremes.  
That problem will be pursued elsewhere.

\section{Adiabatic Index $\Gamma$ of Neutron Star Matter}

A measure of the stiffness of matter described by the EOS
$P = P(\tilde{\rho})$, is the adiabatic index $\Gamma$:
\beq
  \Gamma = \frac{\tilde{\rho}}{P} \frac{\;\partial P}{\!\partial\tilde{\rho}}
   = \frac{\tilde{\rho}}{P} c_s^2.
  \label{gamma}
\eeq
If $\Gamma$ were constant, then the EOS would become
$P \propto \tilde{\rho}^{\Gamma}$.  In this form, it is called
a {\em polytrope}, an idealized EOS on which many pioneering
studies of stellar structure were based \cite{Cha39}.
The values of $\Gamma$ in several limiting cases are well known.
At lower densities, the rest mass of the constituents dominates the 
energy density of matter, {\em i.e.}
$\tilde{\rho} \approx \rho \: m$. 
In this limit, for matter without correlations, 
$\Gamma = $ 5/3, 2 or 3 when the pressure is provided by Fermi kinetic
energy, static two-body or three-body interactions respectively. 
Repulsive momentum dependent interactions lead to larger values of $\Gamma$.
In the opposite extreme of very high density, the energy per particle 
is much larger than the particle's rest mass.  Neglecting rest mass, 
it is given by $\lambda \rho^n$, where $n=1/3, 1$ or 2 for the above 
three cases, and $\lambda$ is a constant.  
Thus $\tilde{\rho} = \lambda \rho^{n+1}$, $P = n \lambda \rho^{n+1}$, 
and $\Gamma = 1$ for all values of $n$.
This does not contradict the well-known value of $\Gamma \rightarrow 4/3$
in white dwarfs \cite{Cha39}.
In white dwarf matter the nuclei are nonrelativistic and the electrons 
are relativistic.  
The energy density of this matter $\tilde{\rho}c^2 = \rho Mc^2$, where
$\rho$ is the 
number density of nuclei, and $M$ is their mass. The pressure  
given by relativistic electrons is proportional to $\rho^{4/3}$ or 
equivalently $\tilde{\rho}^{4/3}$, giving $\Gamma = 4/3$.
 
Plots of $\Gamma$ and the sound speed for various cases 
previously described are given in Fig.~\ref{fig:fig20}.  The two 
curves corresponding to pure nucleons behave as expected from 
the above limits.  From the crust, $\Gamma$ rises to 
$\sim 3$: of the two maxima, the greater value 
corresponds to the model with the explicit three-body 
interaction UIX$^*$; $\Gamma$ then decreases at higher densities. 
The small peaks followed by cusps in these curves at 
$\tilde{\rho}/m \sim 0.15$~fm$^{-3}$ mark the softening
of the EOS due to the opening of the muon channel.  
The deep discontinuity in both $\Gamma$ and $c_s$ for the
A18+$\delta v$+UIX$^*$
model are due to the LDP-HDP transition.  The indicated values
$\Gamma = 0.075$ and $c_s/c = 0.045$ shown at the bottom of the
declevity are averages resulting from the treatment of that
transition in Sec.V-C.  (When treated in terms of a mixture
of neutral fluids, in which case the pressure is constant,
these quantities would be zero.)

It is simple to express in terms of $c_s^2$, the extent 
of the stellar radius $\Delta r_T$
occupied by this phase transition  from Eq.~(\ref{dpdr}):
\beq
 \Delta r_T \approx \frac{c_s^2}{\tilde{\rho} g_T} 
\delta \tilde{\rho}, \qquad
   g_T \approx \frac{G m(r_T)\Lambda(r_T)}{r_T^2},
\eeq
where  $\tilde{\rho}$ is an average quantity for the two phases, 
$m(r_T)$ is the mass interior
to the radius $r_T$ at which the transition ocurs, 
$\Lambda(r_T)$ is the redshift factor there, and $\delta \tilde{\rho}
\approx \delta \rho\: m$
is the density range of the transition, given earlier.
The gravitational acceleration $g_T$ at the transition region
is greater for the more massive star,  and explains the trend of
our values obtained from the star profiles.

The use of maximally incompressible matter beyond $\rho = \rho_{sl}$,
occurring in Fig.~\ref{fig:fig20} at 
$\tilde{\rho}_{sl}/m = 1.06$~fm$^{-3}$, produces
a sharp reduction in $\Gamma$.
This modification produces so little effect on the stellar mass limits
(shown in Fig.~\ref{fig:fig13}) because it occurs at a density close to
the central density in the star with maximum mass.
The curves corresponding to nucleon-quark mixed phases discussed
in Sec.V-D reduce $\Gamma$ and $c_s$ more gradually: for an ultra-relativistic
quark gas $c_s = c/\sqrt{3}$, and the sound velocities of models with 
quarks are seen to be approaching that value.  Note that 
the partial derivative involved in the calculation of $\Gamma$ 
and the sound velocity of nucleon-quark mixed phases
must be carried out holding the quark fraction $f_Q$ constant.

\section{Conclusions}

The results obtained with the modern A18 NNI for the gross properties of 
neutron stars, such as the mass limit and the radius of 1.4\msol~stars, are 
not very different from those obtained from the earlier U14 and A14 
models \cite{WFF88}.  The main difference is that with A18 and A14 NNI 
there is the indication of the neutral pion condensation phase transition, 
while there is none with U14.  However, the fact that A18 provides an
exact fit to the NN scattering data while U14 does not weighs against the U14 
results.

As is well known, the cooling of neutron stars is accelerated 
by pion condensation \cite{Pet92}. The detailed mechanism associated
with the spin-ordered condensate that we find is as yet unexplored.
In addition, we estimate that due to this phase transition, there may be 
regions in the star where the matter density changes rapidly over a ten 
meter distance scale.  The matter in this thin layer 
of the star can have interesting structures on a ten fermi length scale.

We find that inclusion of the relativistic boost 
correction, $\delta v$, to the NNI 
increases the mass limit from 1.67\msol to 1.8\msol without any TNI, 
while it is reduced from 2.38\msol to 2.20\msol upon inclusion of the Urbana 
models of TNI.  The reduction occurs because the TNI needed to fit nuclear 
data have weaker repulsive parts after including $\delta v$.  For the 
same reason, the effect of Urbana models of TNI on the mass limit is diminished 
from 0.71\msol to 0.4\msol when the $\delta v$ is included.  Note that these 
models of TNI have only two terms; the long range two-pion exchange term, 
and a short range term with no spin-isospin dependence.  Their strengths
are determined from the density of SNM and the triton energy. However the 
triton, which has isospin 1/2, is insensitive to the interaction between
three neutrons.  Improved models of TNI must consider data such as the 
binding energy of $^8$He, which are sensitive to the interaction between three 
neutrons.

If the effective value of the bag constant B is larger than 122 MeV/fm$^3$ 
it appears that only the heaviest neutron stars may have small drops of 
quark matter in their interior.  The quark composition of these drops is 
similar to that of aggregates of $\Sigma^-$ hyperons.  If the interaction 
of $\Sigma^-$ with dense nucleon matter is not repulsive, there may exist 
$\Sigma^-$ hyperons in nucleon matter at densities below the threshold 
for the appearence of quark drops.  It is necessary to build $N\Sigma^-$ 
and $NN\Sigma^-$ interaction models to more fully explore this possibility.

Using maximally incompressible matter at high densities we find that 
the upper limit for the maximum mass
consistent with nuclear data is $\sim$2.5\msol ; this is 
not far from the prediction of 2.21\msol of our 
A18+$\delta v$+UIX$^*$ model.  The lower limit for the maximum mass 
is more difficult to establish due to the unknown interactions between 
hyperons and nucleons.  It may be as low as 1.74 $M_{\odot}$ if the 
quark bag constant B has a value of $\sim 122$ MeV/fm$^3$, 
and if the three neutron interaction is not as 
repulsive as the Urbana model IX at high densities, even without 
admixture of hyperons in matter.  Recently several authors \cite{nx1,nx2,nx3} 
have argued that there are indications of the existence of neutron stars 
with $M \sim 2M_{\odot}$.  If these are confirmed, then models without 
TNI will be ruled out. However, such a possibility is still the
subject of active debate \cite{nx4}.

\acknowledgements
The authors thank F. Arias de Saavedra, G. Baym, D. Markovic, 
C. J. Pethick, J. Wambach and R.B. Wiringa for useful discussions.
This work was supported by the U.S.\ National Science
Foundation through grant PHY 94-21309. VCS calculations of nucleon matter 
were performed on the Cray C90 at the Pittsburgh Supercomputing Center.

\appendix

\section{Form of Effective Hamiltonian}

The effective interactions that fit the models examined in
the first part of the paper all have the form
\beqa
H & = & \left(\frac{\hbar^2}{2m}+ (p_3 + (1-x_p)p_5)\rho e^{-p_4 \rho}
                           \right) \tau_n
 +  \left(\frac{\hbar^2}{2m}+ (p_3 + x_p p_5)\rho e^{-p_4 \rho}
                           \right) \tau_p \nonumber \\
 & + &  g(\rho, x_p=.5) \left(1 - (1 - 2 x_p)^2\right)
             + g(\rho, x_p=0) (1 - 2 x_p)^2   , \label{happ}
\eeqa
where $\rho = \rho_n + \rho_p$, and at zero temperature
\beq
\tau_p = \frac{3}{5} (3\pi^2\rho)^{2/3} x_p^{5/3}, \qquad
\tau_n = \frac{3}{5} (3\pi^2\rho)^{2/3} (1-x_p)^{5/3}.
\eeq
The parameters defining the $\tau$-dependent
terms are the same for all of the models, and are given in the
caption of Table~\ref{tab:t12}.  For the A18 and A18+$\delta v$ models at 
all densities and for the LDP of models with TNI, the parametrization is
\beqa
g_L(\rho, x_p=0.5) & = & -\rho^2\left(p_1+p_2\rho+p_6\rho^2
     + (p_{10}+p_{11}\rho) \;{\rm e}^{-p_9^2\rho^2}\right), \nonumber \\
g_L(\rho, x_p=0) & = & -\rho^2\left(p_{12}/\rho+p_7+p_8 \rho
              +p_{13}\;{\rm e}^{-p_9^2\rho^2}\right), \label{gl}
\eeqa
while for the HDP of models with TNI,
\beqa
g_H(\rho,x_p=0.5)  =  g_L(\rho, x_p=0.5)
& -& \rho^2\Big( p_{17}(\rho-p_{19}) \nonumber \\
  &+&p_{21}(\rho-p_{19})^2\Big){\rm e}^{p_{18}(\rho-p_{19})}, \nonumber \\
g_H(\rho,x_p=0)  =  g_L(\rho, x_p=0)
& -& \rho^2\Big( p_{15}(\rho-p_{20}) \nonumber \\
       & + & p_{14}(\rho-p_{20})^2\Big){\rm e}^{p_{16}(\rho-p_{20})}, \label{gh}
\eeqa
The values of the parameters are as given in Table~\ref{tab:t12}.

\section{Moment of Inertia of a Slowly Rotating Star}

We use the general relativistic equations for a slowly
rotating star as described by Hartle \cite{Har67}.
The metric for the nonrotating star is
\beq
ds^2 = - e^{\nu(r)} dt^2 + e^{\lambda(r)} dr^2 + r^2 (d\theta^2 +
\sin^2 \theta d\phi^2).  \label{metric}
\eeq
It involves the radial functions $\nu(r)$ and $\lambda(r)$.
The Oppenheimer-Volkoff equations for the pressure $P(r)$, mass function
$m(r)$ and $\Lambda (r)$ are given in Sec.~V, 
and $\lambda (r) = ln(\Lambda (r))$. 
The function $\nu(r)$ is defined by
\beq
\frac{d \nu}{d r} = \frac{2 G (m+4 \pi r^3 P/c^2) \Lambda(r)}{r^2} ,
\eeq
with the boundary condition $e^{\nu (R)} = 1/\Lambda(R)$,
and there is also an equation for the rotational drag, $\bar{\omega}(r)$,
\beq
\frac{d}{d r}\left( r^4 j \frac{d \bar{\omega}}{d r}\right) =
- 4 r^3 \frac{d j}{d r} \bar{\omega}.
\eeq
Here $j(r) = e^{-(\nu(r)+\lambda(r))/2}$; it has the boundary
value $j(R)$= 1.  In the limit of slow rotation, such that the angular
velocity $\Omega \ll G M/ R^2 c$, $\bar{\omega}(r)$ has the boundary
condition $ \bar{\omega}(R)/\Omega = 1 - 2 G I/R^3 c^2$.
$I$ is the total moment of inertia, given by either of the integrals
\beq
I = - \frac{2 c^2}{3 G} \int_0^R r^3 \frac{d j(r)}{d r}
\frac{\bar{\omega}(r)}{\Omega} dr
  = \frac{8\pi}{3} \int_0^R r^4 \left(\tilde{\rho}+\frac{P}{c^2}\right) \Lambda(
r) j(r)
\frac{\bar{\omega}(r)}{\Omega} dr.   \label{iexact}
\eeq
This set of equations, together with (\ref{dpdr}), is integrated from $r=0$
to the value $r=R$ where the pressure becomes negligible,
with a given equation of state $P = P(\tilde{\rho})$, and a central density
$\tilde{\rho}(0)$ chosen to give the desired neutron star mass.
One then has also the radius and, after satisfying the boundary conditions,
the moment of inertia.

\section{Details of the Nucleon-Quark Equilibrium}

The nuclear matter-quark matter strong interaction
equilibrium discussed in Sec.V-B
is illustrated graphically in Fig.~\ref{fig:fig21}.  
The $x_p > 0$ (right side) of the plot relates to nucleon matter and
the left relates to quarks. Although states of quark matter can extend to
$x_c = 2$ only the $x_c \la 0$ part is relevant here.
The chemical potentials of isobars corresponding to 
ten equally spaced pressures are shown as dashed lines.
Equilibria corresponding to Eq.~(\ref{munq}) and to pressure 
are represented by the rectangles.
The top (bottom) horizontal sides of the 
rectangles join states with the same $\mu_n$ $(\mu_p)$ 
on quark and nucleon isobars corresponding to the
same pressure.  (Only in a few cases do the pressures involved
correspond to those of the displayed grid.)
The vertical sides ensure that the chemical potentials, $\mu_n$ 
and $\mu_p$ on nucleon side, and $2 \mu_d + \mu_u$ and $2 \mu_u + \mu_d$ 
on the quark side, belong to matter with the same charge per baryon.
Other properties of interest, such as densities, 
are not represented on the plot.
These equilibria are a property purely of the strong 
interactions, irrespective of leptons
and charge neutrality.  The latter property then determines
the quark fraction $f_Q$, defined in Eq.~(\ref{fq}).  The rectangles 
labelled {\it b, c, d} and {\it e}
have values of $f_Q$ going from 0.001 to 0.999 and represent physical
states, while those labelled {\it a} and {\it f} give $f_Q = -0.169$ and 
$f_Q = 2.38$ corresponding to unphysical states.
However, they all correspond to equilibrium
under the strong interactions.

\newpage

\begin{figure}
\caption{
Central and tensor healing distances ($d_c$ and $d_t$),
for SNM and PNM, with A18 interaction alone (upper graph), and with
A18+UIX interaction (lower graph.)}
\label{fig:fig1}
\end{figure}

\begin{figure}
\caption{The energy per nucleon, $E(\rho)$, of SNM for various  
interaction models.}
\label{fig:fig2}
\end{figure}

\begin{figure}
\caption{The energy per nucleon, $E(\rho)$, of PNM for various 
interactions models.}
\label{fig:fig3}
\end{figure}

\begin{figure}
\caption{
Kinetic and interaction energies in A18+$\delta v$+UIX$^*$ 
model of SNM.}
\label{fig:fig4}
\end{figure}

\begin{figure}
\caption{ 
Kinetic and interaction energies in A18+$\delta v$+UIX$^*$ 
model of PNM.}
\label{fig:fig5}
\end{figure}

\begin{figure}
\caption{ 
The PNM and SNM energies for the
A18+$\delta v$+UIX$^*$ 
model, and the fits to them using an
effective interaction.  The full lines represent the stable
phases, and the dotted lines are their
extrapolations.}
\label{fig:fig6}
\end{figure}

\begin{figure}
\caption{
On a plot of proton fraction $x_p$ {\it vs.} baryon density, for the
A18+$\delta v$+UIX$^*$ 
model, the boundary between the LDP and HDP, obtained in the
manner described in the text.  The dashed curve is the 
proton fraction of beta-stable matter, and the dotted lines
mark the boundary of the mixed phase region.}
\label{fig:fig7}
\end{figure}

\begin{figure}
\caption{
For beta-stable matter, the proton fractions $x_p$ for the
four models discussed in the text {\em vs.} baryon density.
The dashed curve, U14-DDI (FPS), is from \protect\cite{LRP93}, 
the dotted line, A18(LOB) is from \protect\cite{Eng97},
the points are from \protect\cite{WFF88,Wir98b} for the U14+UVII model,
and the dotted line is the threshold for the direct Urca cooling 
process, as a function of $\rho$.}
\label{fig:fig8}
\end{figure}

\begin{figure}
\caption{
For beta-stable matter, the electron chemical potentials 
$\mu_e$ for the
four models discussed in the text {\it vs.} baryon density.
The dashed curve, U14-DDI (FPS), is from \protect\cite{LRP93}, 
and the horizontal line, at $\mu_e = m_\mu c^2$, is the threshold 
for a muon contribution to the lepton fraction.}
\label{fig:fig9}
\end{figure}

\begin{figure}
\caption{
Comparison of the energies of PNM and SNM obtained for the A18 model 
with the VCS method in the present work, and with LOB calculations by 
Engvik {\em et al.} \protect\cite{Eng97}.}
\label{fig:fig10}
\end{figure}

\begin{figure}
\caption{
Neutron star gravitational mass, in solar masses, {\it vs.} central 
baryon density, for the four models described in the text.
The full curves are for beta-stable matter, and the
dotted lines are for pure neutron matter.  The vertical lines show
the density above which the matter is superluminal.
The dashed curve, FPS, is from \protect\cite{LRP93}.}
\label{fig:fig11}
\end{figure}

\begin{figure}
\caption{
Neutron star gravitational mass, in solar masses, {\it vs.} 
radius, in kilometers, for the four models described in the text.
The full curves are for beta-stable matter, and the
dotted ones are for pure neutron matter.  
The dashed curve, FPS, is from 
\protect\cite{LRP93}.}
\label{fig:fig12}
\end{figure}

\begin{figure}
\caption{
For beta-stable matter according to the
A18+$\delta v$+UIX$^*$
model, the total energy per baryon {\it vs.} baryon density (full curve).
The dashed curves are for the assumption that matter
is maximally incompressible for densities greater than
the indicated value, and the lower full curve is for the A18+$\delta v$ model.}
\label{fig:fig13}
\end{figure}

\begin{figure}
\caption{
Neutron star gravitational mass, in solar masses, {\it vs.} radius 
for the A18+$\delta v$+UIX$^*$  
model (upper full curve), for the maximally incompressible modifications
of this model at densities beyond choosen values of $\rho_I$
(dashed curves), and for the A18+$\delta v$ model (lower full curve).}
\label{fig:fig14}
\end{figure}

\begin{figure}
\caption{
The neutron and electron chemical potentials in beta stable matter 
according to models A18+$\delta v$+UIX$^*$ (full line) and A18+$\delta v$ 
(dashed line).  Threshold densities for the 
appearance of noninteracting hyperons are marked by 
horizontal line segments.}
\label{fig:fig15}
\end{figure}

\begin{figure}
\caption{
For beta-stable matter, the energies per unit volume for the 
A18+$\delta v$+UIX$^*$ and the A18+$\delta v$ 
models, and the quark bag models with B = 122 and 200 MeV/fm$^3$
are shown by full and dashed lines;
the dotted lines correspond to neutral mixtures of charged 
nuclear and quark matter.}
\label{fig:fig16}
\end{figure}

\begin{figure}
\caption{
For the A18+$\delta v$+UIX$^*$ 
models and the quark bag model with B = 200 MeV/fm$^3$,
the proton fraction of nucleon matter $x_p$, the  
charge $(x_c)$ and strangeness $(x_s)$ per baryon of quark 
matter, the volume fraction $f_Q$ occupied by quark matter,
and the charge $x_{\ell}$ per baryon carried by the leptons in the mixture.}
\label{fig:fig17}
\end{figure}

\begin{figure}
\caption{
Neutron star gravitational mass, in solar masses, {\it vs.} 
central baryon density, for 
the A18+$\delta v$+UIX$^*$ and A18+$\delta v$ models 
with (dashed curves) and without (solid lines) quark matter 
admixture. The two dashed curves for the A18+$\delta v$ model 
are for B = 122 and 200 respectively.}
\label{fig:fig18}
\end{figure}

\begin{figure}
\caption{
Density profiles of 2.1M$_{\odot}$ and 1.41M$_{\odot}$ 
stars of beta-stable matter
using the A18+$\delta v$+UIX$^*$ 
model without quark matter admixture (solid lines), and of 
2.00M$_{\odot}$ star with quark matter admixture (dashed line),
assuming a bag constant $B=200$~MeV~fm$^{-3}$.  }
\label{fig:fig19}
\end{figure}

\begin{figure}
\caption{
The sound speed and the adiabatic index $\Gamma$ 
{\it vs.} $\tilde{\rho}/m$ 
for beta-stable matter according to the
A18+$\delta v$+UIX$^*$ (solid curves) and A18+$\delta v$ (long dashed curves)
models. The short dashed curves are for the assumption that matter
is maximally incompressible for densities greater than
the indicated value. The dotted curves are for the nucleon-quark
mixed phase.}
\label{fig:fig20}
\end{figure}

\begin{figure}
\caption{
Graphical representation of the strong interaction equilibrium
between the A18+ $\delta v$+UIX$^\ast$ model of
nucleon matter and the $B = 200$~MeV quark matter for the
density region of the mixing (see text).   
Dashed curves labelled 1 through 10 are strong interaction isobars, 
{\em i.e.} without considering charge effects and leptons.  They are 
for pressures from 100~MeV~fm$^{-3}$
to 550~MeV~fm$^{-3}$ in intervals of 50~MeV~fm$^{-3}$.
The upper (lower) set of curves on the right side show the neutron 
(proton) chemical  
potential {\it vs} proton fraction (right side), and 
the quark-matter counterparts on the left side show
$\mu_u + 2\mu_d$ ($2\mu_u + \mu_d$) {\it vs} charge per baryon $x_c$.
The corners of the rectangles correspond to the equilibrium values
of the chemical potentials and charge fractions.}
\label{fig:fig21}
\end{figure}

\newpage

\begin{table}
\caption{Cluster contributions to A18 SNM $E(\rho)$ in MeV.}
\begin{tabular}{cccccccccc}
$\rho$&$T_F$&$\langle T\rangle_{2B}$&$\langle T\rangle_{MB}$&
$\Delta T$&
$\langle v^s\rangle_{2B}$&
$\langle v^{MD}\rangle_{2B}$&
$\langle v^s\rangle_{MB}$&
$\langle v^{MD}\rangle_{MB}$&
$\delta E_{2B}$\\
\hline

0.04&  8.77&  5.33&-0.29& 0.16& -19.46&   0.03&  1.43&  0.17&-0.27\\
0.08& 13.93& 10.30&-1.06& 0.47& -36.84&   0.43&  4.44&  0.64&-0.57\\
0.16& 22.11& 19.64&-1.79& 1.11& -66.48&   2.04&  8.44&  2.58&-1.13\\
0.24& 28.97& 26.95&-2.47& 1.49& -91.46&   4.69& 12.06&  5.39&-1.73\\
0.32& 35.09& 34.45&-3.92& 1.63&-115.82&   7.89& 17.17&  9.37&-2.37\\
0.40& 40.72& 40.55&-3.23& 1.53&-135.38&  11.96& 17.57& 14.48&-3.03\\
0.48& 45.99& 46.22&-1.95& 1.17&-152.83&  16.69& 16.55& 20.76&-3.63\\
0.56& 50.96& 51.95&-1.82& 0.46&-170.59&  21.72& 17.92& 28.24&-4.17\\
0.64& 55.71& 58.03&-3.73&-0.99&-188.73&  26.61& 22.56& 36.91&-4.60\\
0.80& 64.64& 68.60& 4.47&-4.07&-202.42&  34.67&  8.42& 51.76&-5.13\\
0.96& 73.00& 79.71&-0.40&-8.27&-236.38&  47.43& 18.47& 80.12&-5.43

\end{tabular}
\label{tab:t1}
\end{table}

\begin{table}
\caption{Cluster contributions to A18 PNM $E(\rho)$ in MeV.}
\begin{tabular}{cccccccccc}
$\rho$&$T_F$&$\langle T\rangle_{2B}$&$\langle T\rangle_{MB}$&
$\Delta T$&
$\langle v^s\rangle_{2B}$&
$\langle v^{MD}\rangle_{2B}$&
$\langle v^s\rangle_{MB}$&
$\langle v^{MD}\rangle_{MB}$&
$\delta E_{2B}$ \\
\hline

0.04&  13.92&  5.08& -1.93& 0.14& -15.79&  0.41&  4.90&  -0.14& -0.39\\
0.08&  22.09&  8.21& -2.49& 0.26& -26.87&  1.34&  6.92&  -0.22& -0.46\\
0.16&  35.07& 15.00& -3.02& 0.43& -48.10&  4.15&  9.75&   0.28& -0.80\\
0.24&  45.95& 21.67& -2.89& 0.20& -67.45&  7.67& 11.01&   1.92& -1.20\\
0.32&  55.67& 28.50& -3.65&-0.54& -86.52& 11.95& 12.81&   5.16& -1.73\\
0.40&  64.60& 31.71& -0.28&-1.07& -98.77& 17.57&  6.97&   9.95& -2.33\\
0.48&  72.95& 37.39&  4.57&-1.23&-113.39& 23.89&  0.93&  15.57& -3.00\\
0.56&  80.84& 43.86&  3.15& 7.25&-141.33& 35.90& -1.98&  32.50& -3.86\\
0.64&  88.37& 56.20&  0.99& 8.55&-168.18& 40.75&  0.50&  45.53& -4.80\\
0.80& 102.54& 67.08&  2.35& 5.72&-201.26& 57.99& -6.50&  73.07& -7.00\\
0.96& 115.80& 80.44& -2.75& 2.83&-237.50& 74.78& -8.05& 112.66&-10.08

\end{tabular}
\label{tab:t2}
\end{table}

\begin{table}
\caption{Cluster contributions to A18+UIX SNM $E(\rho)$ in MeV.}
\begin{tabular}{cccccccccc}
$\rho$&$T_F$&$\langle T\rangle_{2B}$&$\langle T\rangle_{MB}$&
$\Delta T$&
$\langle v^s\rangle_{2B}$&
$\langle v^{MD}\rangle_{2B}$&
$\langle v^s\rangle_{MB}$&
$\langle v^{MD}\rangle_{MB}$&
$\delta E_{2B}$ \\
\hline
0.04&  8.77&   5.70& -0.34&  0.19& -20.00& -0.01&   1.60&   0.19&-0.28\\
0.08& 13.93&  10.16& -0.97&  0.46& -36.51&  0.44&   4.21&   0.64&-0.60\\
0.12& 18.25&  15.27& -0.74&  0.89& -51.63&  1.24&   5.25&   1.60&-1.10\\
0.16& 22.11&  20.97& -0.81&  1.21& -66.74&  2.12&   6.52&   3.01&-1.80\\
0.20& 25.65&  25.84& -0.46&  1.51& -80.18&  3.47&   7.12&   4.77&-2.54\\
0.24& 28.97&  30.51&  0.22&  1.87& -93.16&  5.38&   7.51&   7.10&-3.33\\
0.32& 35.09&  39.38&  1.32&  2.46&-117.90&  9.41&   8.60&  12.56&-5.21\\
0.40& 40.72&  53.32&  0.54&  6.24&-154.44& 15.18&  23.65&  25.65&-6.93\\
0.48& 45.99&  61.13&  1.08&  7.80&-178.46& 18.79&  27.12&  34.34&-7.93\\
0.56& 50.96&  69.68&  2.27&  8.51&-201.66& 24.75&  27.70&  45.69&-8.67\\
0.64& 55.71&  79.46&  1.92& 10.20&-227.19& 28.60&  32.93&  58.95&-9.20\\
0.80& 64.64&  93.07&  1.71& 12.89&-269.82& 39.35&  39.57&  87.26&-9.73\\
0.96& 73.00& 111.43& -6.19& 18.52&-321.23& 48.48&  61.29& 128.26&-9.93
\end{tabular}
\label{tab:t3}
\end{table}

\begin{table}
\caption{Cluster contributions to A18+UIX PNM $E(\rho)$ in MeV.}
\begin{tabular}{cccccccccc}
$\rho$&$T_F$&$\langle T\rangle_{2B}$&$\langle T\rangle_{MB}$&
$\Delta T$&
$\langle v^s\rangle_{2B}$&
$\langle v^{MD}\rangle_{2B}$&
$\langle v^s\rangle_{MB}$&
$\langle v^{MD}\rangle_{MB}$&
$\delta E_{2B}$ \\
\hline
0.02&   8.77&   3.89& -1.66& 0.08&  -10.20&  0.11&  3.70&  -0.06&-0.23\\
0.04&  13.92&   5.15& -1.77& 0.14&  -15.60&  0.38&  4.53&  -0.14&-0.42\\
0.08&  22.09&   8.43& -2.25& 0.30&  -26.69&  1.27&  6.37&  -0.23&-0.83\\
0.12&  28.95&  11.97& -2.70& 0.42&  -37.77&  2.51&  8.22&  -0.14&-1.13\\
0.16&  35.07&  15.90& -4.00& 0.18&  -49.36&  3.87& 11.44&   0.04&-1.26\\
0.20&  40.69&  22.37&  0.42& 2.78&  -59.04& 10.65&  3.40&   4.87&-1.29\\
0.24&  45.95&  25.45&  1.15& 3.25&  -69.41& 13.03&  2.89&   6.32&-1.65\\
0.32&  55.67&  35.82&  2.49& 4.44&  -93.04& 18.04&  2.48&  11.04&-2.45\\
0.40&  64.60&  42.87&  4.34& 5.38& -113.78& 25.07&  0.31&  16.94&-3.20\\
0.48&  72.95&  51.57&  6.35& 6.27& -135.45& 32.61& -2.18&  24.67&-4.00\\
0.56&  80.84&  61.73&  8.57& 7.07& -157.94& 40.68& -4.92&  34.25&-4.75\\
0.64&  88.37&  80.34& 12.00& 7.62& -185.72& 53.35& -8.56&  47.64&-5.45\\
0.80& 102.54& 101.95& 18.33& 7.90& -230.35& 77.63&-16.19&  74.42&-6.50\\
0.96& 115.80& 120.83& 21.91& 7.62& -273.05& 99.09&-22.17& 109.75&-7.35
\end{tabular}
\label{tab:t4}
\end{table}

\begin{table}
\caption{Contributions of pion-exchange and phenomenological parts of
nuclear interactions to the $E(\rho)$ in MeV.}
\begin{tabular}{c|cc|cccc|cc|cccc}
&SNM&&&SNM&&&PNM&&&PNM&& \\
&A18&&&A18+&UIX&&A18&&&A18+&UIX \\
$\rho$&
$\langle v^\pi_{ij}\rangle$&
$\langle v^R_{ij}\rangle$&
$\langle v^\pi_{ij}\rangle$&
$\langle v^R_{ij}\rangle$&
$\langle V^{2\pi}_{ijk}\rangle$&
$\langle V^R_{ijk}\rangle$&
$\langle v^\pi_{ij}\rangle$&
$\langle v^R_{ij}\rangle$&
$\langle v^\pi_{ij}\rangle$&
$\langle v^R_{ij}\rangle$&
$\langle V^{2\pi}_{ijk}\rangle$&
$\langle V^R_{ijk}\rangle$\\
\hline
0.04&-11.4& -6.4& -11.8& -6.4&  -0.4&  0.3& -1.7& -9.0& -1.8& -9.1&   0.1&  0.1\\
0.08&-19.2&-12.1& -19.1&-12.1&  -0.8&  1.5& -2.7&-16.1& -3.0&-16.3&   0.3&  0.8\\
0.12&     &     & -27.5&-16.0&  -2.1&  3.4&     &     & -3.9&-23.2&   0.6&  2.2\\
0.16&-32.6&-20.9& -35.3&-19.8&  -3.6&  6.4& -4.6&-29.3& -4.1&-29.9&   1.2&  4.5\\
0.20&     &     & -42.0&-22.8&  -5.5& 10.6&     &     &-17.2&-22.9&  -8.7& 10.1\\
0.24&-41.3&-28.1& -48.5&-24.6&  -8.1& 15.9& -5.8&-41.1&-19.1&-28.0& -10.1& 15.2\\
0.32&-47.7&-33.7& -59.8&-27.6& -13.3& 30.9& -6.3&-50.4&-25.1&-36.4& -17.4& 30.6\\
0.40&-54.7&-36.7& -72.6&-17.4& -38.4& 53.0& -7.4&-56.9&-30.1&-41.4& -24.2& 50.9\\
0.48&-61.2&-37.6& -81.4&-16.8& -50.2& 80.3& -9.6&-63.4&-35.7&-44.7& -34.1& 78.0\\
0.56&-65.7&-37.0& -91.5&-12.0& -65.2&114.5&-33.0&-41.9&-41.6&-46.3& -47.2&112.7\\
0.64&-66.9&-35.8& -99.4& -7.3& -82.0&155.9&-39.2&-42.2&-52.2&-41.1& -76.9&160.3\\
0.80&-67.1&-40.5&-112.1&  8.5&-117.7&260.2&-42.3&-34.4&-63.3&-31.2&-116.3&267.9\\
0.96&-70.9&-19.5&-121.2& 38.0&-169.4&397.7&-46.2&-11.9&-70.9&-15.5&-155.0&402.5

\end{tabular}
\label{tab:t5}
\end{table}

\begin{table}
\caption{The $E(\rho)$ of SNM in MeV.}
\begin{tabular}{cccccc}
$\rho$&A18&A18+$\delta v$&A18+UIX&A18+$\delta v$+UIX$^\ast$&corrected\\
\hline
0.04&  -4.28& -4.08&  -4.39&  -4.31&  -6.48\\
0.08&  -8.72& -8.07&  -8.06&  -7.97& -12.13\\
0.12&       &      & -10.52& -10.54& -15.04\\
0.16& -14.59&-12.54& -11.85& -12.16& -16.00\\
0.20&       &      & -11.28& -12.21& -15.09\\
0.24& -17.61&-13.69&  -8.99& -10.89& -12.88\\
0.32& -18.13&-11.87&   0.84&  -4.21&  -5.03\\
0.40& -16.37& -7.70&  12.23&   2.42&   2.13\\
0.48& -12.21& -1.01&  32.18&  15.56&  15.46\\
0.56&  -5.79&  8.16&  59.99&  34.42&  34.39\\
0.64&   2.76& 19.54&  95.05&  58.36&  58.35\\
0.80&  25.01& 45.24& 188.51& 121.25& 121.25\\
0.96&  56.51& 82.63& 313.46& 204.02& 204.02\\
\end{tabular}
\label{tab:t6}
\end{table}

\begin{table}
\caption{The $E(\rho)$ of PNM in MeV.}
\begin{tabular}{ccccc}
$\rho$&A18&A18+$\delta v$&A18+UIX&A18+$\delta v$+UIX$^\ast$\\
\hline
0.02&       &       &   4.35&   4.45\\
0.04&   6.06&   6.32&   6.23&   6.45\\
0.08&   8.53&   9.26&   9.21&   9.65\\
0.12&       &       &  12.71&  13.29\\
0.16&  12.33&  14.51&  17.38&  17.94\\
0.20&       &       &  23.47&  22.92\\
0.24&  16.69&  20.76&  28.85&  27.49\\
0.32&  22.19&  28.59&  43.28&  38.82\\
0.40&  29.41&  38.10&  63.79&  54.95\\
0.48&  38.91&  50.35&  90.46&  75.13\\
0.56&  49.08&  66.00& 123.93&  99.75\\
0.64&  59.37&  81.15& 165.40& 127.58\\
0.80&  88.27& 119.46& 273.37& 205.34\\
0.96& 125.29& 167.02& 412.30& 305.87\\
\end{tabular}
\label{tab:t7}
\end{table}

\begin{table}
\caption{Contributions of relativistic boost interactions to the
$E(\rho)$ in MeV.}
\begin{tabular}{c|cccc||cccc}
&A18&&&&A18+UIX\\
&SNM&&PNM&&SNM&&PNM\\
$\rho$&
$\langle\delta v\rangle_{2B}$&
$\langle\delta v\rangle_{MB}$&
$\langle\delta v\rangle_{2B}$&
$\langle\delta v\rangle_{MB}$&
$\langle\delta v\rangle_{2B}$&
$\langle\delta v\rangle_{MB}$&
$\langle\delta v\rangle_{2B}$&
$\langle\delta v\rangle_{MB}$\\
\hline
0.02&     &     &     &     &     &     & 0.08& 0.03\\
0.04& 0.15& 0.05& 0.21& 0.05& 0.15& 0.05& 0.21& 0.05\\
0.08& 0.45& 0.20& 0.59& 0.14& 0.44& 0.20& 0.58& 0.15\\
0.12&     &     &     &     & 0.85& 0.40& 1.10& 0.29\\
0.16& 1.32& 0.74& 1.71& 0.46& 1.33& 0.73& 1.72& 0.49\\
0.20&     &     &     &     & 1.88& 1.09& 2.60& 0.58\\
0.24& 2.42& 1.50& 3.16& 0.91& 2.49& 1.50& 3.42& 0.84\\
0.32& 3.76& 2.49& 4.92& 1.48& 3.86& 2.51& 5.46& 1.41\\
0.40& 5.18& 3.49& 6.64& 2.06& 6.04& 3.75& 7.73& 2.25\\
0.48& 6.69& 4.52& 8.67& 2.77& 7.92& 5.17&10.34& 3.20\\
0.56& 8.35& 5.59&12.28& 4.64& 9.96& 6.81&13.27& 4.23\\
0.64&10.16& 6.62&15.81& 5.97&12.29& 8.69&16.69& 4.81\\
0.80&12.61& 7.62&22.12& 9.07&17.12&11.90&23.83& 7.25\\
0.96&16.95& 9.17&29.62&12.11&23.11&14.61&31.91&10.59\\
\end{tabular}
\label{tab:t8}
\end{table}

\begin{table}
\caption{A18+UIX$^\ast$: Contributions to $\langle \delta v^{RE}\rangle$ 
in MeV.}
\begin{tabular}{c|cccc|cccc}
&SNM&&&&PNM\\
$\rho$&
$C_{2B}$&
$W_s(K)$&
$W_s(F)$&
$W_{cch}$&
$C_{2B}$&
$W_s(K)$&
$W_s(F)$&
$W_{cch}$\\
\hline
0.02&     &     &     &     & 0.04& 0.00& 0.02& 0.00\\
0.04& 0.09& 0.00& 0.04& 0.00& 0.11& 0.00& 0.04& 0.00\\
0.08& 0.24&-0.02& 0.15& 0.00& 0.31&-0.01& 0.12& 0.00\\
0.12& 0.46&-0.04& 0.31& 0.00& 0.58&-0.03& 0.23& 0.00\\
0.16& 0.73&-0.06& 0.54& 0.00& 0.93&-0.03& 0.39&-0.01\\
0.20& 1.03&-0.09& 0.81& 0.00& 1.32&-0.16& 0.50&-0.03\\
0.24& 1.36&-0.14& 1.11& 0.00& 1.73&-0.19& 0.68&-0.03\\
0.32& 2.11&-0.22& 1.85& 0.00& 2.77&-0.32& 1.14&-0.06\\
0.40& 3.25&-0.70& 3.17&-0.04& 3.90&-0.40& 1.69&-0.08\\
0.48& 4.26&-0.94& 4.35&-0.04& 5.23&-0.51& 2.34&-0.09\\
0.56& 5.36&-1.15& 5.61&-0.05& 6.71&-0.66& 3.05&-0.14\\
0.64& 6.63&-1.53& 7.20&-0.05& 8.51&-1.00& 3.68&-0.24\\
0.80& 9.02&-1.96& 9.58&-0.05&12.14&-1.43& 5.41&-0.32\\
0.96&12.44&-3.58&13.37&-0.16&16.23&-1.87& 7.65&-0.40
\end{tabular}
\label{tab:t9}
\end{table}

\begin{table}
\caption{A18+UIX$^\ast$: Contributions to $\langle \delta v^{LC}\rangle$ 
in MeV.}
\begin{tabular}{c|cccc|cccc}
&SNM&&&&PNM\\
$\rho$&
$C_{2B}$&
$W_s(K)$&
$W_s(F)$&
$W_{cch}$&
$C_{2B}$&
$W_s(K)$&
$W_s(F)$&
$W_{cch}$\\
\hline
0.02&     &     &     &     & 0.03& 0.00& 0.01& 0.00\\
0.04& 0.07& 0.00& 0.02& 0.00& 0.09& 0.00& 0.03& 0.00\\
0.08& 0.20&-0.02& 0.08& 0.00& 0.27&-0.01& 0.06& 0.00\\
0.12& 0.39&-0.04& 0.17& 0.00& 0.51&-0.01& 0.11&-0.01\\
0.16& 0.59&-0.06& 0.32&-0.01& 0.80&-0.02& 0.18&-0.02\\
0.20& 0.85&-0.07& 0.48&-0.02& 1.28&-0.12& 0.42&-0.03\\
0.24& 1.13&-0.11& 0.65&-0.03& 1.69&-0.13& 0.55&-0.04\\
0.32& 1.76&-0.16& 1.09&-0.06& 2.68&-0.20& 0.92&-0.07\\
0.40& 2.79&-0.52& 1.96&-0.14& 3.82&-0.23& 1.37&-0.11\\
0.48& 3.66&-0.64& 2.67&-0.23& 5.11&-0.29& 1.93&-0.16\\
0.56& 4.60&-0.72& 3.45&-0.31& 6.55&-0.37& 2.55&-0.21\\
0.64& 5.66&-0.89& 4.39&-0.43& 8.17&-0.59& 3.19&-0.25\\
0.80& 7.79&-0.96& 5.77&-0.78&11.69&-0.85& 4.81&-0.37\\
0.96&10.67&-1.49& 7.80&-1.34&15.68&-1.06& 6.82&-0.55
\end{tabular}
\label{tab:t10}
\end{table}

\begin{table}
\caption[Incompressible core and quark models.]
{Maximum gravitational masses, in $M_{\odot}$,
and moments of inertia $I$, in $M_{\odot}$~km$^2$,
for stars with beta-stable
matter and with PNM;
with incompressible (INC) matter at $\rho > \rho_I$~(fm$^{-3}$) for the 
A18+$\delta v$+UIX$^*$ model;
and with mixed nuclear-quark matter (NM+QM) 
phase (bag constant $B$ in MeV fm$^{-3}$).}

\vspace{0.1in}
\begin{tabular}{l|c|c|c||l|c|c}
  NM &  Max. mass & Max. mass & Max. $I$ &
 $\rho_I$ of INC    &  Max. mass & Max. $I$ \\
 models &   beta-stable & PNM  & beta-stable &Models & beta-stable & beta-stable \\
   \hline
A18+$\delta v$+UIX$^*$ &  2.20   &  2.21 & 115. &
 0.32 &  2.92 & 261. \\
A18+UIX             &  2.38   &  2.39 & 143. &
 0.48 &  2.46  & 157. \\
A18+$\delta v$             &  1.80 &  1.81 & 67. &
 0.64 &  2.26 & 123. \\
A18                        &  1.67   & 1.68 & 55. &
 0.86 &  2.19 & 115.  \\
 FPS                             &  1.80  & - & 73. & & & \\
  \hline   \hline
 NM+QM   & {\em B} & Max. mass & Max. $I$&
 {\em B} &  Max. mass & Max. $I$ \\
 models  & &  beta-stable & beta-stable &  & beta-stable & beta-stable \\
  \hline
A18+$\delta v$+UIX$^*$ &122 & 1.91 & 96. &
 200 & 2.02 & 107. \\
A18+$\delta v$ &   122 & 1.74 & 66. &
  200 & 1.76 & 67. \\
\end{tabular}
\label{tab:t11}
\end{table}

\begin{table}
\caption
[Parameter values for effective Hamiltonian]
{Parameter values for effective Hamiltonian for different models.
The parameters $p_3=89.8$~MeV fm$^{5}$, $p_4=0.457$~fm$^{3}$ and
$p_5=-59.0$~MeV fm$^{5}$ (see Eq.~(\ref{happ})) are common to all four models.
The dimensions of the parameters, involving MeV and/or powers of fm,
can be worked out from Eqs.~(\ref{happ}), (\ref{gl}) or (\ref{gh}).}
\vspace{0.1in}
\begin{tabular}{l|cccccccccc}
  model   &  $p_1$ &  $p_2$ &  $p_6$ &  $p_7$ &  $p_8$ &  $p_9$  &  $p_{10}$
&  $p_{11}$  &  $p_{12}$  & $p_{13}$ \\
\hline
A18+$\delta v$+UIX$^*$
&  337.2 & -382. & -19.1 & 214.6 & -384. & 6.4 & 69. & -33. & 0.35 & 0. \\
A18+UIX
 &  328.8 & -404.6 & -34. & 217.5 & -385.6 & 6.35 & 25.4 & 0. & 0.47 & -0.9  \\
A18+$\delta v$
 &  281.0 & -151.1 & -10.6 & 210.1 & -158. & 5.88 & 58.8 & -15. & -0.2 & -0.9 \\
A18
 &  297.6 & -134.6 & -15.9 & 215.0 & -116.5 & 6.42 & 51. & -35. & -0.2 & 0. \\
\hline
  model   &  $p_{14}$ &  $p_{15}$ &  $p_{16}$ &  $p_{17}$
 &  $p_{18}$  &  $p_{19}$  &  $p_{20}$  &  $p_{21}$ & & \\
\hline
A18+$\delta v$+UIX$^*$
 & 0. & 287. & -1.54 & 175.0 & -1.45 & 0.32 & 0.195 & 0.& & \\
A18+UIX
 & -452. & 217.1 & -1.0 & 100.3 & -1.19 & 0.32 & 0.2 & -275.& & \\
\end{tabular}
\label{tab:t12}
\end{table}

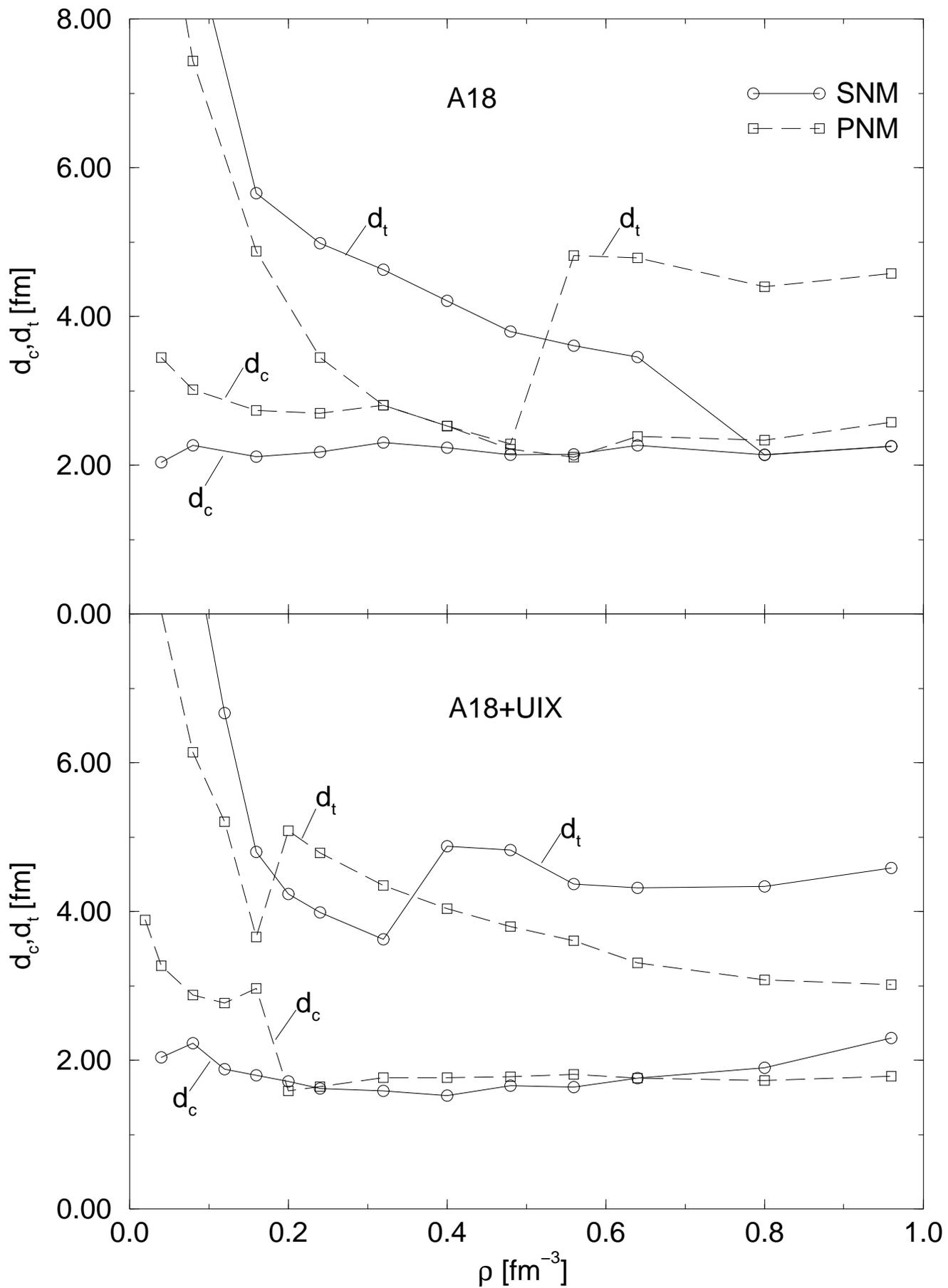
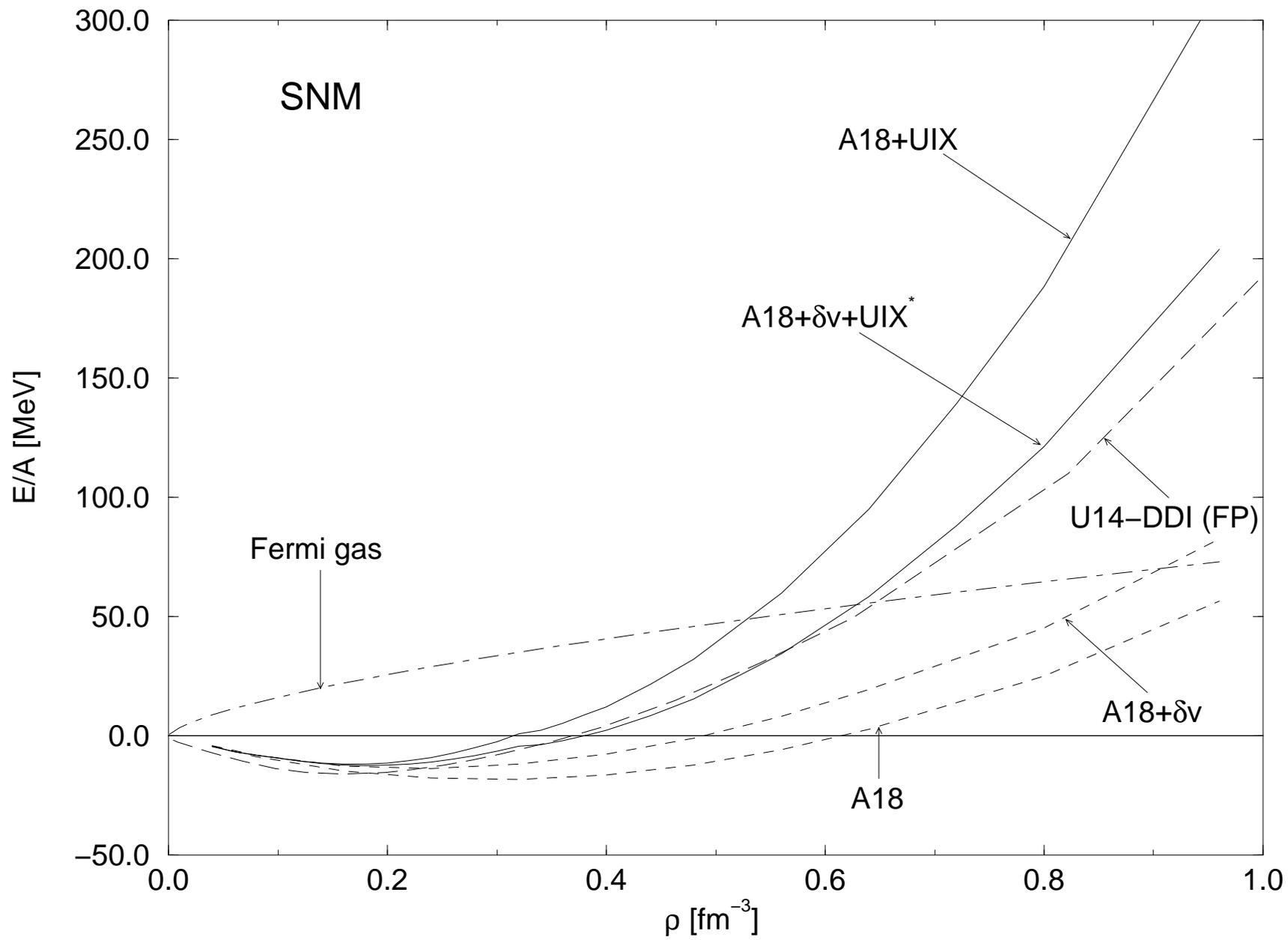
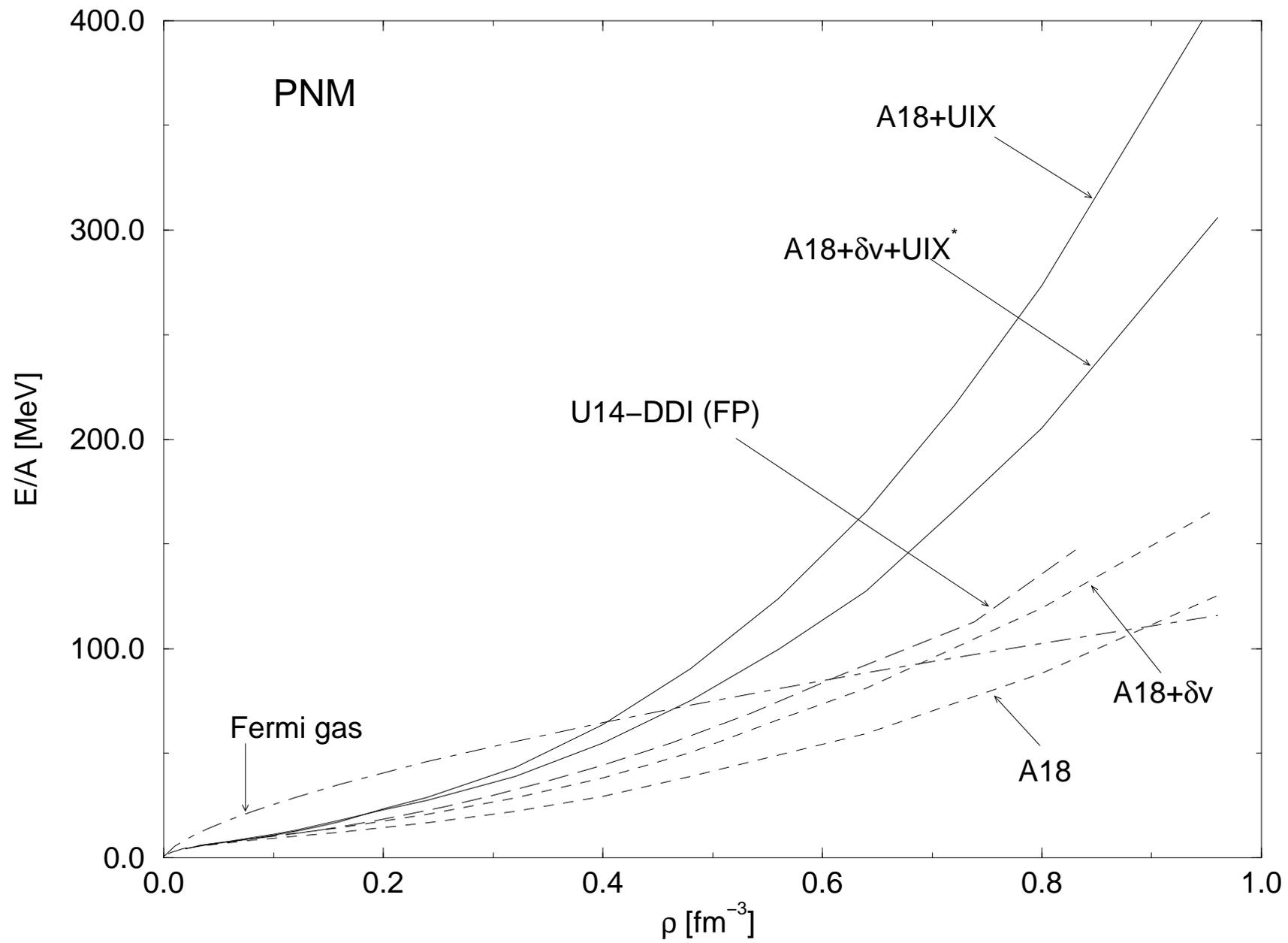
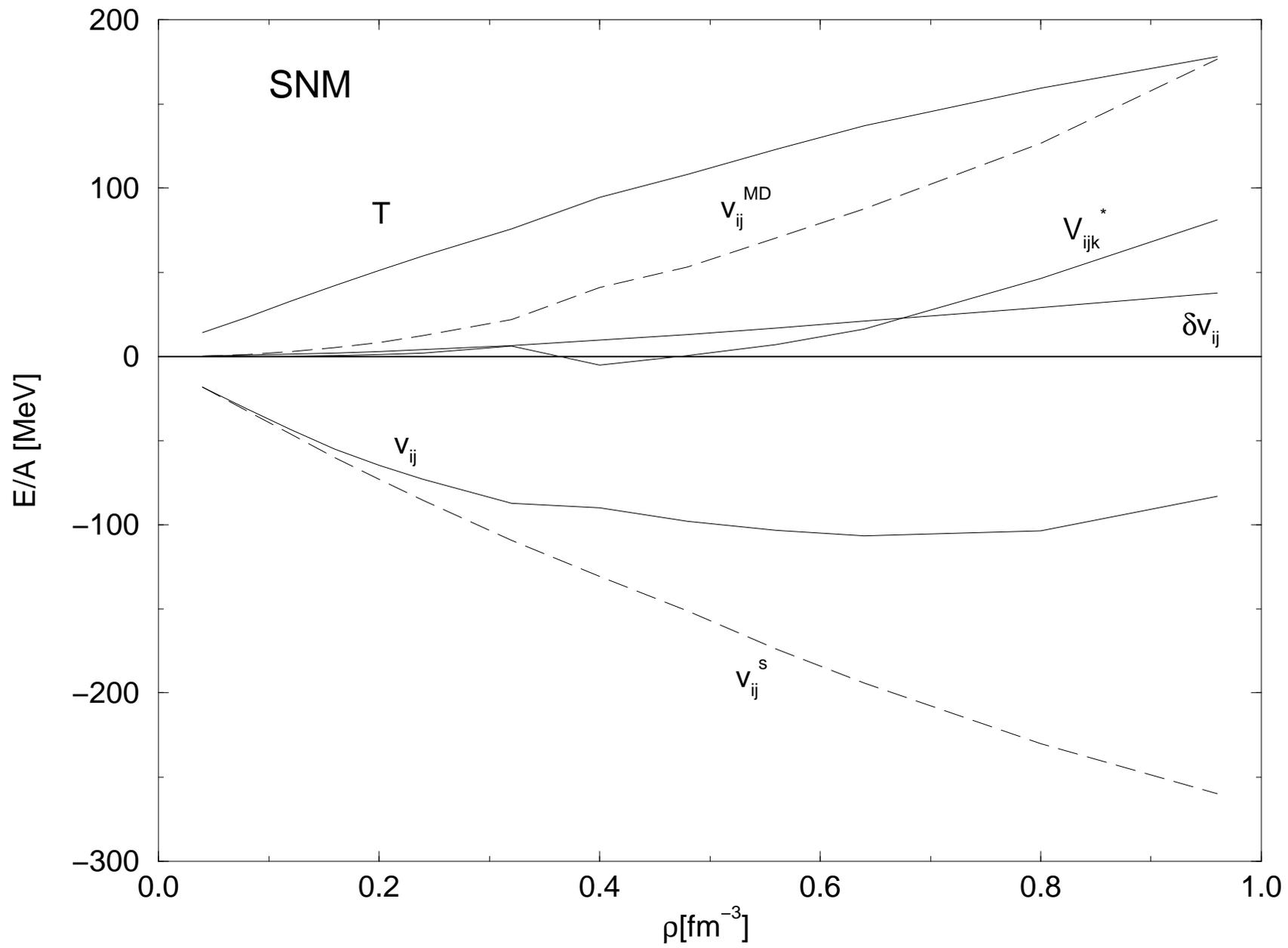
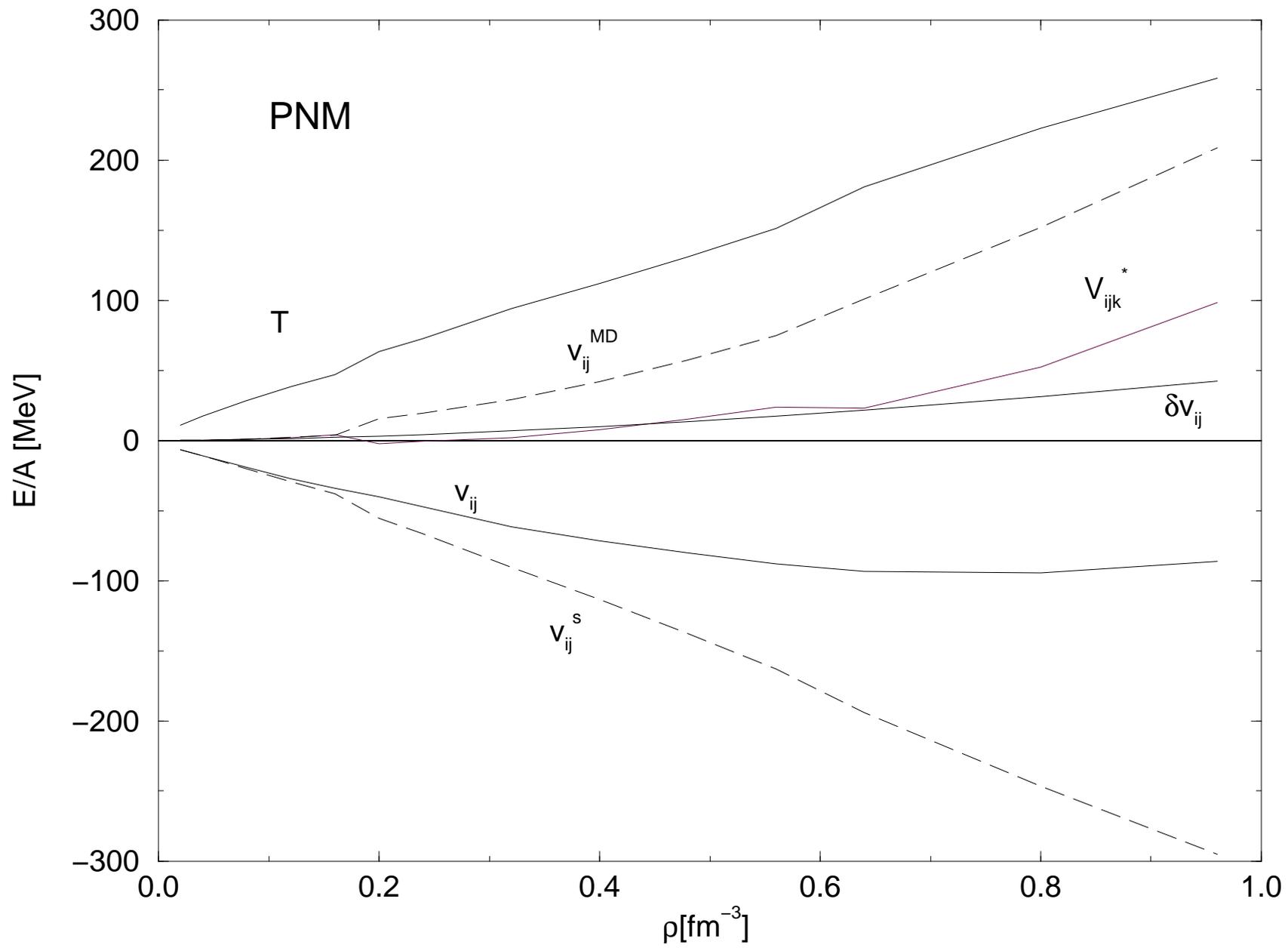
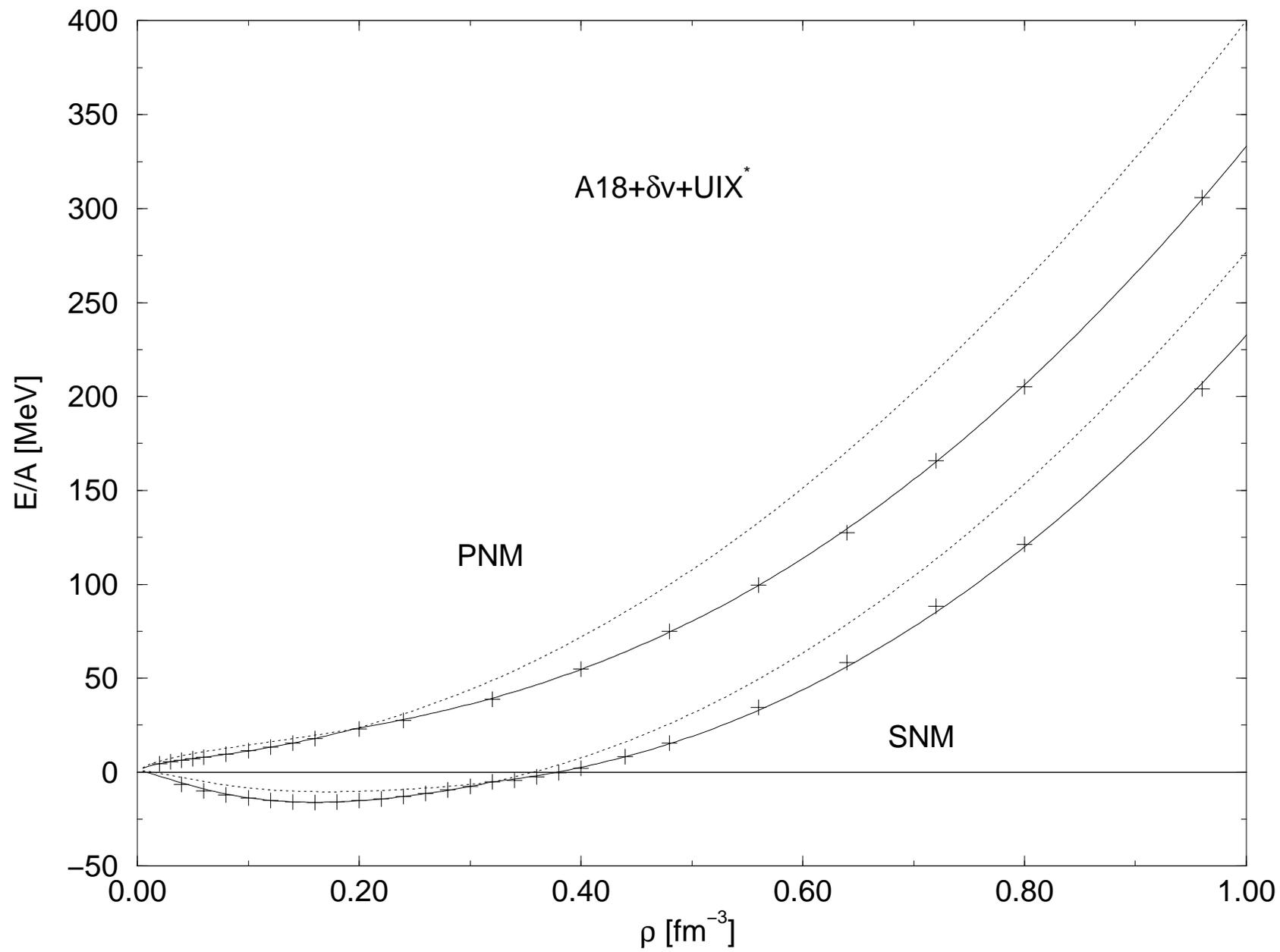
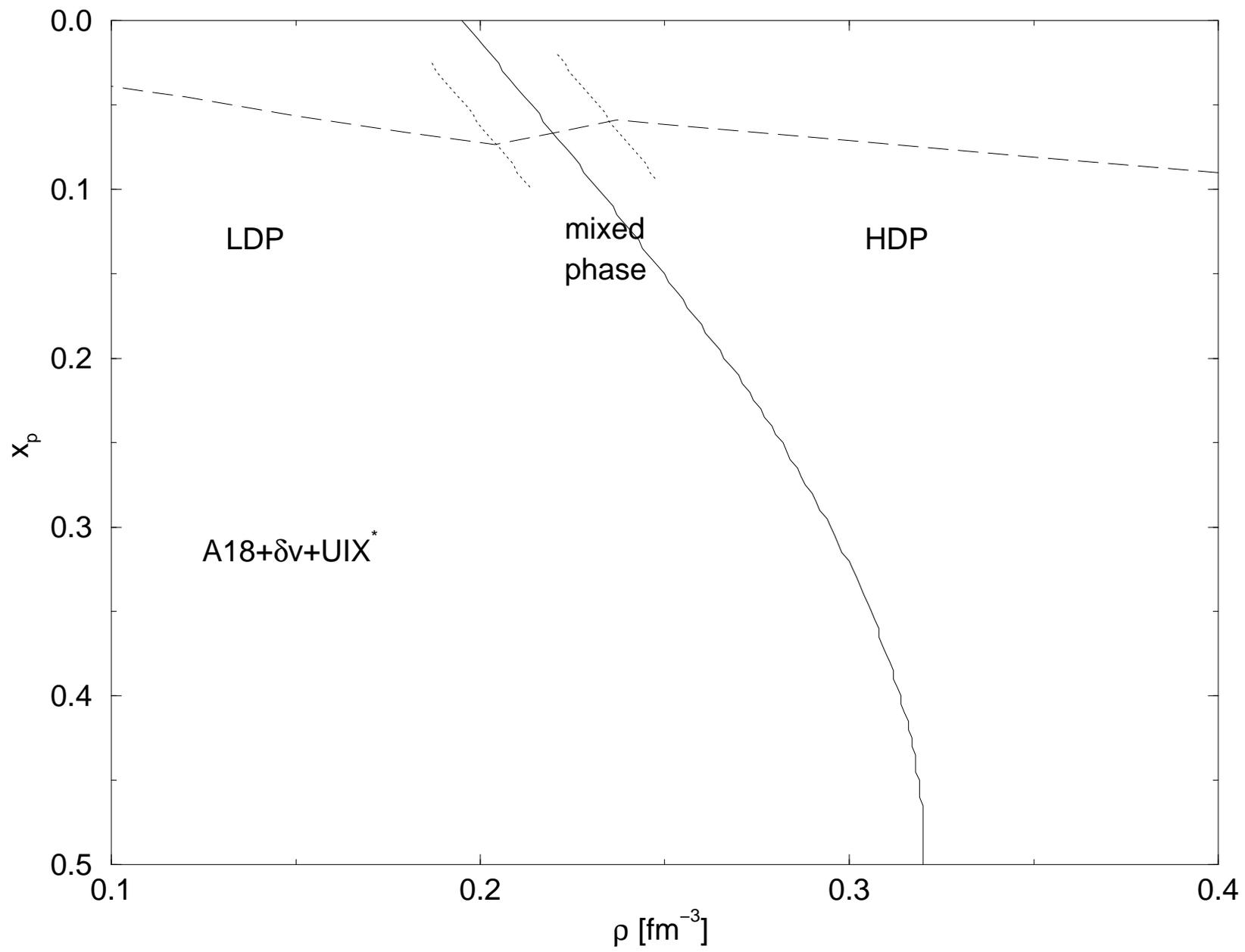
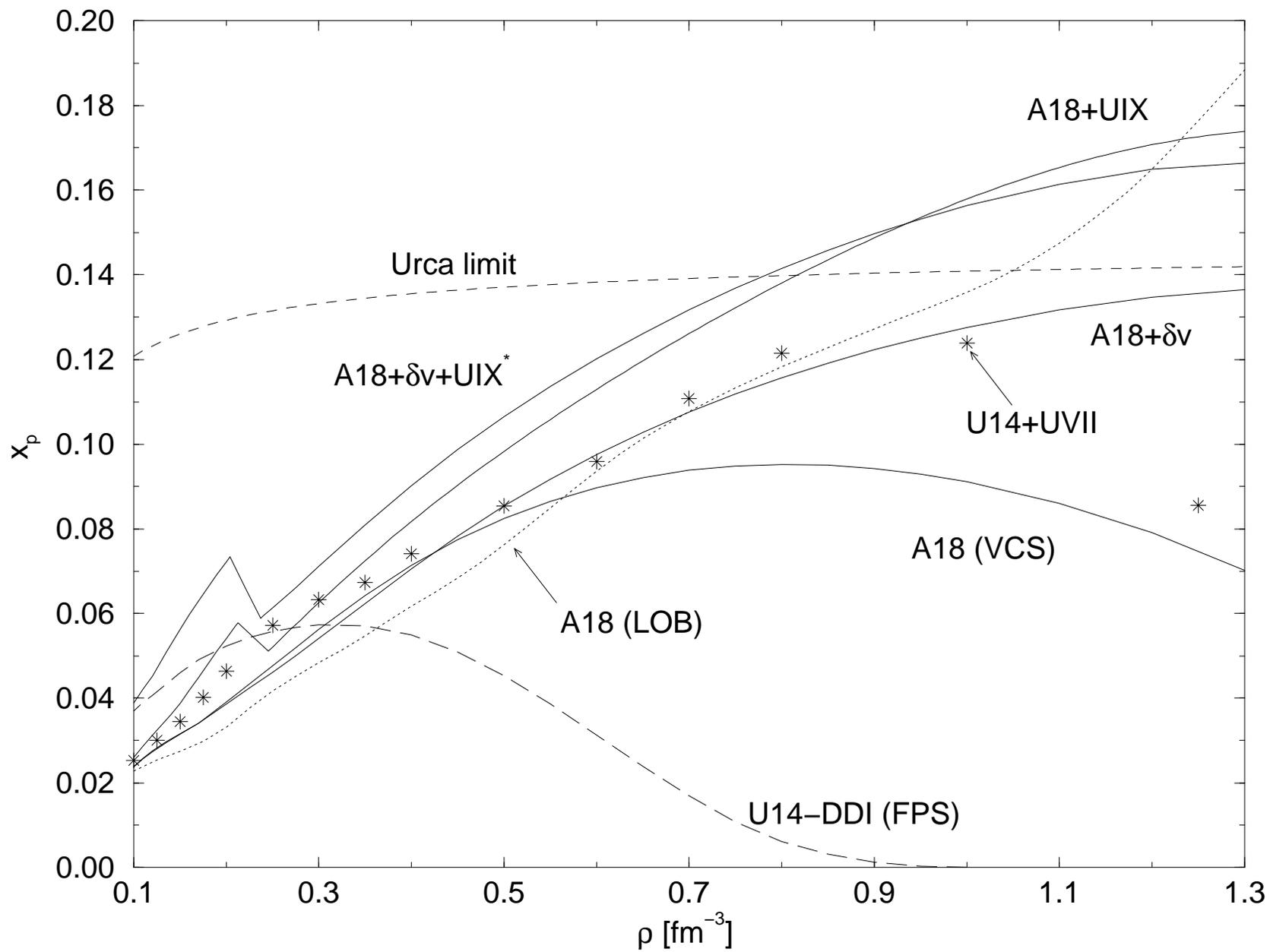
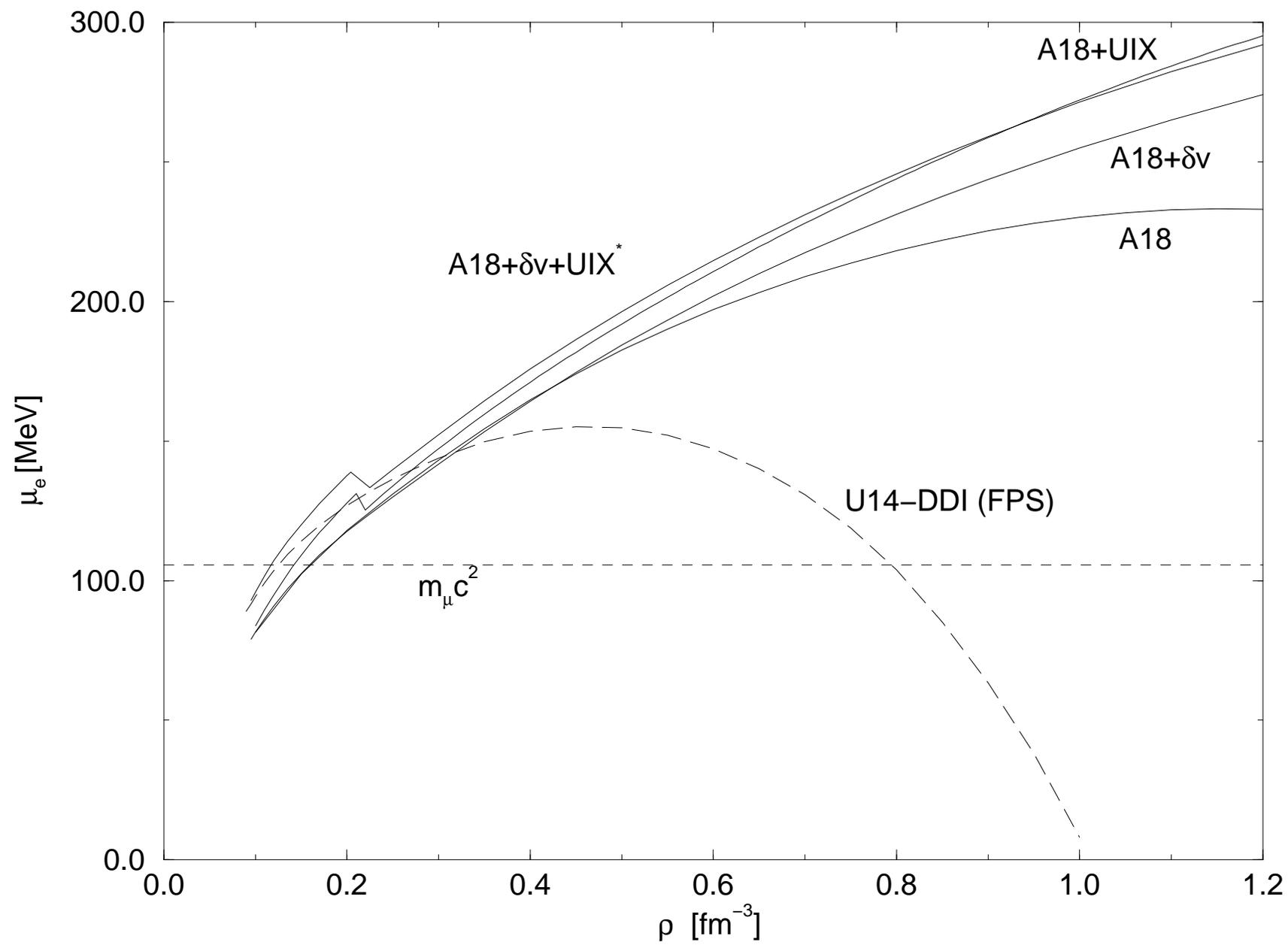
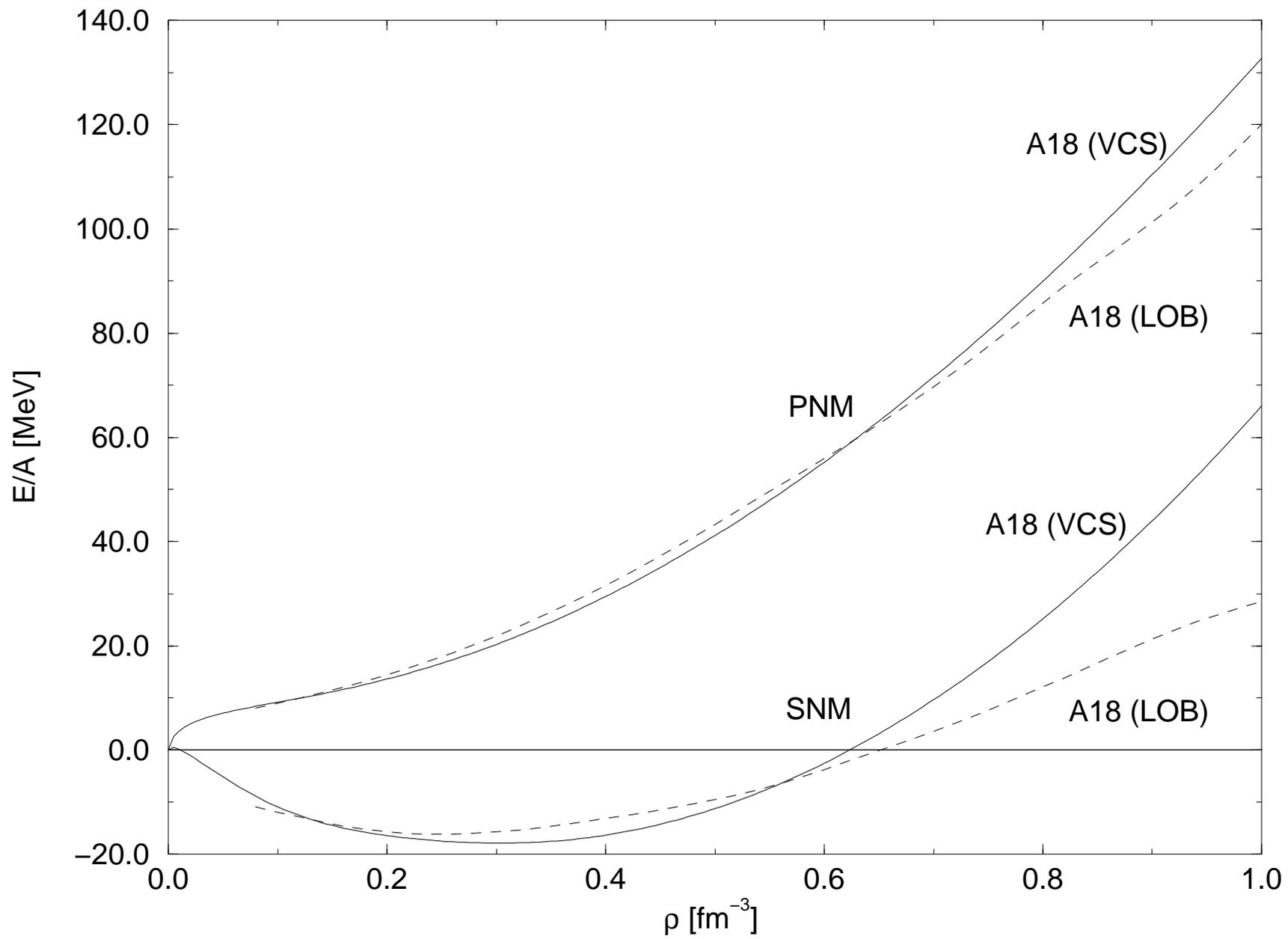
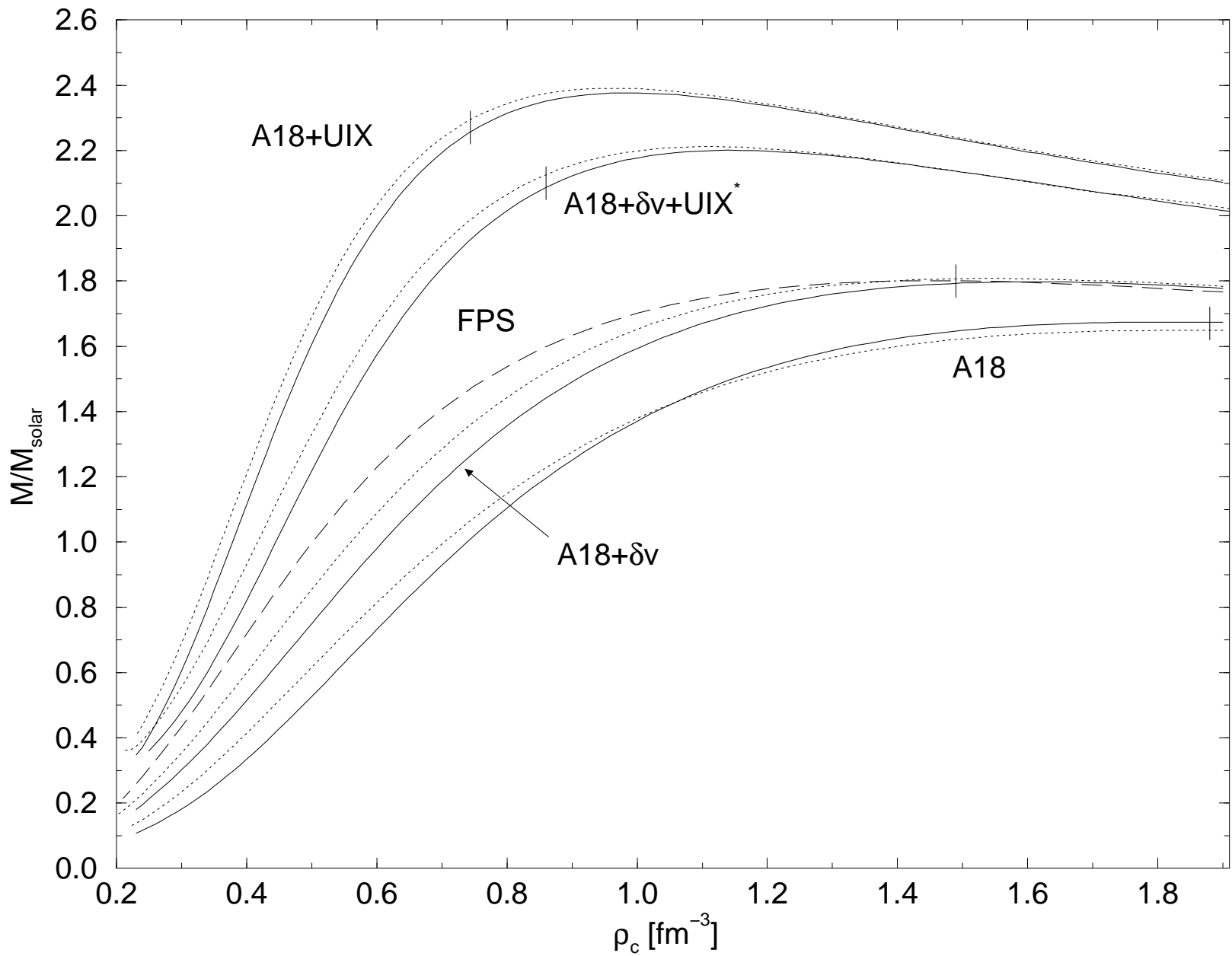
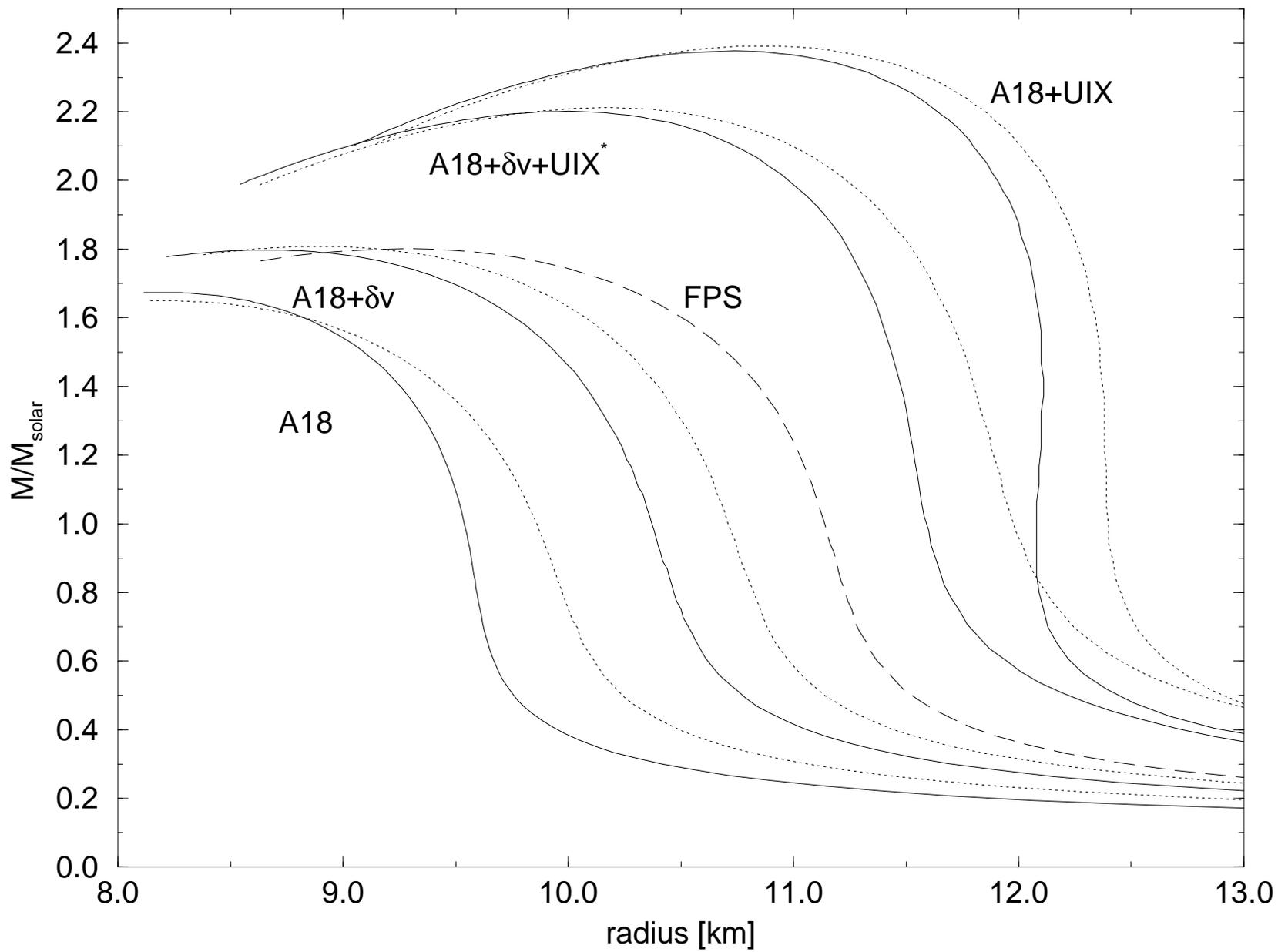
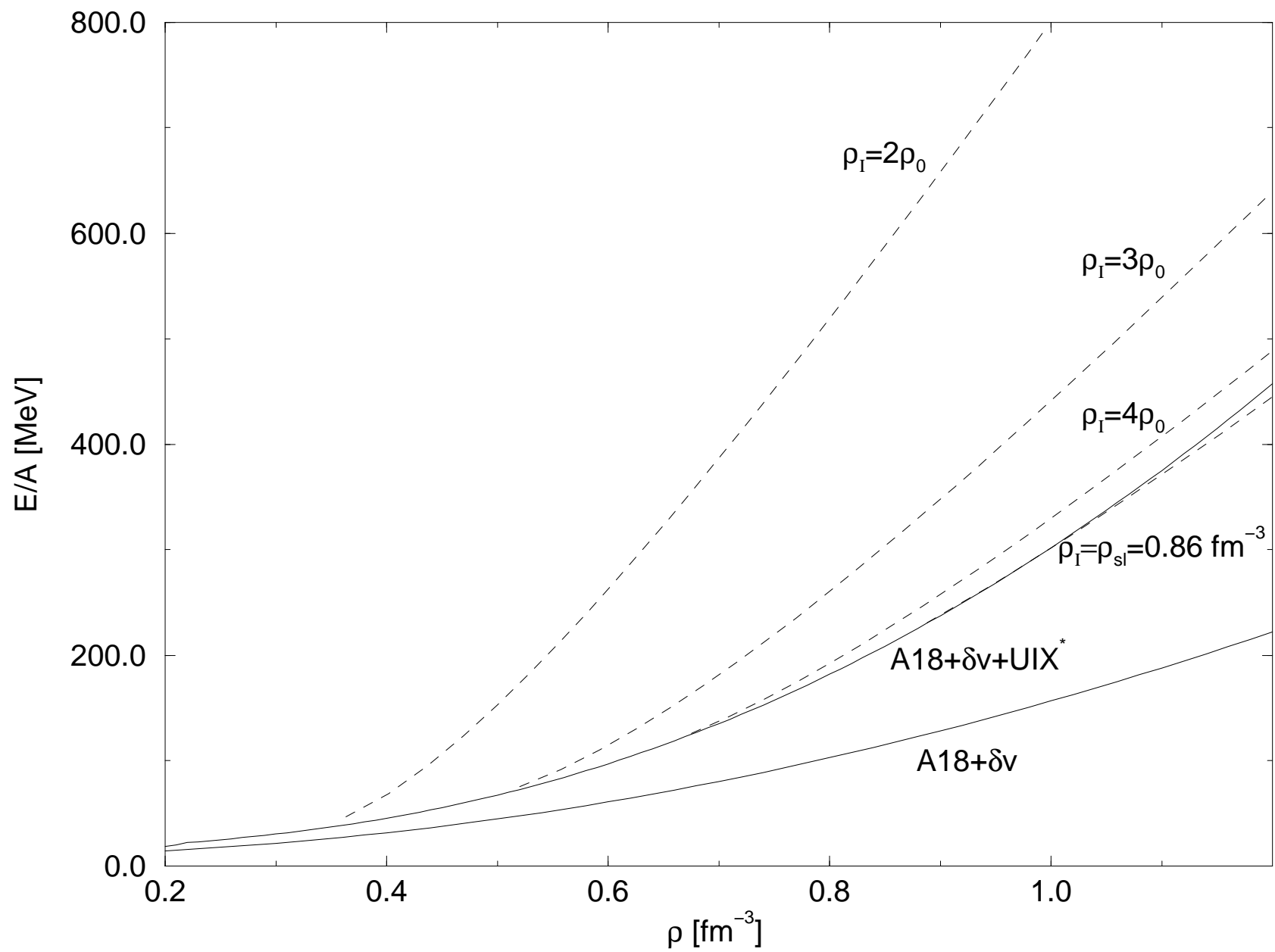
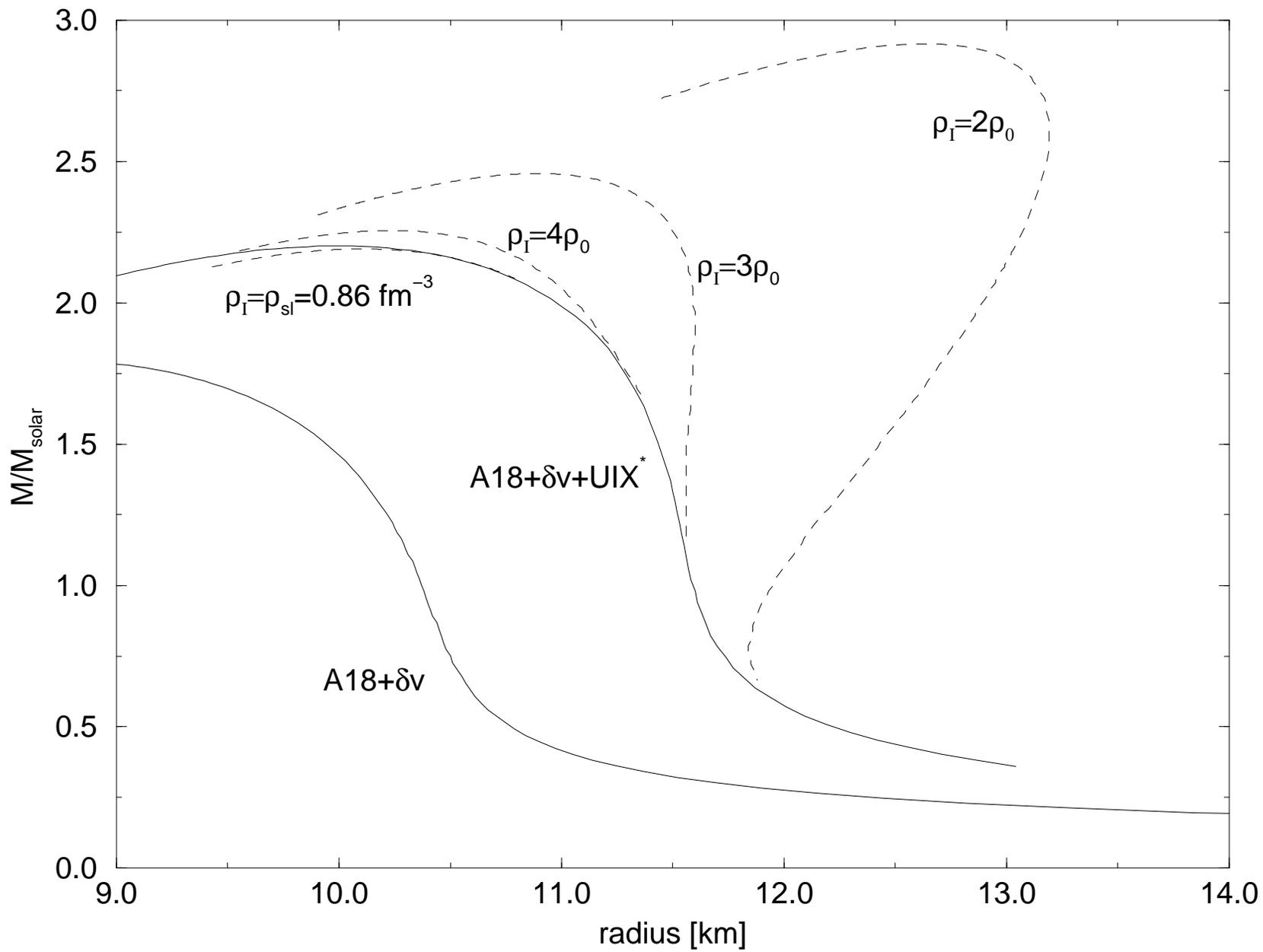
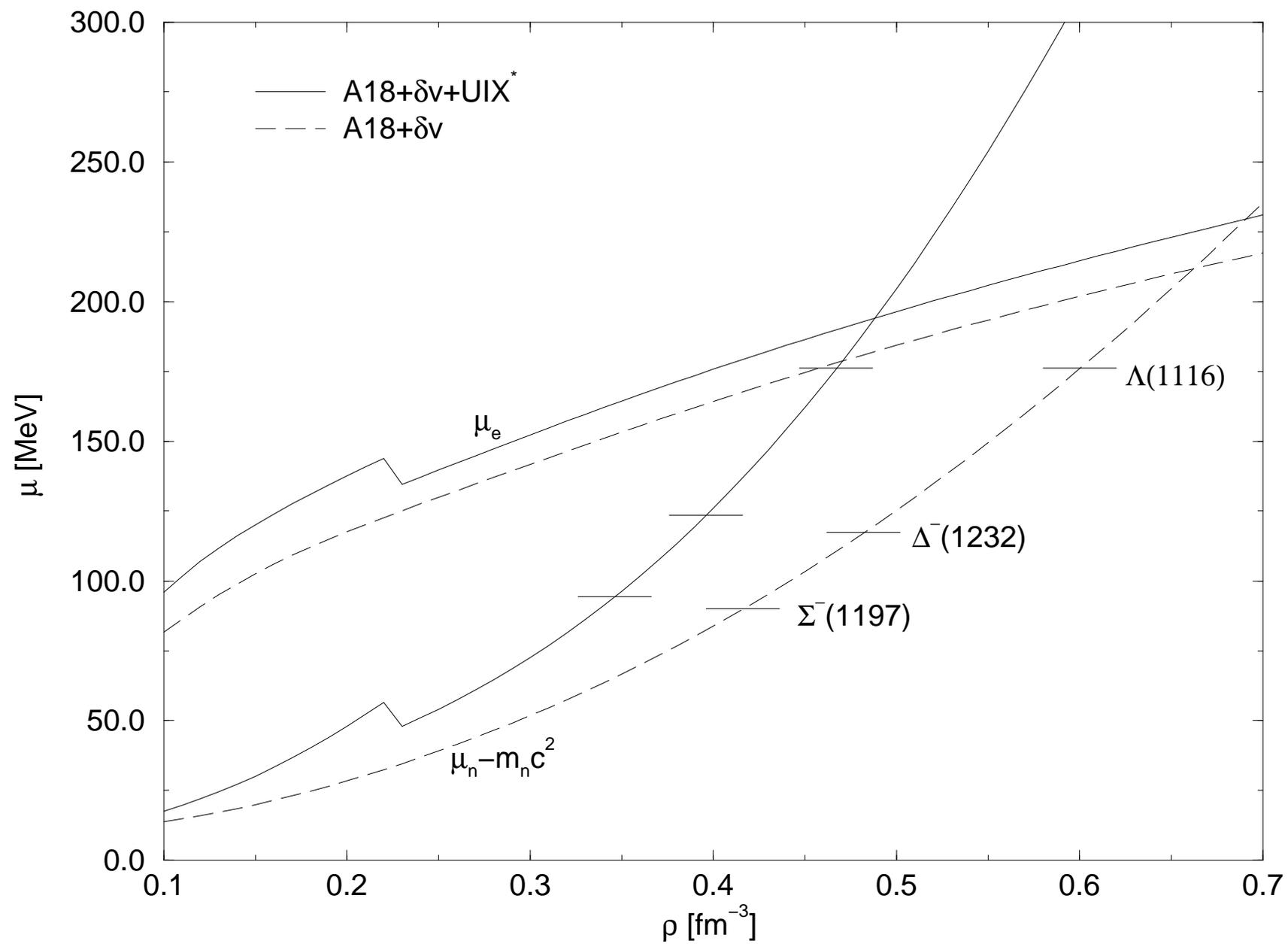
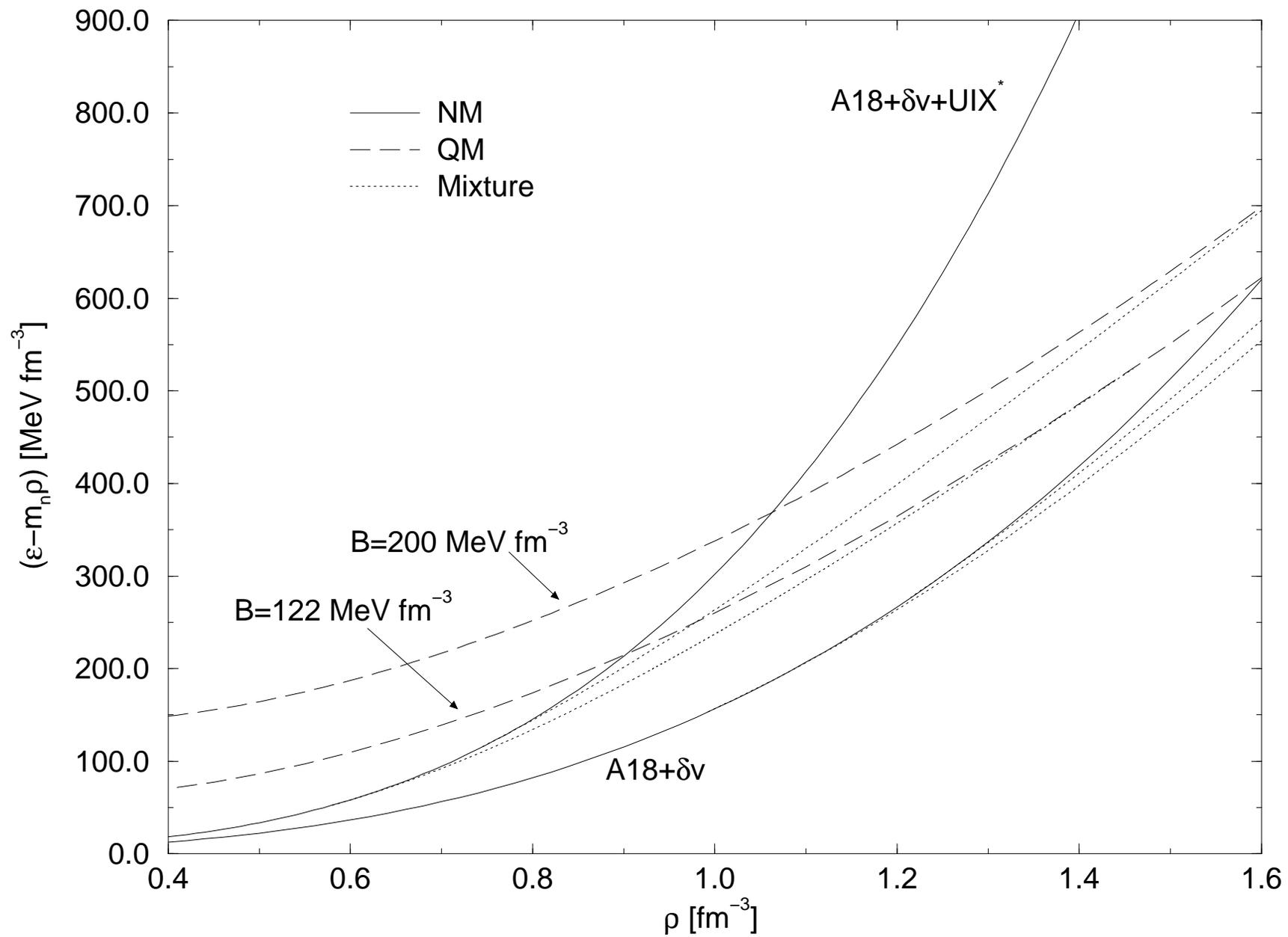
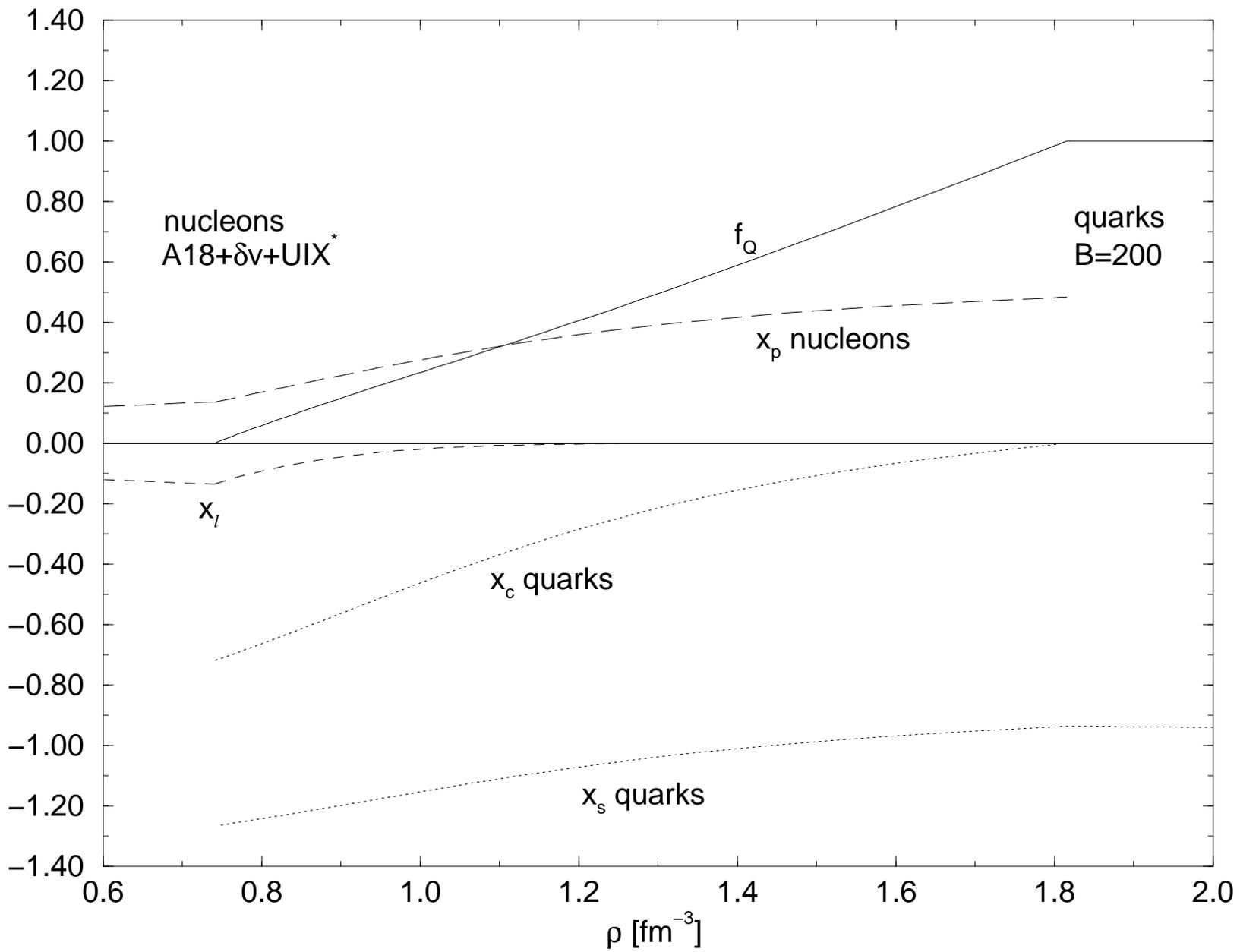
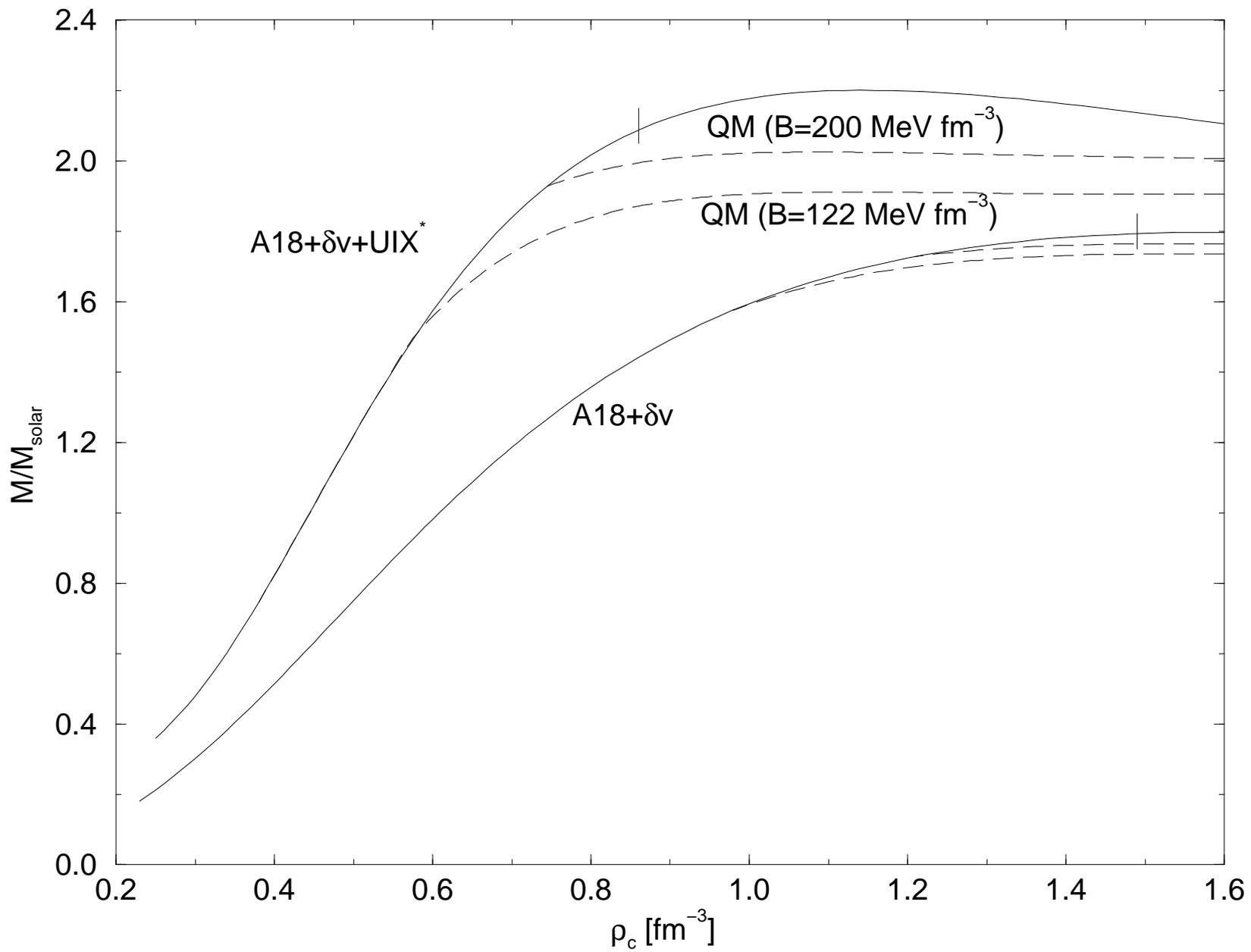
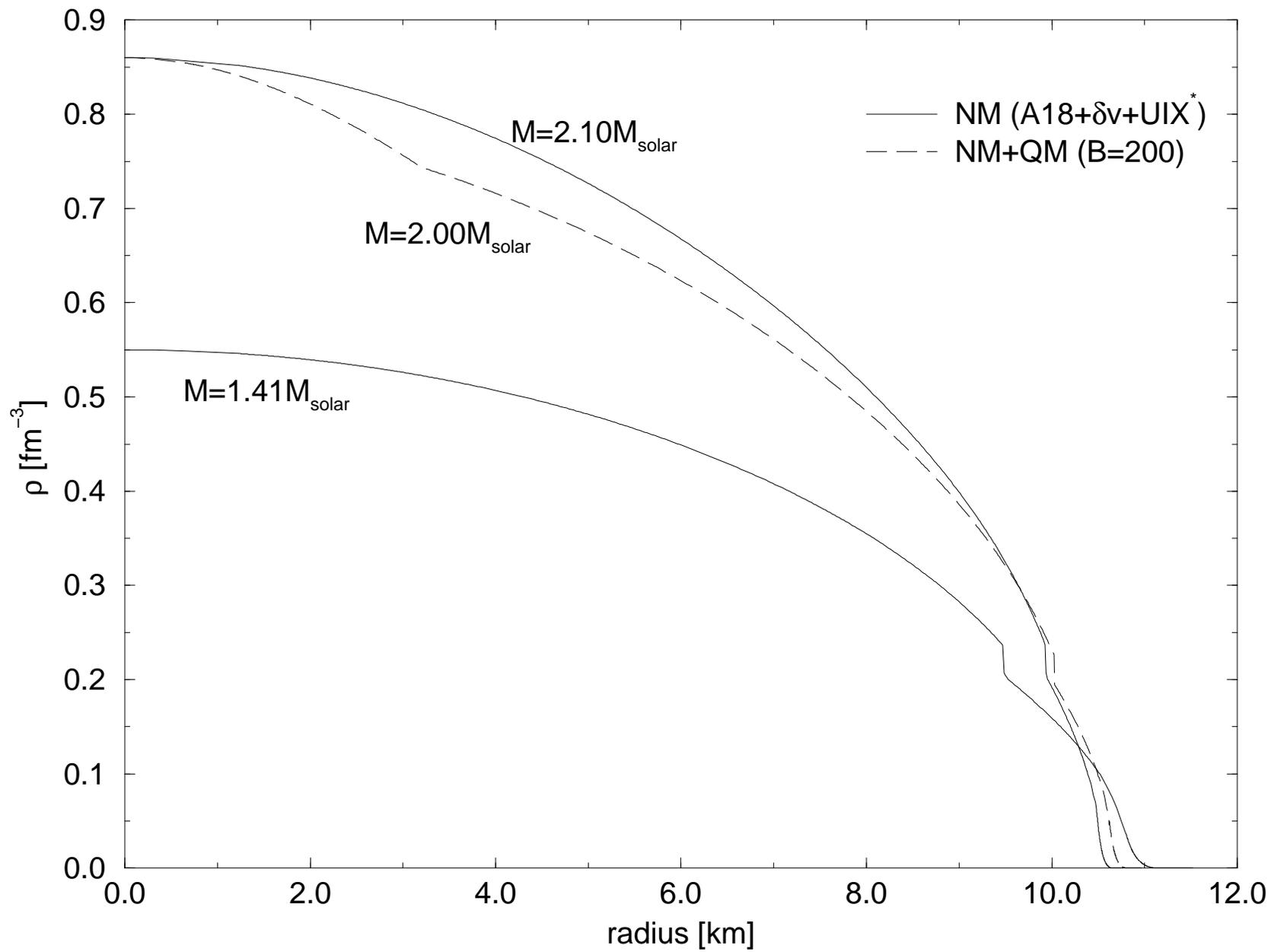
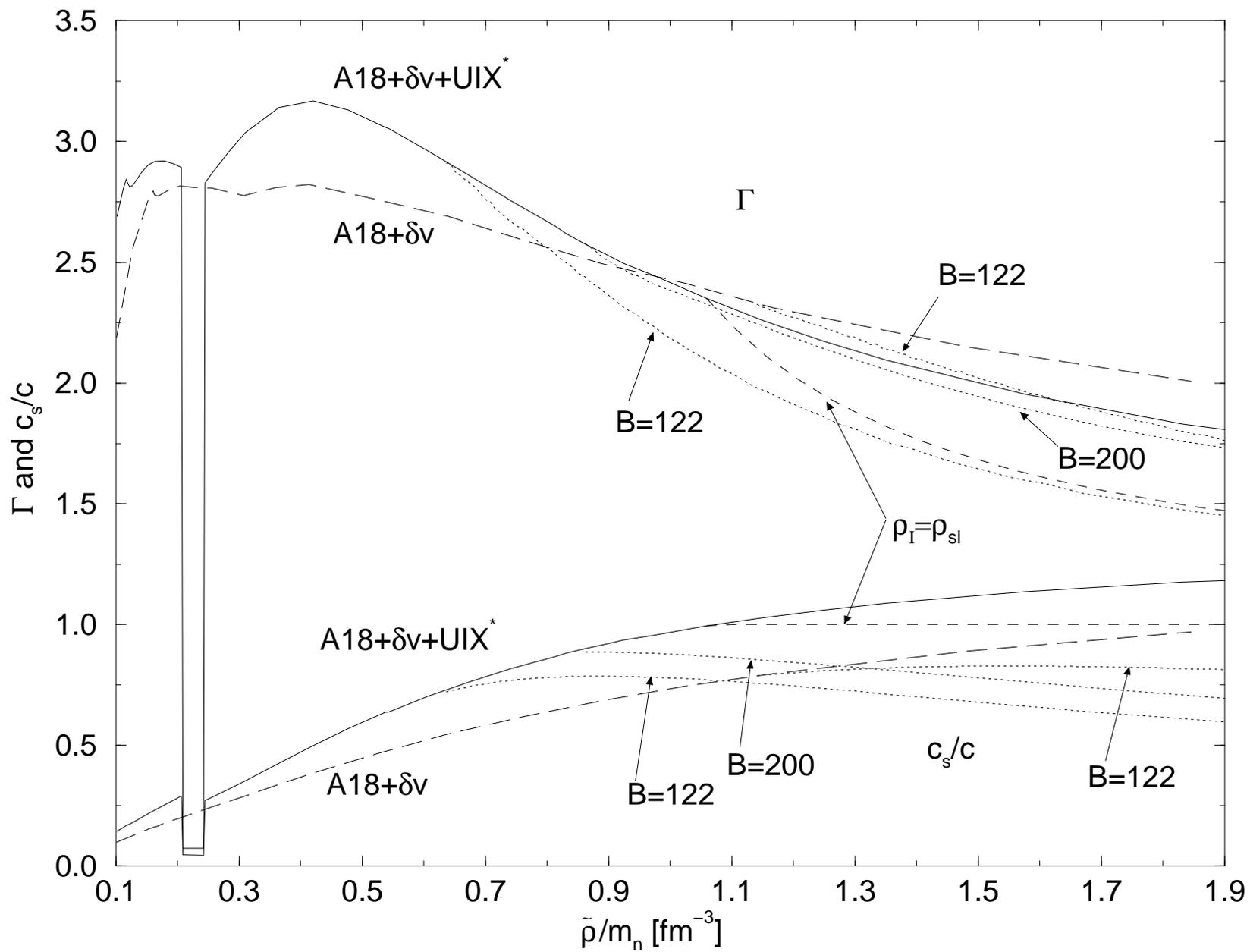
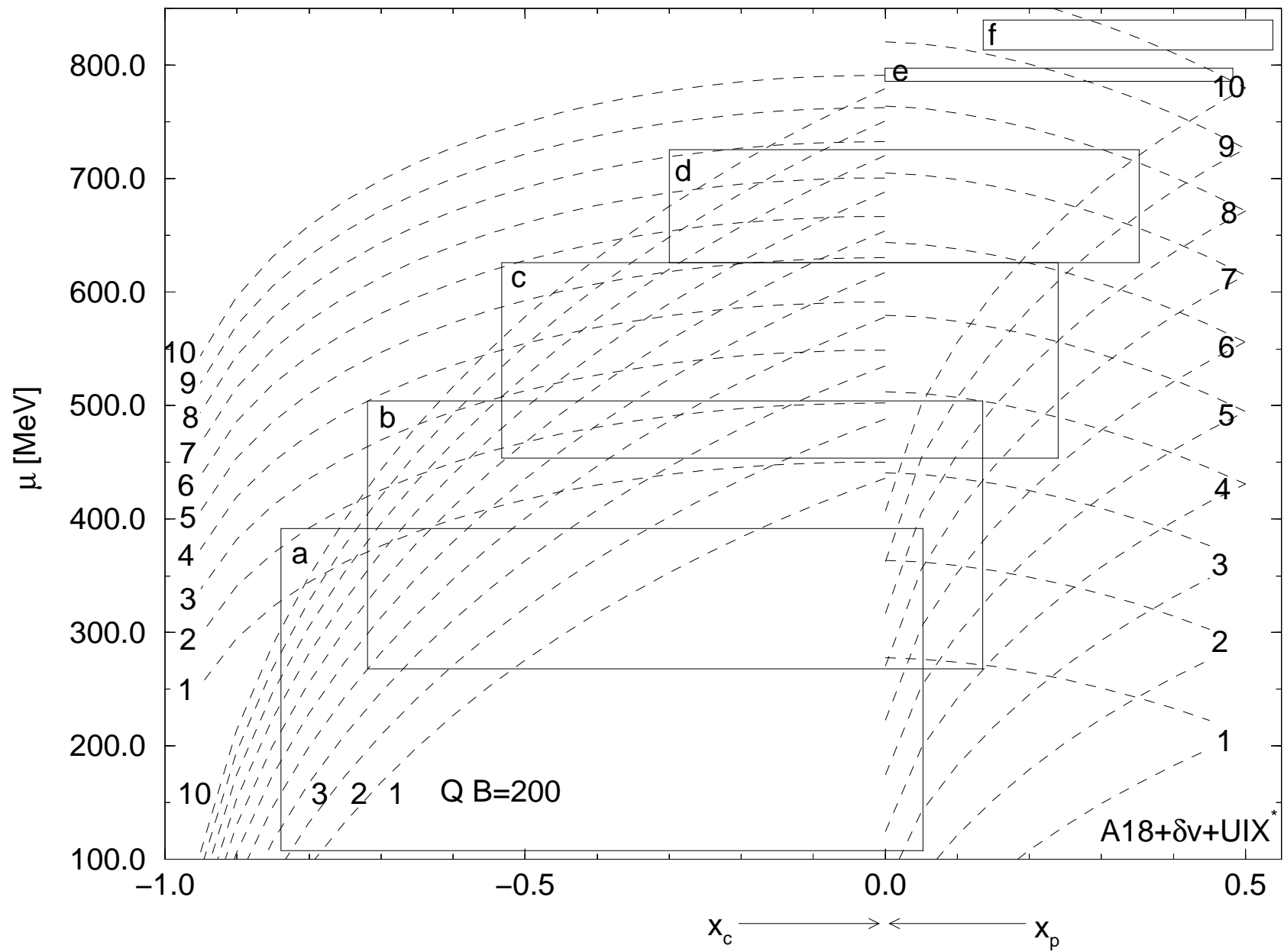
\end{document}